\documentclass[acmlarge,screen,authorversion]{acmart}            

\AtBeginDocument{%
  }

\setcopyright{cc}      
\setcctype{by}
\acmJournal{HEALTH}
\acmYear{2026} 
\acmVolume{} 
\acmNumber{} 
\acmDOI{10.1145/3810243}

\usepackage{subcaption}
\usepackage{url}
\usepackage{multirow}
\usepackage[inline]{enumitem}
\usepackage{bbding}
\usepackage{rotating}   
\usepackage{acro}
\usepackage[misc]{ifsym}

\usepackage{lscape}

\usepackage{chngcntr}                

\acsetup{list/display=used}  
\acsetup{patch/maketitle=false} 
\DeclareAcronym{hcp}{
  short = HCP,
  long  = healthcare professional
}

\newcommand{\AppNameNoSpace}{\textit{WoundAIssist}}
\newcommand{\AppName}[1][ ]{\textit{WoundAIssist}#1}
\newcommand{\tablesubsize}{\fontsize{7.2pt}{8.5pt}\selectfont}

\begin{document}

\title[WoundAIssist: An AI-Based Mobile Application for Remote Chronic Wound Care in Elderly Patients]{WoundAIssist: Development and Evaluation of an AI-Based Mobile Application for Remote Chronic Wound Care in Elderly Patients}

\author{Vanessa Borst {\small (\Letter)}}
\email{vanessa.borst@uni-wuerzburg.de}
\orcid{0009-0004-7123-7934}
\affiliation{%
  \institution{Institute of Computer Science, University of W{\"u}rzburg }
  \department{Chair of Computer Science II - Software Engineering}
  \city{W{\"u}rzburg}
  \country{Germany}
}
\author{Anna Riedmann}
\email{anna.riedmann@uni-wuerzburg.de}
\orcid{0000-0002-1969-8563}
\affiliation{%
  \institution{Institute of Computer Science, University of W{\"u}rzburg }
  \department{Chair of Computer Science V - Socially Interactive Agents}
  \city{W{\"u}rzburg}
  \country{Germany}
}
\author{Tassilo Dege}
\email{Dege\_T@ukw.de}
\orcid{0000-0001-6158-9048}
\affiliation{%
  \institution{University Hospital of W{\"u}rzburg }
  \department{Department of Dermatology, Venereology and Allergology}
  \city{W{\"u}rzburg}
  \country{Germany}
}
\author{Konstantin M{\"u}ller}
\email{konstantin.mueller@uni-wuerzburg.de}
\orcid{0000-0001-6540-3124}
\affiliation{%
  \institution{Institute of Computer Science, University of W{\"u}rzburg }
  \department{Chair of Computer Science II - Software Engineering}
  \city{W{\"u}rzburg}
  \country{Germany}
}
\author{Astrid Schmieder}
\email{Schmieder\_A@ukw.de}
\orcid{0000-0002-6421-9699}
\affiliation{%
  \institution{University Hospital of W{\"u}rzburg }
  \department{Department of Dermatology, Venereology and Allergology}
  \city{W{\"u}rzburg}
  \country{Germany}
}
\author{Birgit Lugrin}
\email{birgit.lugrin@uni-wuerzburg.de}
\orcid{0000-0002-2362-0080}
\affiliation{%
  \institution{Institute of Computer Science, University of W{\"u}rzburg }
\department{Chair of Computer Science V - Socially Interactive Agents}
  \city{W{\"u}rzburg}
  \country{Germany}
}
\author{Samuel Kounev}
\email{samuel.kounev@uni-wuerzburg.de}
\orcid{0000-0001-9742-2063}
\affiliation{%
  \institution{Institute of Computer Science, University of W{\"u}rzburg }
  \department{Chair of Computer Science II - Software Engineering}
  \city{W{\"u}rzburg}
  \country{Germany}
}

\renewcommand{\shortauthors}{Borst et al.}

\begin{abstract} 

The rising prevalence of chronic wounds, especially in aging populations, presents a significant healthcare challenge due to prolonged hospitalizations, elevated costs, and reduced patient quality of life. Traditional wound care is resource-intensive, requiring frequent in-person visits that strain both patients and \acp{hcp}.
Thus, we present \mbox{\textbf{\AppNameNoSpace}}, a patient-centered, AI-driven mobile application supporting telemedical wound care. \AppName enables patients to document wounds at home via photographs and questionnaires, while physicians remain engaged in the care process through remote monitoring and video consultations.
A distinguishing feature is an integrated lightweight deep learning model for on-device wound segmentation, guiding users during image capture. Combined with patient-reported data and server-side AI analysis, this enables continuous monitoring of wound healing progression. 
Developed through an iterative, user-centered process involving patients and domain experts, \AppName prioritizes an user-friendly design, particularly for elderly patients.
A conclusive usability study with patients and dermatologists reported excellent usability, good app quality, and favorable perceptions of the AI-driven wound recognition.
Our main contribution is two-fold: (I) the development and (II) evaluation of \AppNameNoSpace, an easy-to-use yet comprehensive telehealth solution designed to bridge the gap between patients and \acp{hcp}. 
Additionally, we synthesize design insights for remote patient monitoring apps, derived from over three years of interdisciplinary research, that may inform the development of similar digital health tools across clinical domains.

\end{abstract}



\maketitle

\makeatletter
\fancyfoot[RO,LE]{%
	\footnotesize
	\textcopyright\ 2026 Owner/author(s). Author's version. Definitive Version of Record: ACM Transactions on Computing for Healthcare, \href{https://doi.org/10.1145/3810243}{10.1145/3810243}%
}
\makeatother

\thispagestyle{empty}

\section{Introduction} 
\label{sec:intro}

Demographic changes and the aging population in industrialized nations pose numerous challenges for society and healthcare systems. 
A key concern is the growing prevalence of chronic wounds, particularly in elderly patients with multiple comorbidities~\cite{gould2015chronic}, as well as in individuals with conditions like diabetes and obesity~\cite{chandan2023humanWounds}. 
In developed countries, it has been estimated that 1-2\% of the population will experience a chronic wound during their lifetime~\cite{sen2009human}. In the U.S. alone, nearly seven million individuals live with a chronic wound, corresponding to approximately one in four families being affected~\cite{Healogics_wound_care_awareness}. Similar high prevalence patterns are reported internationally; for instance, in the United Kingdom, an estimated 3.8 million patients receive wound care by the National Health Service each year, representing nearly 7\% of the UK adult population~\cite{isla_NHS_statistics}. 

Chronic wounds impose a substantial burden on patients due to accompanying pain, exudate, and limited mobility, which can result in diminished quality of life and psychological distress, including stress, anxiety, and depression~\cite{herberger2011quality,renner2017depression}. 
When inadequately treated, chronic wounds may progress to severe complications such as infection and amputation, with reported amputation risks of up to 30\% in untreated cases~\cite{anx_fast_facts_on_wound_care}.
Beyond their impact on patients' physical and mental well-being, chronic wounds require continuous, time-intensive treatment and frequent clinical assessments.
Such assessments commonly rely on manual measurement techniques, most often using rulers or metric tapes to estimate wound area by multiplying length and width---an approach that systematically overestimates wound size by assuming a regular rectangular geometry that does not reflect the typically irregular shape of wounds~\cite{langemo2008measuring}.
The challenge of imprecise wound assessment is further compounded by the substantial economic burden associated with chronic wound management,  driven by personnel-related costs, including wound care specialists and patient transportation, as well as material expenses such as dressings and systemic drugs~\cite{purwins2010cost}. 
For example, chronic wound care is estimated to cost the U.S. healthcare system up to 20 billion USD annually, with treatment costs for advanced wounds exceeding 50 billion USD~\cite{anx_fast_facts_on_wound_care}.
Lastly, comprehensive and professional treatment may not be consistently available in all regions, particularly in remote areas or due to growing staff shortages, which further exacerbates patient suffering and widens the gaps in care.

These challenges underscore the urgent need for innovative solutions that can alleviate the burden on healthcare systems while enhancing both accessibility and quality of care. In this context, the rise of mobile health (mHealth) technologies presents a promising opportunity to transform wound care through digital applications. Recent years have seen substantial advancements in smartphone hardware, particularly in processing power and camera capabilities. In parallel, breakthroughs in Artificial Intelligence (AI), including Deep Residual Neural Networks~\cite{he2016deep}, the Transformer architecture~\cite{vaswani2017attention}, and Vision Transformers (ViTs)~\cite{dosovitskiy21ViT}, have significantly broadened the scope of AI-driven healthcare solutions.
%
Despite this potential, there remains a notable gap in patient-centered wound care applications that foster active patient involvement while supporting seamless collaboration with healthcare professionals (\acp{hcp}). Most existing digital wound care tools are tailored primarily for clinical use by \acp{hcp}~\cite{dege2024,kabir2024mobile}, offering limited support for patient self-management. Moreover, they often lack user-friendly interfaces, which is particularly challenging for older adults who represent a large portion of the target population. 
From a technical standpoint, comparatively little research has focused on developing robust yet computationally efficient AI models for automated wound segmentation that can be run directly on mobile devices. Despite this, such models have the potential to guide users during image acquisition and, in principle, support remote monitoring by providing automated wound size estimation and longitudinal tracking.


To address these gaps, we present the \mbox{(I) development} and the \mbox{(II) evaluation} of \AppNameNoSpace, a patient-centered mobile application for AI-assisted chronic wound monitoring at home, designed with a focus on elderly patients.

%
Unlike many existing mHealth apps, \AppName was developed through an iterative, user-centered design process. 
Specifically, \acp{hcp} were engaged from the outset to ensure clinical relevance, and the system was subsequently refined through a usability study involving patients with chronic wounds to systematically align clinical requirements with user needs.
%
The resulting system supports regular wound documentation through smartphone-based image capture and targeted questionnaires. A central feature is the integration of a lightweight AI model for on-device wound segmentation, designed to assist users in capturing consistent, high-quality images that enable subsequent server-side wound size estimation.
\AppName further combines these automated image analysis and structured self-reporting functionalities with additional features, including longitudinal wound and progress tracking as well as direct clinician communication via appointment scheduling and video calls.
%
Collectively, these components form an integrated telemedical workflow that links AI-assisted wound documentation with continuous remote monitoring and clinical oversight.
By prioritizing usability and approachability for the intended end users, including older adults, the system seeks to enhance patient engagement and self-management. Through reliable assessment of wound progression beyond manual measurements, it aims to reduce costs and logistical burdens associated with in-person visits and patient transportation, thereby supporting improved continuity of care.
%
To evaluate \AppNameNoSpace, we conducted two successive usability studies. 
\begin{enumerate}[label=(\Roman*), leftmargin=*, align=left, labelsep=0.2em]
    \item \textbf{Preliminary usability study (Study A)}: This study explored patient experiences with an early prototype of the app, co-developed with dermatologists and HCI researchers, to address the following questions: 
    \begin{itemize}[leftmargin=*, align=left]
        \item \textit{RQ-A1:} Are the core features of the app prototype easy to use and intuitive?
        \item \textit{RQ-A2:} How do patients perceive the app’s impact on care, feature scope, and potential for future use?
    \end{itemize}
    \item \textbf{Conclusive usability study (Study B)}: We evaluated the final version of \AppName with both patients and \acp{hcp} to investigate the following research questions, where RQ-B3 specifically examines whether the benefits anticipated by clinicians during the initial requirements phase align with patients’ experiences:
    \begin{itemize}[leftmargin=*, align=left]
        \item \textit{RQ-B1}: How do stakeholders perceive the overall usability and app quality of \AppNameNoSpace?
        \item \textit{RQ-B2}: How do stakeholders perceive AI-driven wound segmentation during image acquisition (live) and after image capture (a posteriori)?
        \item \textit{RQ-B3}: Do patients and physicians differ in their attitudes toward the app and its AI component?
    \end{itemize}
\end{enumerate}

\noindent 
Overall, our contributions are two-fold, corresponding to the (I) development and (II) evaluation of \AppNameNoSpace:
\begin{enumerate}[label=(\Roman*), leftmargin=*, align=left, labelsep=0.2em]
    \item The \textbf{\AppName app} itself, representing a methodological contribution to AI-assisted healthcare design through the iterative development of a patient-centered, AI-driven mobile app for telemedical wound care.
    \item \textbf{Empirical insights} from two usability studies, analyzing the app’s usability, acceptance, and potential barriers to adoption, with particular attention to older adults’ interaction with the system. This includes user perceptions---both patients’ experiences and physicians’ evaluations---of AI-guided wound imaging.
\end{enumerate}

As an ancillary contribution, we summarize lessons learned for the design of patient-centered remote disease monitoring apps, based on two usability studies and over three years of development experience. Drawing on feedback from dermatologists and patients, these insights aim to guide the design of comparable mHealth tools.

\section{Related Work}
\label{sec:related_work}
\subsection{Mobile Health Applications: Current Landscape and Usability Considerations for Elderly Users} \label{sec:rw-mapps}
The adoption of mHealth apps in healthcare has surged over the past decade, accelerated by the COVID-19 pandemic and the increasing prevalence of electronic devices in daily life~\cite{haggag2022large}. 
Becoming increasingly ubiquitous, they now cover a wide range of areas, including fitness, lifestyle management, nutrition, medication adherence, disease management, women's health, and healthcare providers/payors~\cite{haggag2022large}. 
The latter categories are particularly valuable for supporting disease management~\cite{hamine2015impact}, enabling patient self-care~\cite{anderson2016mobile}, and allowing remote monitoring by \acp{hcp}, thereby reducing the need for in-person consultations~\cite{kabir2024mobile}. 
Consequently, a variety of mHealth apps have been developed targeting conditions such as heart failure~\cite{athilingam2018mobile}, diabetes~\cite{geirhos2022standardized}, and depression~\cite{molloy2021engagement}. In dermatology, applications address areas like skin cancer diagnosis~\cite{zaidan2018review}, psoriasis~\cite{lull2022german}, and wound care~\cite{kabir2024mobile,dege2024}. 

With the growing diversity and clinical relevance of mHealth solutions, usability---regarded as a key factor for evaluating the quality of interaction with a particular system~\cite{ISO9241-11}---has emerged as a critical determinant of their successful adoption. In mobile contexts, usability specifically encompasses an application's efficiency, effectiveness, and ability to provide a satisfying experience~\cite{Weichbroth2020}.
Developing mHealth apps for a heterogeneous target group requires careful consideration of target group-specific design peculiarities to ensure usability. This is particularly important for older adults, a demographic often characterized by low adoption and usage rates---frequently attributed to inappropriate design choices~\cite{liu2021} and age-related barriers to mHealth engagement~\cite{ramdowar2024}.

The latter can be grouped into four categories: perceptual (visual and auditive), physical, cognitive, and motivational~\cite{elguerapaez2019,ramdowar2024,wildenbos2018}.
Visual barriers occur when reduced visual acuity impairs the perception and interpretation of small or unclear interface elements (e.g., icons, images, charts, buttons)~\cite{elguerapaez2019,ramdowar2024}. 
Auditory barriers arise from diminished hearing, which may negatively affect user experience and satisfaction with mHealth apps~\cite{ramdowar2024,wildenbos2018}. 
Physical barriers stem from age-related declines in motor skills, including slower or stiffer movements and reduced reflexes~\cite{ramdowar2024,wildenbos2018}, which makes precise interactions and fast or repetitive actions such as scrolling more difficult~\cite{elguerapaez2019,ramdowar2024}. Cognitive barriers result from declines in memory, attention, and processing speed, making it harder for older users to navigate mHealth apps~\cite{ramdowar2024,wildenbos2018}. These difficulties are amplified by unclear or ambiguous graphics and icons, highlighting the need for clear navigation, consistent terminology, and minimal interaction steps~\cite{elguerapaez2019,ramdowar2024}. 
Motivational barriers arise from limited digital literacy or low confidence in using the app, which can reduce users' engagement~\cite{ramdowar2024,wildenbos2018}. Although such age-related barriers may limit mHealth app adoption, tailoring applications to overcome these challenges can enhance older adults’ willingness to use them~\cite{ramdowar2024}.

To address this, app features, including button design, help and explanation functions, and visual aspects such as color contrast, must be tailored to the needs of the intended users~\cite{isakovic2016}.
As summarized by Liu et al.~\cite{liu2021}, interface design for mHealth apps targeting older adults should account for vision impairment (e.g., adjusted font size), motor coordination problems (e.g., easy input mechanics such as large buttons), and cognitive and memory decline (e.g., simple navigation and consistent layout).
Accordingly, mHealth usability should be systematically assessed in the target population, ideally combining subjective and objective measures with task-based metrics (e.g., task completion) for a comprehensive assessment~\cite{broekhuis2019}.
However, a review by Wang et al.~\cite{wang2022usability} found that 37\% of studies relied on a single evaluation method, contrary to this recommendation.
%
Beyond these methodological limitations in usability assessment, existing mHealth solutions often fall short in the following key areas:

\begin{enumerate}[leftmargin=*]
    \item \textbf{Limited involvement of \acp{hcp}}: 
    Active participation of \acp{hcp} in mHealth app development can be crucial for their adoption in practice. 
    For instance, involving urologists in urology app development is likely to increase download rates~\cite{pereira2016expert}, yet expert participation remains low: only 13.4\% of apps for the general population involved HCPs and 1.9\% involved urological associations~\cite{pereira2015mhealth}.
    Similar trends occur in other fields:
    Only 35\% (7/20) of rheumatoid arthritis apps involved \acp{hcp} during development~\cite{luo2019mobile}, and only 48.8\% (81/166) of apps for cancer patients were developed by healthcare organizations, raising concerns about patient safety~\cite{collado2016smartphone}. 
    \item \textbf{Limited involvement of patients}:     
    User-centered design approaches actively engage patients throughout development~\cite{mao2005}. They are therefore crucial for effective mHealth technologies, as they can increase intention to use, improve care effectiveness and efficiency, and enhance user satisfaction~\cite{or2022human}.
    Nonetheless, many mHealth apps still fall short in engaging patients during the design phase. A secondary analysis of 1,595 health apps revealed that only 8.71\% (139/1,595) incorporated user participation during development~\cite{frey2023association}. 
    For example, only 15\% (5/32) of apps for rheumatic and musculoskeletal conditions involved patients~\cite{najm2019mobile}, resulting in tools that may be difficult to use and fail to meet patient needs~\cite{verhoeven2010asynchronous}.
    Limited involvement is particularly problematic for older adults, as actively engaging them in technology design can enhance understanding of their needs and daily lives, inform design adjustments, and increase their sense of participation~\cite{fischer2020importance}.

\end{enumerate}

\subsection{Wound Assessment in Practice: Measurement Techniques and mHealth Tools} 

Clinical wound assessment typically relies on manual measurement techniques, each with specific advantages and limitations. Consistent use of the same method over time is generally recommended for longitudinal monitoring and evaluation of treatment response~\cite{nichols2015wound}.
The most common approach estimates wound area by multiplying length and width using a ruler or metric tape, although this method is known to systematically overestimate actual wound size due to irregular wound shapes~\cite{shaw2011wound}.
Additionally, variation in how clinicians define length and width, as reported by Langemo et al.~\cite{langemo2008measuring}, highlights a lack of standardization and introduces subjectivity, which may compromise measurement reliability and comparability across assessments.
Apart from the \textit{ruler} method, \textit{tracing} involves outlining the wound perimeter on transparent acetate film and estimating area by counting enclosed square centimeters on a metric grid.
While this may improve measurement reliability, it remains subjective---particularly in delineating wound boundaries and handling partial grid units~\cite{shaw2011wound,langemo2008measuring}. Moreover, it requires direct physical contact with the wound, which can cause patient discomfort and introduces a risk of contamination and potential infection.
The \textit{Kundin device}, a disposable plastic-coated three-dimensional gauge that quantifies wound area and volume using a mathematical model, is similarly invasive~\cite{langemo2008measuring,shaw2011wound}.

To overcome the limitations of manual wound measurement, various digital and software-based approaches have been introduced to support non-invasive and more standardized wound assessment.
%
One such approach is 
\textit{computerized planimetry of digital images}, which analyzes wound photographs using specialized software.
Calibration is performed by including a reference object of known dimensions (e.g., a ruler) in the same plane as the wound and aligned perpendicularly to the camera lens axis~\cite{foltynski2015wound}.
Once calibrated, the software calculates wound area based on the established scale. However, precise delineation of wound boundaries typically still requires manual interaction via a mouse or touch-enabled input device~\cite{foltynski2015wound,jorgensen2016methods,mayrovitz2009wound}.
%
%
Another technique, \textit{computerized stereophotogrammetry}, uses at least two overlapping stereoscopic photographs taken from known angles to reconstruct a three-dimensional model of the wound surface without 
physical contact. 
Similar to conventional digital planimetry, wound boundaries are typically delineated manually, after which specialized software calculates wound dimensions, including area and volume~\cite{shaw2011wound, langemo2008measuring}.  Although this method is quite accurate, it requires specialized stereographic imaging equipment at the point of care and is time-consuming~\cite{shaw2011wound, langemo2008measuring}.

Driven by the emergence of mHealth, an abundance of wound care apps has been developed, some of which aim to streamline the time-intensive process of traditional wound measurement techniques.
Rather than reviewing all solutions, we focus on key limitations that distinguish existing approaches from our proposed \AppName app:

\begin{enumerate}[leftmargin=*]
    \item \textbf{Lack of patient-centered wound care apps}: 
    Most existing wound care applications, such as the Swiss \mbox{\textit{imitoWound}~\cite{imitoWound}}, Canadian \textit{Swift Skin and Wound}~\cite{swiftWoundCareApp}, and U.S.-based \textit{Tissue Analytics}~\cite{tissueAnalytics}, are primarily designed for \acp{hcp}, not patients. 
    This gap is highlighted by two recent 2024 reviews~\cite{kabir2024mobile,dege2024}.:
    In our analysis of 73 chronic wound care apps available in the German Google Play and Apple App Stores, only ten apps explicitly targeted patients~\cite{dege2024}. Of these, seven included advertisements or required payment, potentially reducing usability and accessibility for financially limited users.
    Similarly, Kabir et al.~\cite{kabir2024mobile} analyzed ten wound assessment and monitoring apps 
    from an initial pool of 170; all targeted practitioners and physicians~\cite{kabir2024mobile}. 
    The authors emphasized the shortage of patient-oriented solutions and their importance for enhancing communication with \acp{hcp}, enabling remote monitoring, and improving patient satisfaction.
    Clinician-focused tools may be overly complex or inaccessible for lay users, creating barriers to patients' self-management~\cite{kabir2024mobile}.

    \item \textbf{Lack of methodological transparency}:
    While Kabir et al. noted that most wound care apps lack automated wound contour detection~\cite{kabir2024mobile}, Griffa et al.~\cite{griffa2024artificial} reviewed ten apps offering automated AI-based ulcer segmentation. They found that most, including commercial solutions such as \textit{imitoWound}~\cite{imitoWound}, \textit{Wound Vision}~\cite{woundVision}, \textit{CARES4WOUNDS}~\cite{cares4wounds}, and \textit{Tissue Analytics}~\cite{tissueAnalytics}, did not disclose essential methodological details.
    These omissions included key aspects of their (proprietary) segmentation and measurement techniques, such as the algorithms used or the composition of training datasets, which limits reproducibility and independent validation.
    
    \newpage
    \item \textbf{Limited scope of existing patient-orientated apps}:
    Among the few apps designed for patients, most target narrow functions, such as documenting wound progress (e.g., \textit{APD Skin Monitoring}~\cite{wu2019APDSkinMonitoring}) or providing wound care information (e.g., \textit{Wound Education}~\cite{yeo2019woundEducation}). Some apps, such as the Austrian \textit{WUND APP}~\cite{WUND_APP}, offer broader features like questionnaire-based documentation and educational resources.
    \textit{Theia}, a post-operative wound surveillance app~\cite{shenoy2018deepwound}, provides AI-based wound \textit{classification} alongside manual tracking of medications, pain, weight, exercise, and dressing frequency, plus the ability to connect with doctors via message, email, or phone. 
    However, to the best of our knowledge, no other patient app offers features comparable to \AppNameNoSpace, including AI-based wound \textit{segmentation} for user guidance and server-side wound size estimation.
\end{enumerate}

\subsection{Mobile AI-Based Wound Segmentation in Practice and User Perception}
Semantic segmentation (SS) is a computer vision technique that classifies each pixel of an image into predefined categories. In the context of wound care, SS enables automated quantification of wound size by distinguishing wound tissue from surrounding skin and background. By assigning pixels to either wound or non-wound regions, SS has the potential to replace labor-intensive manual tracing methods, while enhancing measurement consistency, reproducibility, and objectivity.
However, despite substantial progress in SS, including architectures optimized for mobile inference, applications specifically targeting wound segmentation remain limited. Research in this area is sparse, and there is a notable lack of lightweight models suitable for mobile deployment. 
A comprehensive overview of (mobile) SS and current gaps in AI-based wound segmentation is provided in Section~\ref{appendix:ssec:mobile_semantic_segmentation}.
While our prior comparison of non-wound-specific architectures for mobile wound segmentation~\cite{borst2024early} partially addressed these limitations, further challenges persist in the practical use of mobile AI for app-based wound analysis:

\begin{enumerate}[leftmargin=*]
    \item \textbf{Absence of real-world deployment}: 
    Due to the dynamic nature of home-based wound monitoring, including factors such as varying lighting, camera types, and backgrounds, real-world evaluation is essential to validate the robustness of AI models. However, beyond our aforementioned qualitative comparison using a minimally functional app prototype that was limited to basic image acquisition~\cite{borst2024early}, few wound segmentation models have been tested on mobile devices. A notable exception is the \textit{AutoTrace} tissue segmentation model~\cite{ramachandram2022fully} in the commercial \textit{Swift Skin and Wound} app~\cite{swiftWoundCareApp}. 
    Most other approaches rely on traditional image processing rather than Deep Learning~\cite{rocha2021woundarch,varma2016vision,ferreira2021experimental} or require manual annotation by users~\cite{cazzolato2021utrack} as additional guidance.
    
    \item \textbf{Gap in user perception studies}: 
    There has been limited exploration into how mobile AI-driven wound segmentation is perceived by both patients and physicians. Mohammed et al.~\cite{mohammed2023implementing} conducted an observational cross-sectional study evaluating physicians' satisfaction with the commercial AI-based wound care management app \textit{Swift Skin and Wound}~\cite{swiftWoundCareApp} via an online survey.
    However, as far as we know, no comparable studies have examined patient-centered apps, leaving key questions regarding patient perceptions of AI-based wound recognition as well as a comparative analysis of attitudes between patients and \acp{hcp} largely unexplored.
\end{enumerate}

\subsection{Overall Conclusions from Related Work and Delimitation}

In response to persistent limitations identified above and in recent reviews of mHealth applications for wound care---namely the lack of patient-oriented solutions~\cite{kabir2024mobile, dege2024}, poor performance of existing tools in terms of usability and visual design~\cite{kabir2024mobile}, the frequent absence of Deep Learning–based wound boundary detection~\cite{kabir2024mobile}, and limited methodological transparency in AI-assisted wound assessment~\cite{griffa2024artificial}---this work presents the development and evaluation of a comprehensive, patient-centered wound care app.
To the best of our knowledge, \AppName is unique in its functional scope: it integrates smartphone-captured longitudinal wound images, AI-assisted wound segmentation to guide users, structured patient-reported outcomes (PROs), direct clinician interaction, and server-side AI-based wound size estimation within a single unified system.
The system was developed through an iterative, user-centered design process involving close collaboration with \acp{hcp} and patients with chronic wounds. In particular, age-related barriers to mHealth adoption were explicitly addressed through targeted design decisions. Although the integration of multiple functionalities has the potential to increase interaction complexity, careful attention was paid to balancing clinical utility with high usability.
By embedding a mobile-optimized Deep Learning segmentation model from our prior methodological work~\cite{borst2024early}, we aim to bridge the gap between methodological advances in AI research and routine wound care practice. This study further investigates whether advanced AI techniques can be effectively deployed within an intuitive, patient-centered mobile application. In particular, we examine stakeholder perceptions of the integrated AI component—focusing on perceived usefulness (PU) and perceived ease of use (PEU). For methodological transparency, all details of the AI-based wound segmentation are provided in Section~\ref{appendix:ssec:topformer_architecture}, Section~\ref{appendix:ssec:topformer_training_and_eval}, and Section~\ref{appendix:ssec:topformer_integration_into_app}.

\section{\textit{WoundAIssist}: An AI-Based Mobile Application for Remote Chronic Wound Care in Elderly Patients}
\label{sec:woundAIssist_teaser}

\subsection{Iterative Design and Evaluation Process}
\label{ssec:methodology_iterative_development}

\begin{figure*}[ht]
\centering
\includegraphics[width=\textwidth, clip]{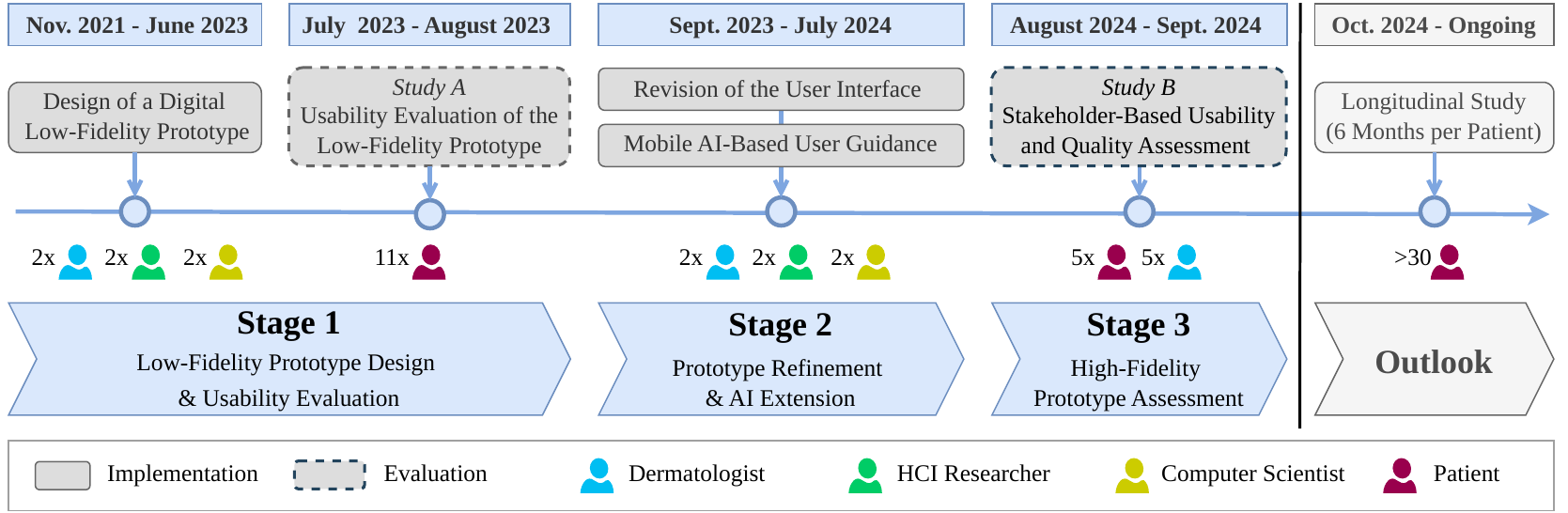}
\caption{Timeline for the derivation and assessment of our \AppName application.}
\label{fig:timeline_app_derivation}
\Description{A figure depicting the three key stages in the development and evaluation of the proposed \AppName mobile app, presented in sequence: 1) the design of a digital low-fidelity prototype and its usability evaluation, 2) the refinement of the prototype through UI revisions and the integration of a mobile AI model for user guidance during image capturing, and 3) the final stakeholder-based assessment of the high-fidelity prototype.}
\end{figure*}

Figure~\ref{fig:timeline_app_derivation} outlines the iterative design and evaluation process of \AppNameNoSpace.
In Stage 1, an initial digital prototype was developed in close collaboration with two dermatologists and two experts in human–computer interaction (HCI) and user-centered mobile design. 
Subsequently, usability was explored with patients with chronic wounds 
through semi-structured interviews (\textit{Study~A}), informing iterative refinements of the user interface. 
Moreover, a lightweight AI model for real-time wound segmentation was integrated to support patients in consistent image capture.
In the final stage (\textit{Study~B}), the high-fidelity prototype was evaluated with key stakeholders---dermatologists and patients---focusing on overall usability, app quality, and user perception of the integrated AI component.
Details on each stage are provided in Sections~\ref{sec:methodology:stage1},~\ref{sec:methodology:stage2}, and~\ref{sec:methodology:stage3}, following the system overview in Section~\ref{sec:technical_components}. 

\subsection{Main Functionalities and Visual Design}
\label{ssec:woundAIssist_teaser:functions}

\begin{figure*}[ht!]
    \centering
    \begin{subfigure}{0.21\textwidth}
        \centering
        \includegraphics[width=\linewidth]{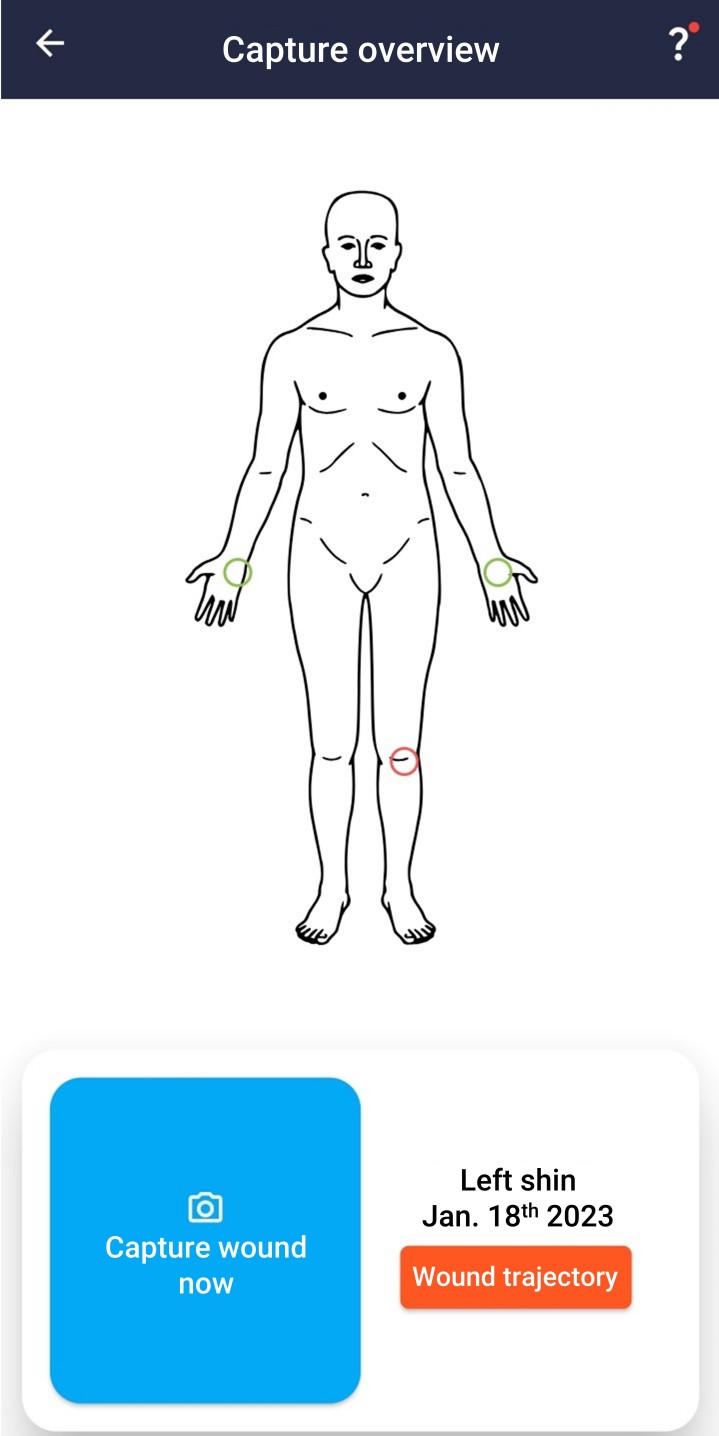}
        \caption{Wound localization}
        \label{fig:wound_localization}
    \end{subfigure} \hfill
    \begin{subfigure}{0.21\textwidth}
        \centering
        \includegraphics[width=\linewidth]{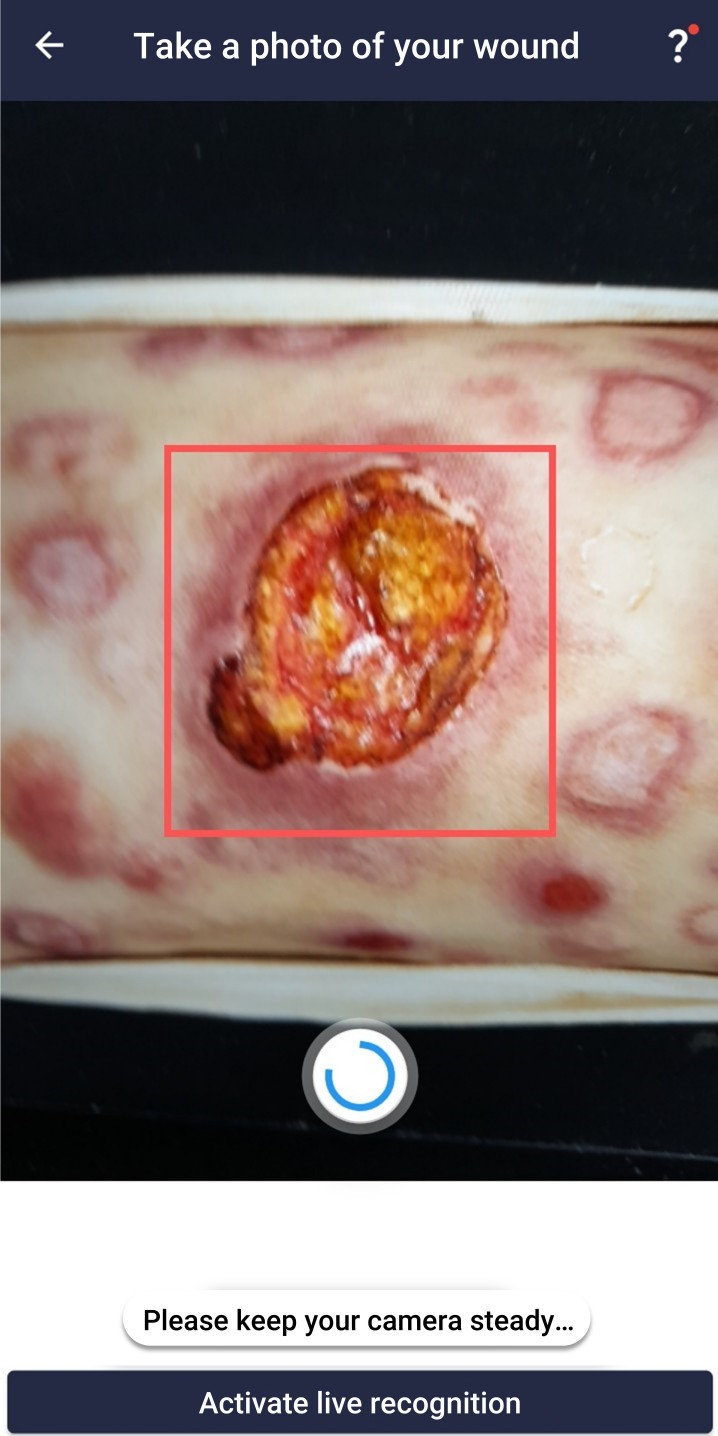}
        \caption{Camera screen}
        \label{fig:camera_screen}
    \end{subfigure} \hfill
    \begin{subfigure}{0.21\textwidth}
        \centering
        \includegraphics[width=\linewidth]{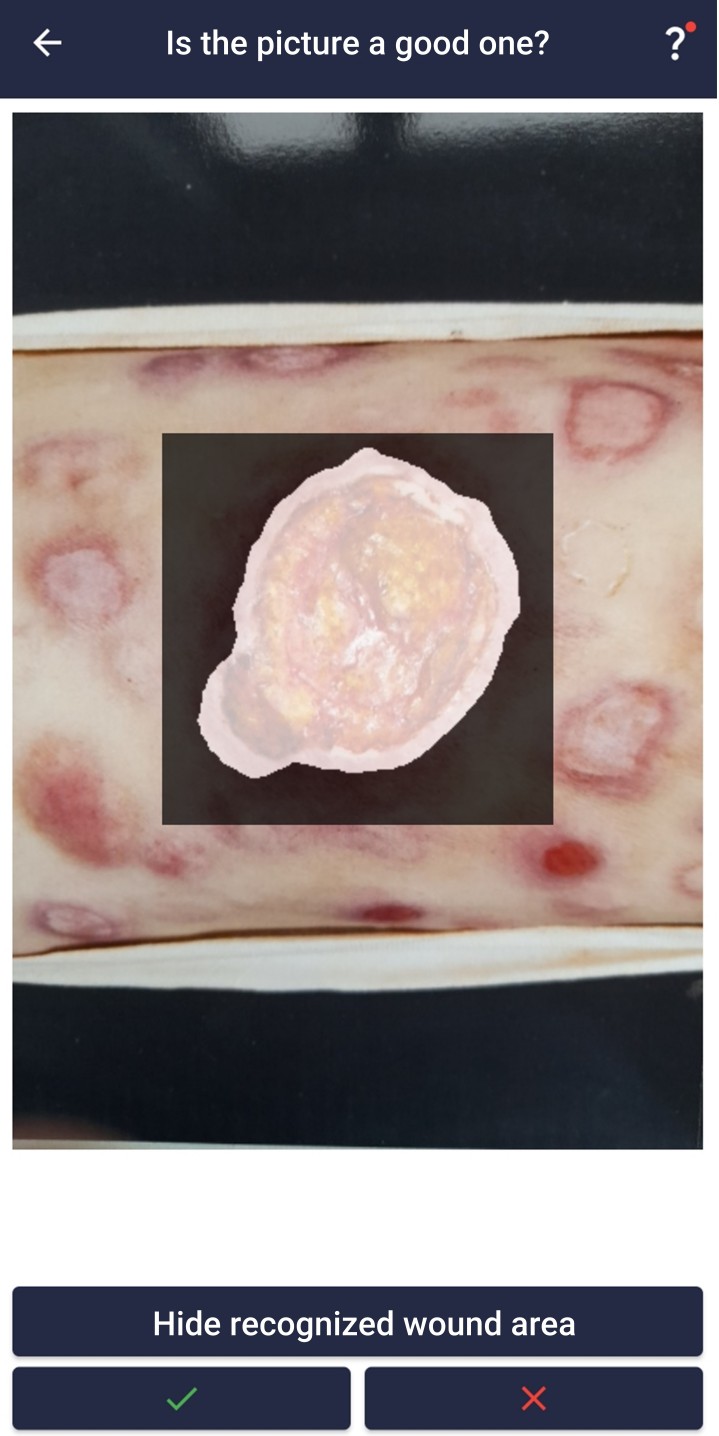}
        \caption{Recognized wound area}
        \label{fig:recognized_wound}
    \end{subfigure} \hfill
    \begin{subfigure}{0.21\textwidth}
        \centering
        \includegraphics[width=\linewidth]{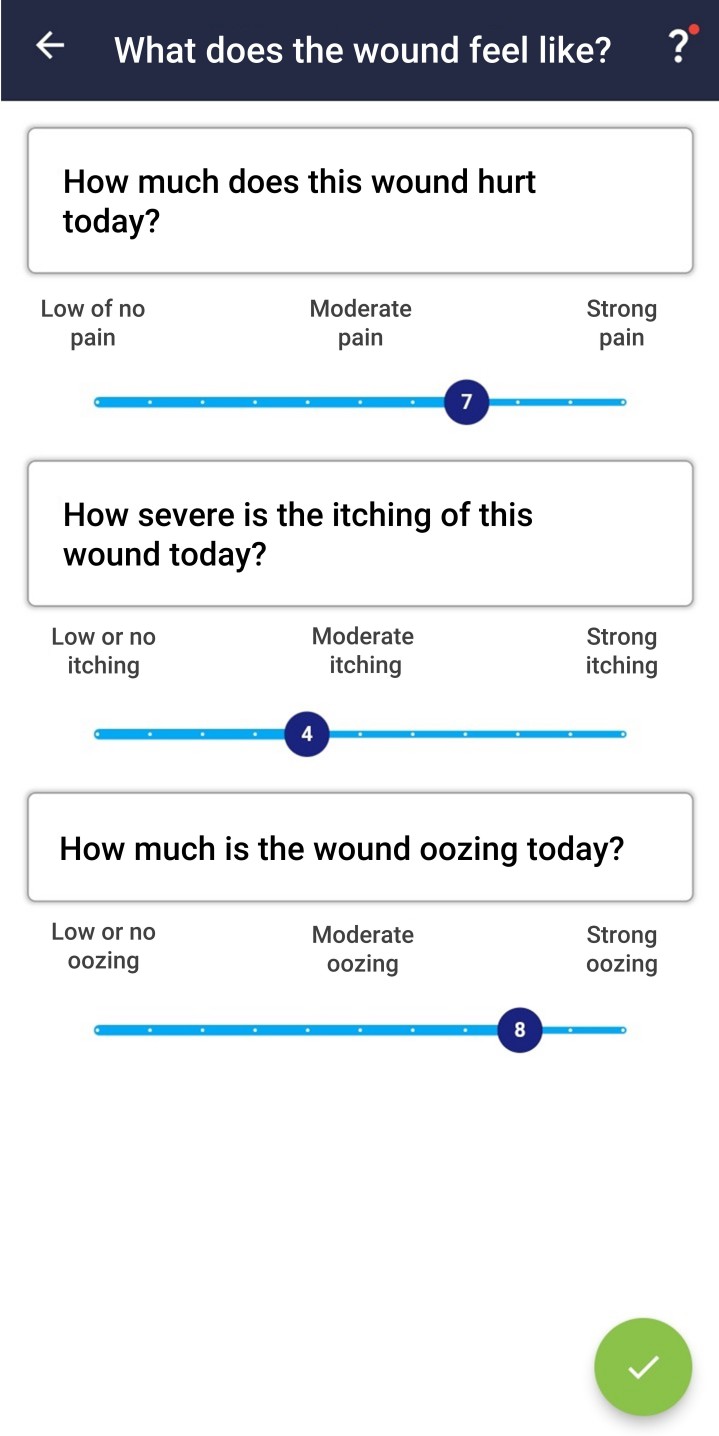}
        \caption{Wound questionnaire}
        \label{fig:wound_questionnaire}
    \end{subfigure} \\
    \vspace{0.5cm} 
    \begin{subfigure}{0.21\textwidth}
        \centering
        \includegraphics[width=\linewidth]{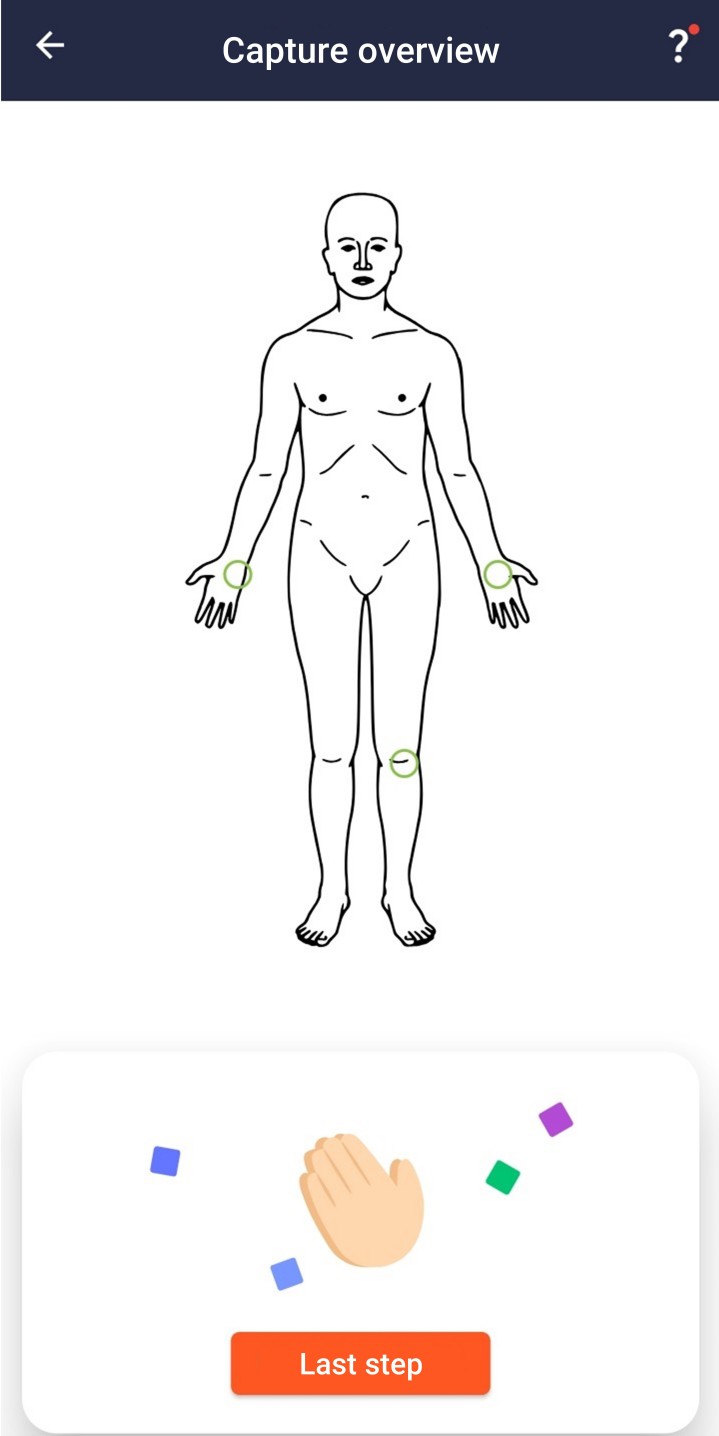}
        \caption{Interim motivation}
        \label{fig:interim_motivation}
    \end{subfigure} \hfill
    \begin{subfigure}{0.42\textwidth}
        \centering
        \includegraphics[width=0.5\linewidth]{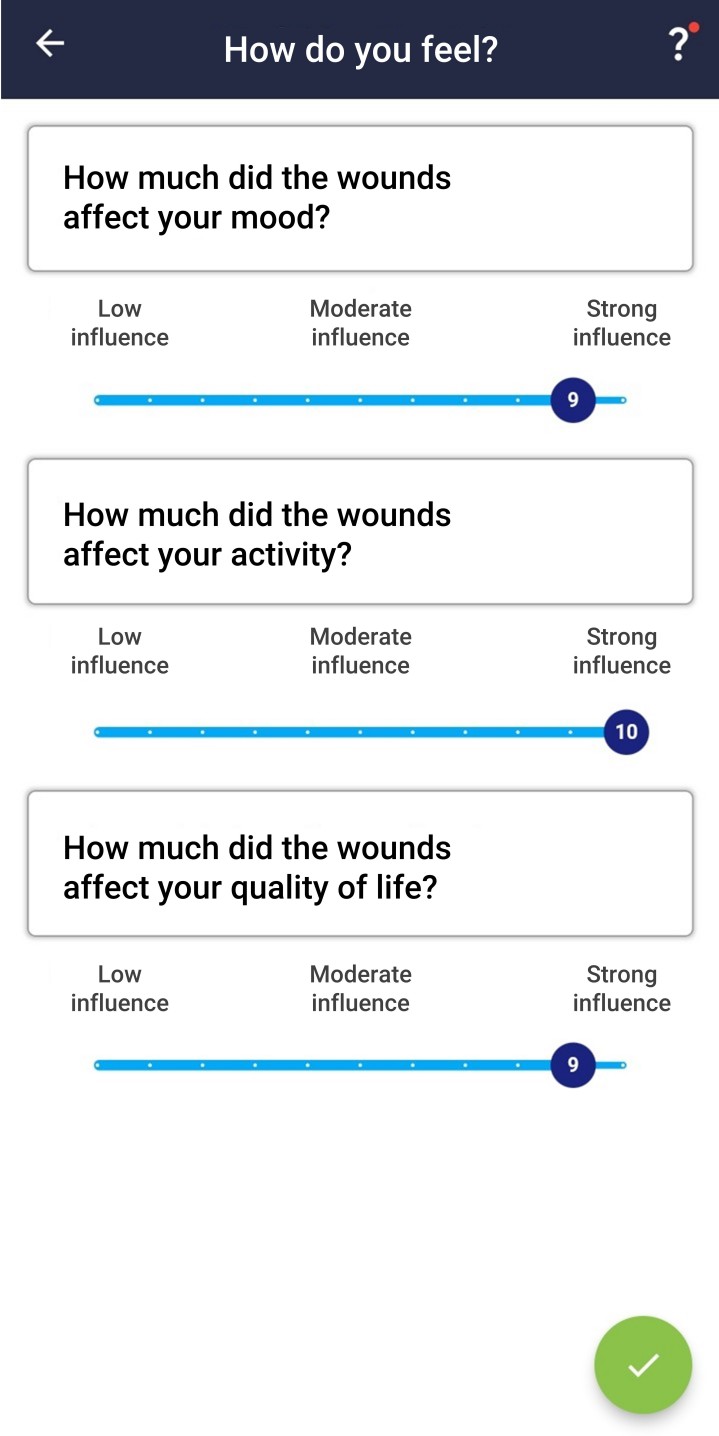}
        \caption{General health questionnaire}
        \label{fig:general_condition_questionnaire}
    \end{subfigure} \hfill
    \begin{subfigure}{0.21\textwidth}
        \centering
        \includegraphics[width=\linewidth]{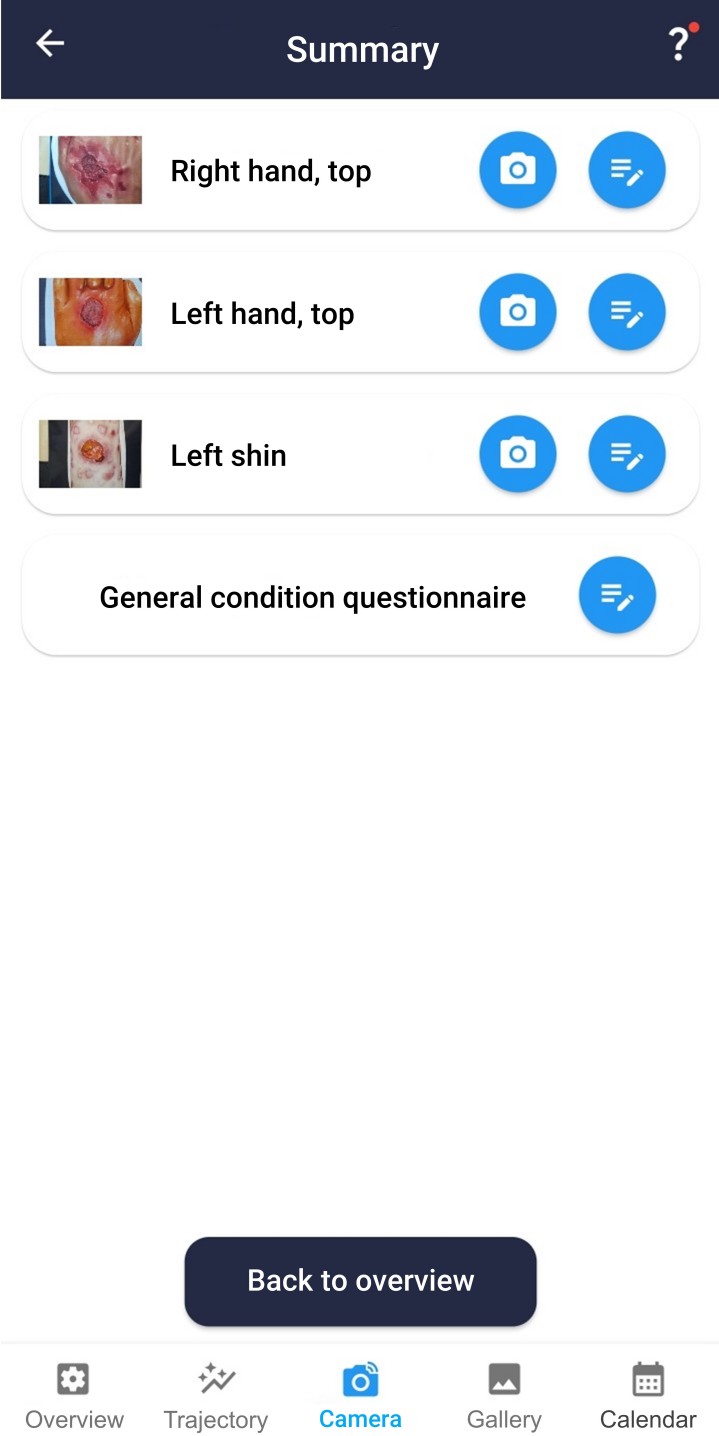}
        \caption{Summary view}
        \label{fig:summary_view}
    \end{subfigure}

    \caption{Screenshots of the \AppName app: Image capturing and questionnaires for patient-reported wound monitoring.}
    \label{fig:screenshots_app:image_capturing_and_questionnaires}
    \Description{Seven app screenshots, including: one showing the wound localization screen; two displaying parts of the camera functionality (one showing the camera screen and the other highlighting the AI-recognized wound area in the captured image); two showcasing the built-in questionnaires (one for wound-specific parameters and the other for general health conditions); one featuring an interim motivation screen; and one displaying the summary screen shown after completing the documentation process.}
\end{figure*}

We next present the core functionalities of the proposed \AppName mobile app, developed to foster self-engagement in wound care for patients with chronic wounds. The app aims to improve accessibility and quality of care by enabling home-based wound monitoring and remote interaction with \acp{hcp}.
To enable automated and pain-free wound documentation, \AppName combines AI-assisted image capture (Figure~\ref{fig:screenshots_app:image_capturing_and_questionnaires}b/c) with structured questionnaires for collecting PROs. 
These encompass wound-specific parameters—such as pain, itching, and exudate---recorded separately for each wound (Figure~\ref{fig:screenshots_app:image_capturing_and_questionnaires}d), as well as the patient’s overall condition, including mood, activity impact, and quality of life (Figure~\ref{fig:screenshots_app:image_capturing_and_questionnaires}f).
Figure~\ref{fig:screenshots_app:image_capturing_and_questionnaires} illustrates the complete documentation workflow with screenshots translated into English for improved accessibility. 

\begin{figure*}[ht]
    \centering
    \begin{subfigure}{0.21\textwidth}
        \centering
        \includegraphics[width=\linewidth]{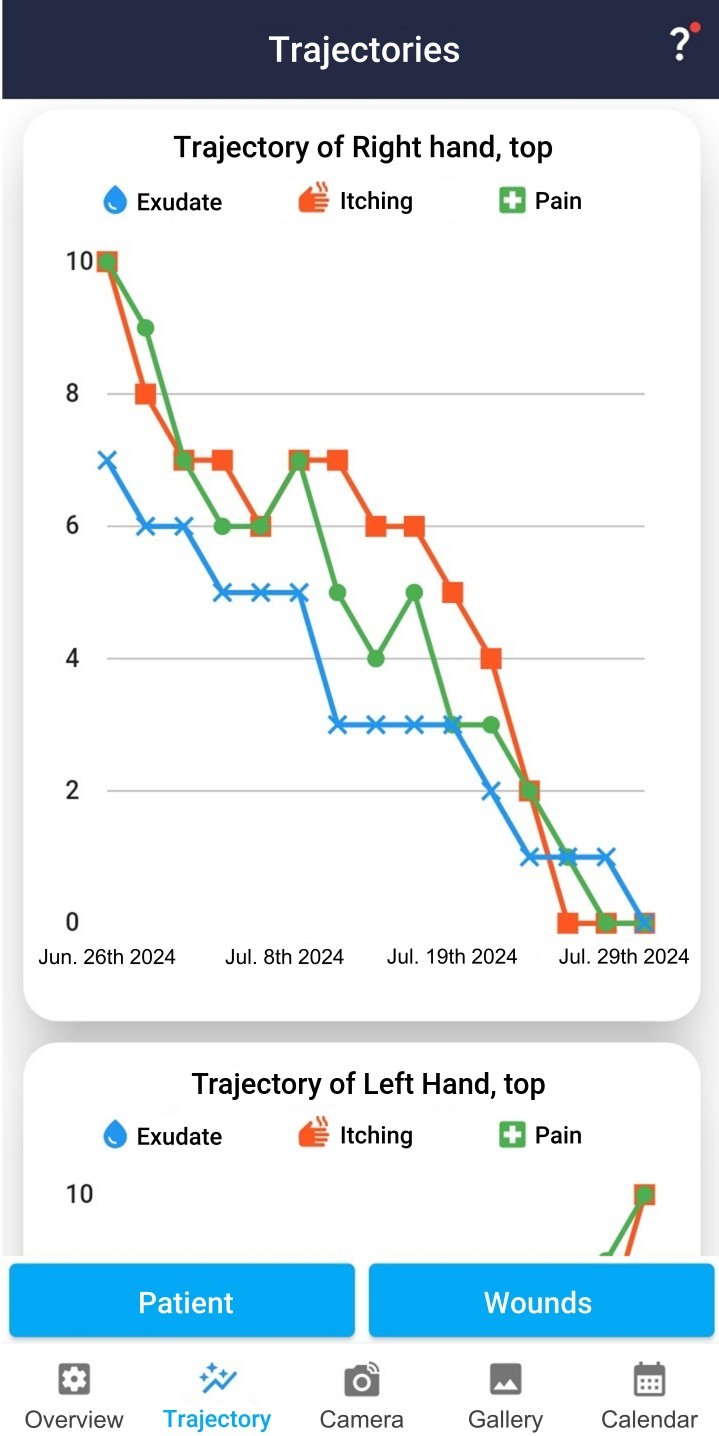}
        \caption{Wound trajectory}
        \label{fig:wound_trajectory}
    \end{subfigure} \hfill
    \begin{subfigure}{0.21\textwidth}
        \centering
        \includegraphics[width=\linewidth]{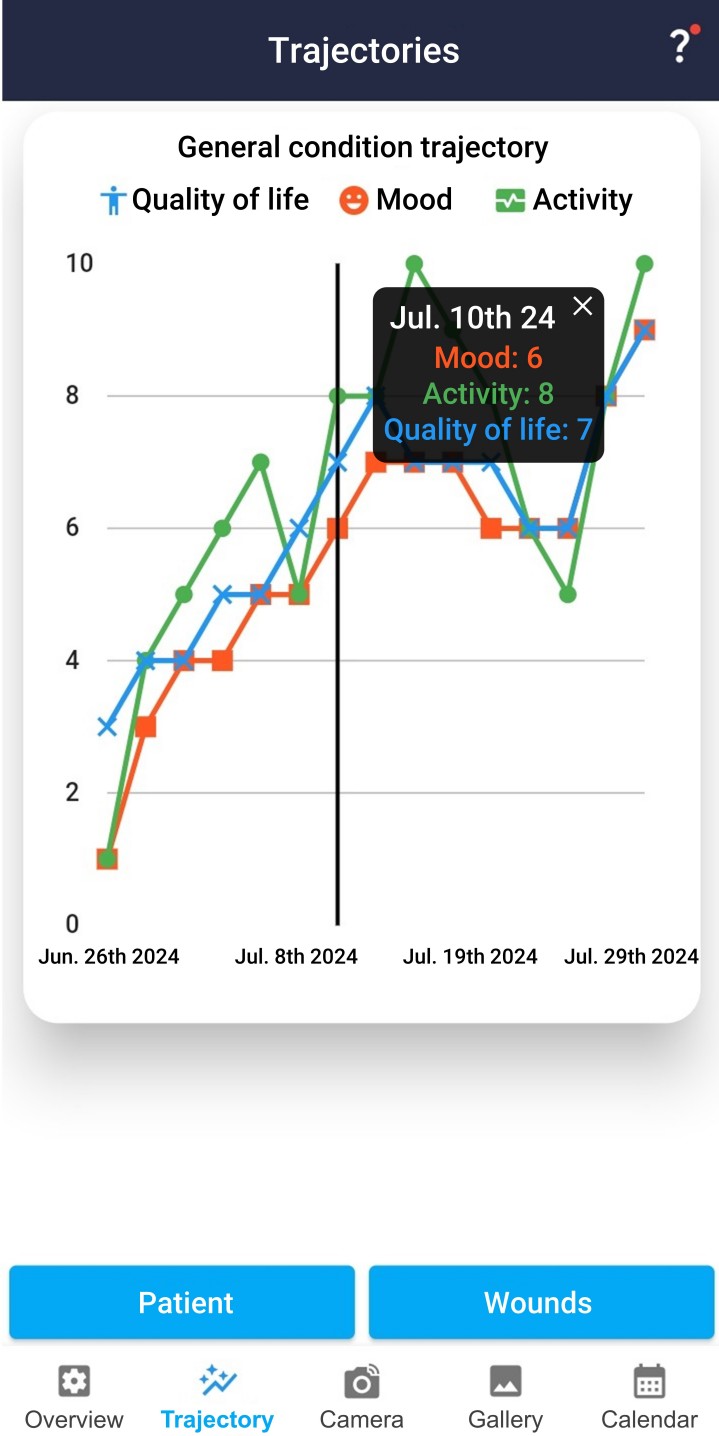}
        \caption{General health traject.}
        \label{fig:general_condition_trajectory}
    \end{subfigure} \hfill
    \begin{subfigure}{0.21\textwidth}
        \centering
        \includegraphics[width=\linewidth]{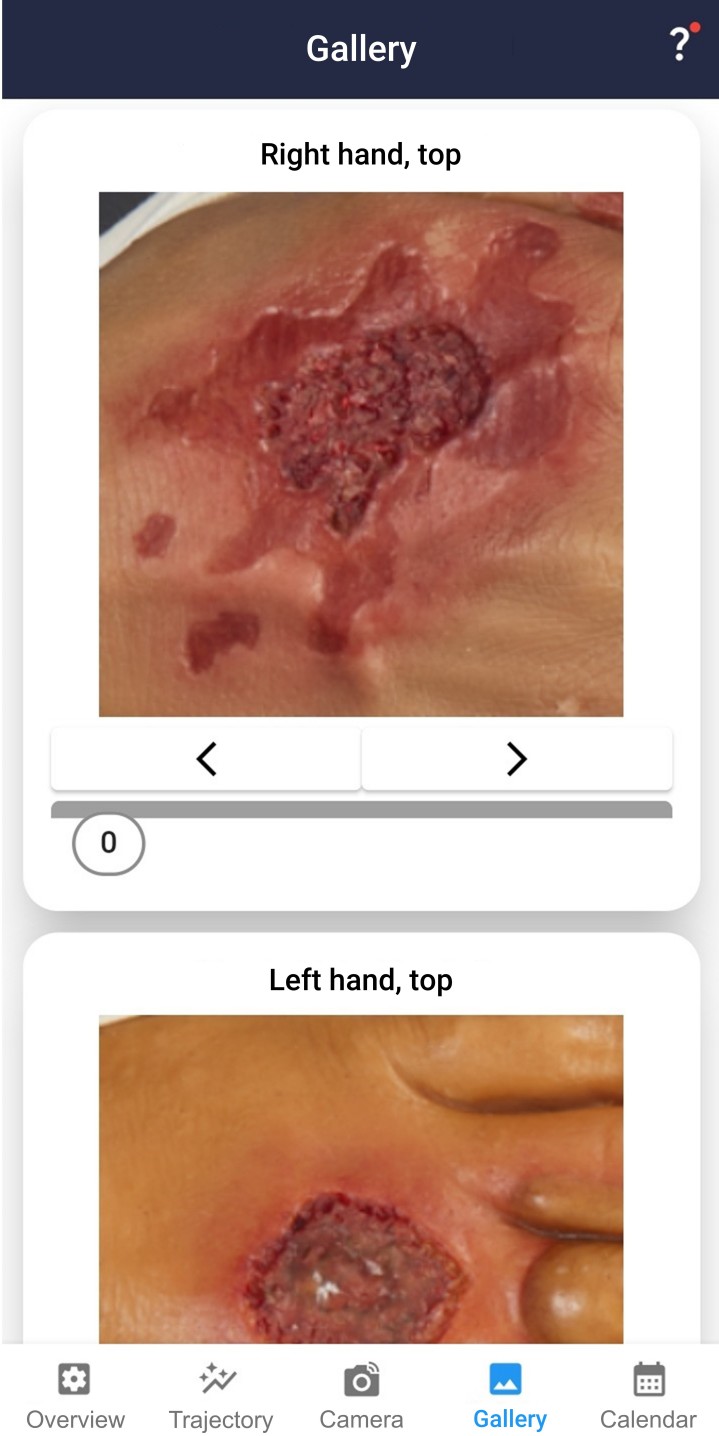}
        \caption{Gallery}
        \label{fig:gallery}
    \end{subfigure} \hfill
    \begin{subfigure}{0.21\textwidth}
        \centering
        \includegraphics[width=\linewidth]{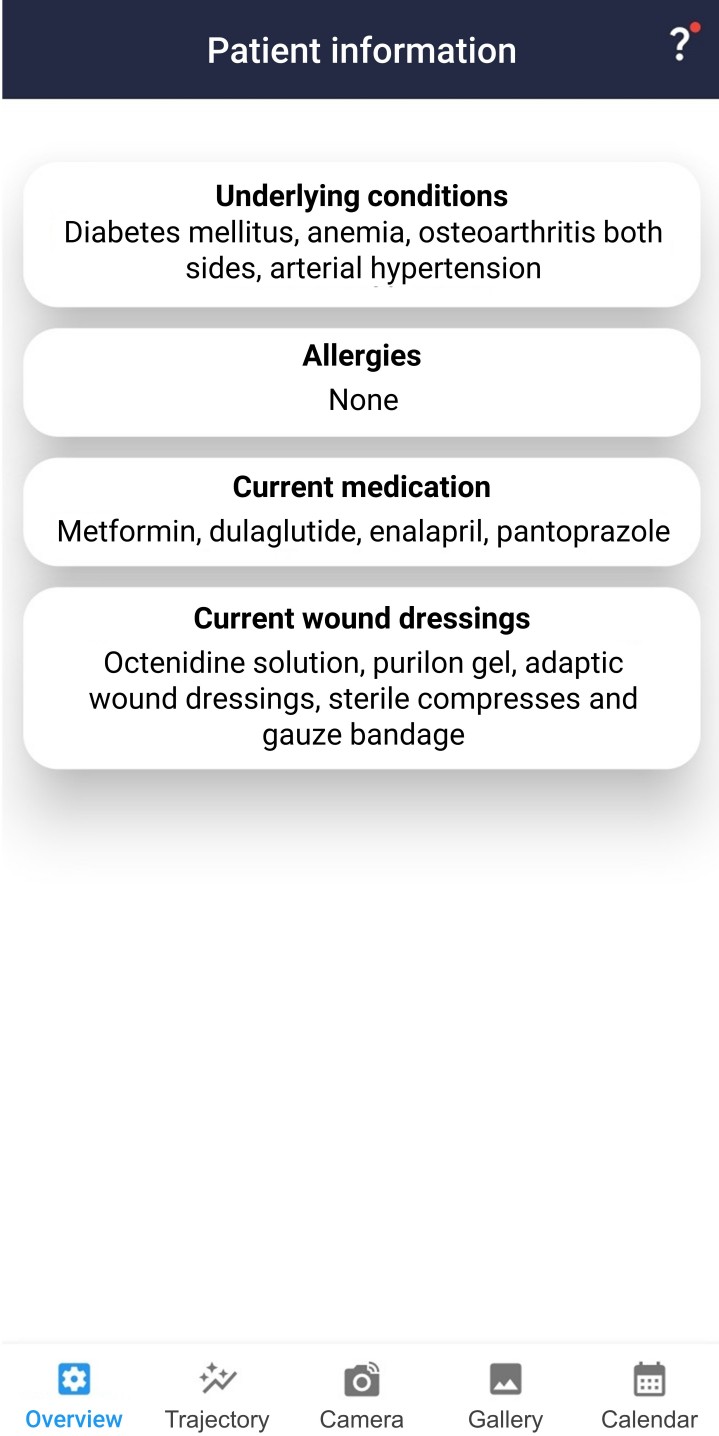}
        \caption{Patient overview}
        \label{fig:patient_info}
    \end{subfigure} \\
    \vspace{0.5cm} 
    \begin{subfigure}{0.21\textwidth}
        \centering
        \includegraphics[width=\linewidth]{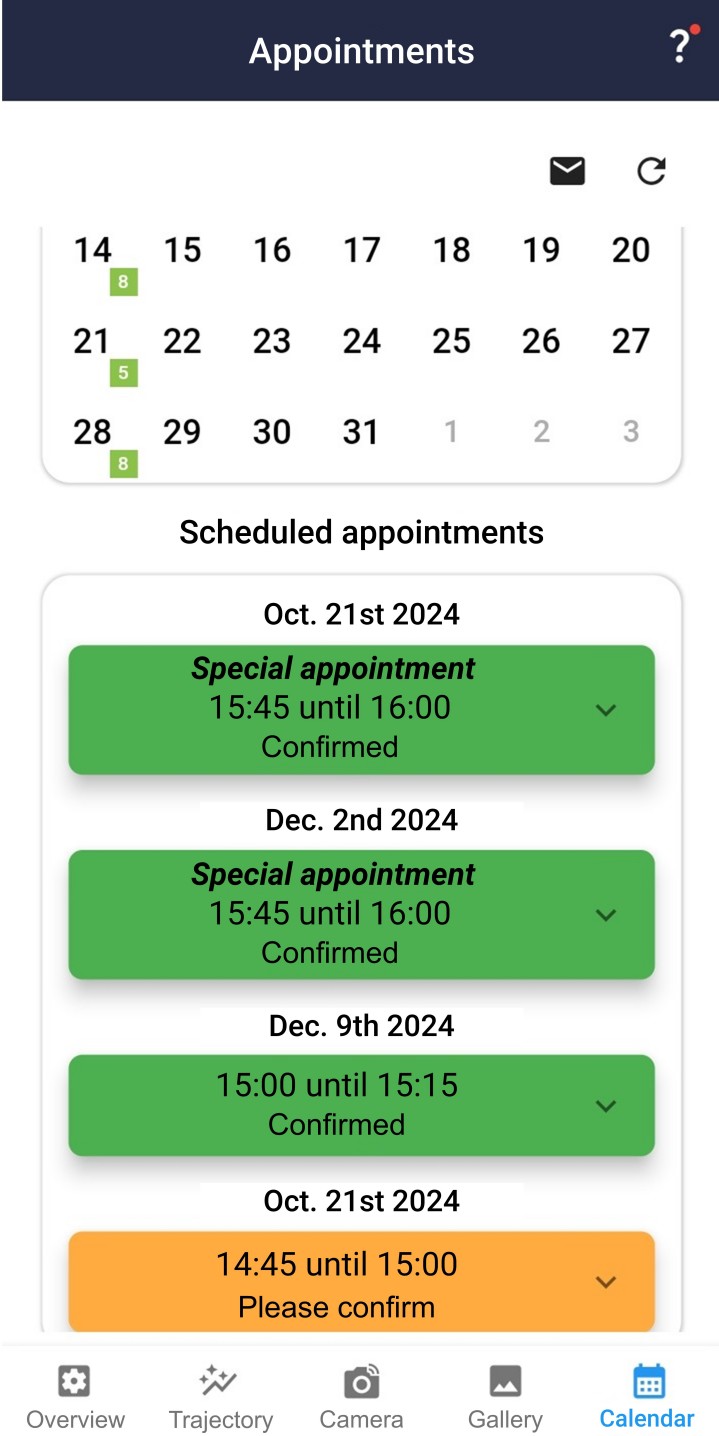}
        \caption{Calendar}
        \label{fig:calendar}
    \end{subfigure} \hfill
    \begin{subfigure}{0.21\textwidth}
        \centering
        \includegraphics[width=\linewidth]{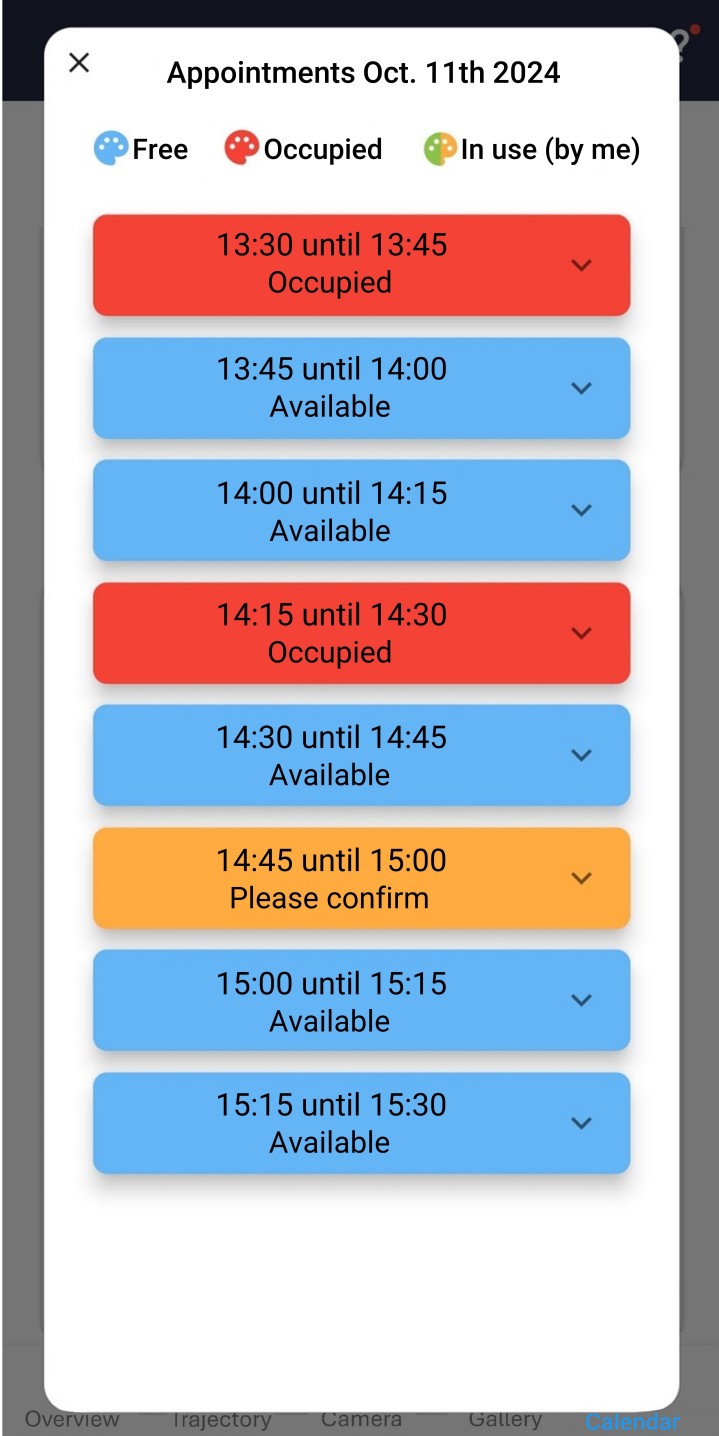}
        \caption{Details per day}
        \label{fig:calendar_details}
    \end{subfigure} \hfill
    \begin{subfigure}{0.21\textwidth}
        \centering
        \includegraphics[width=\linewidth]{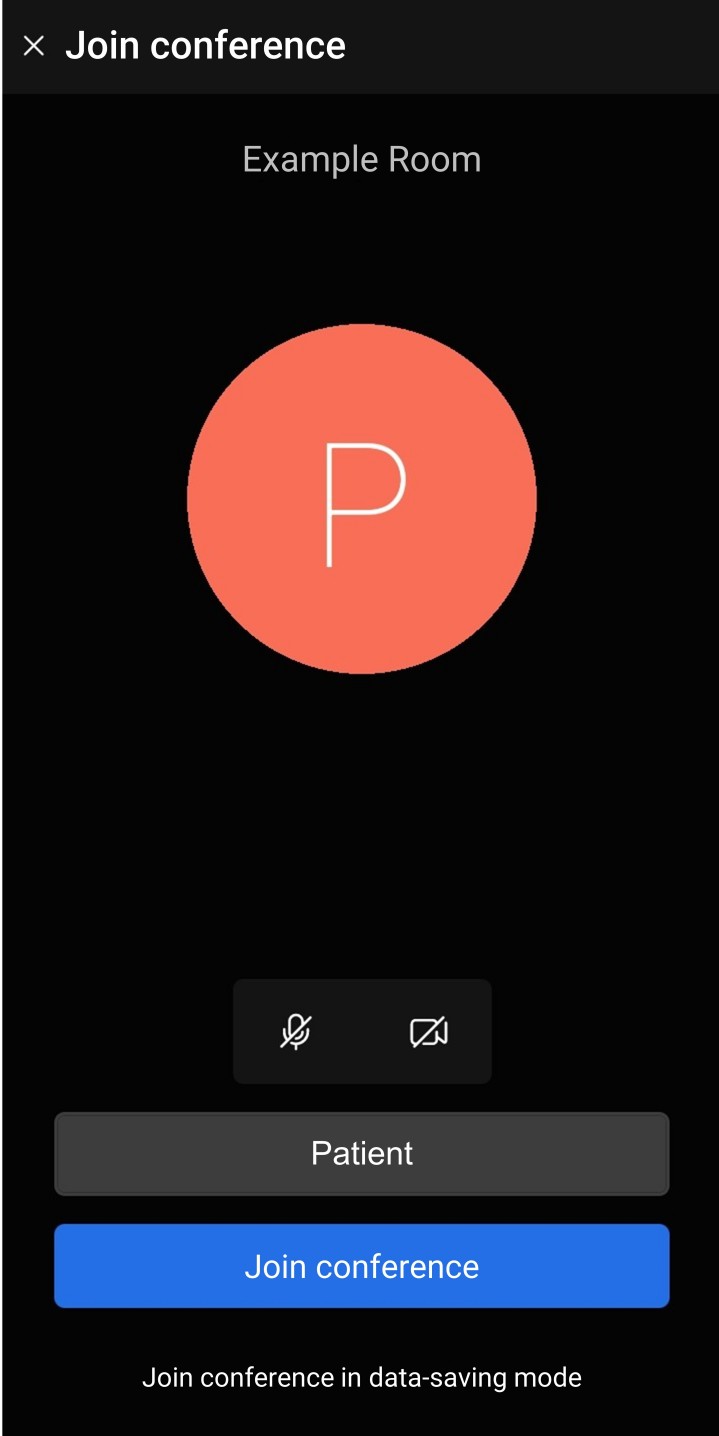}
        \caption{Video chat}
        \label{fig:video_chat} 
    \end{subfigure}\hfill
    \begin{subfigure}{0.21\textwidth}
        \centering
        \includegraphics[width=\linewidth]{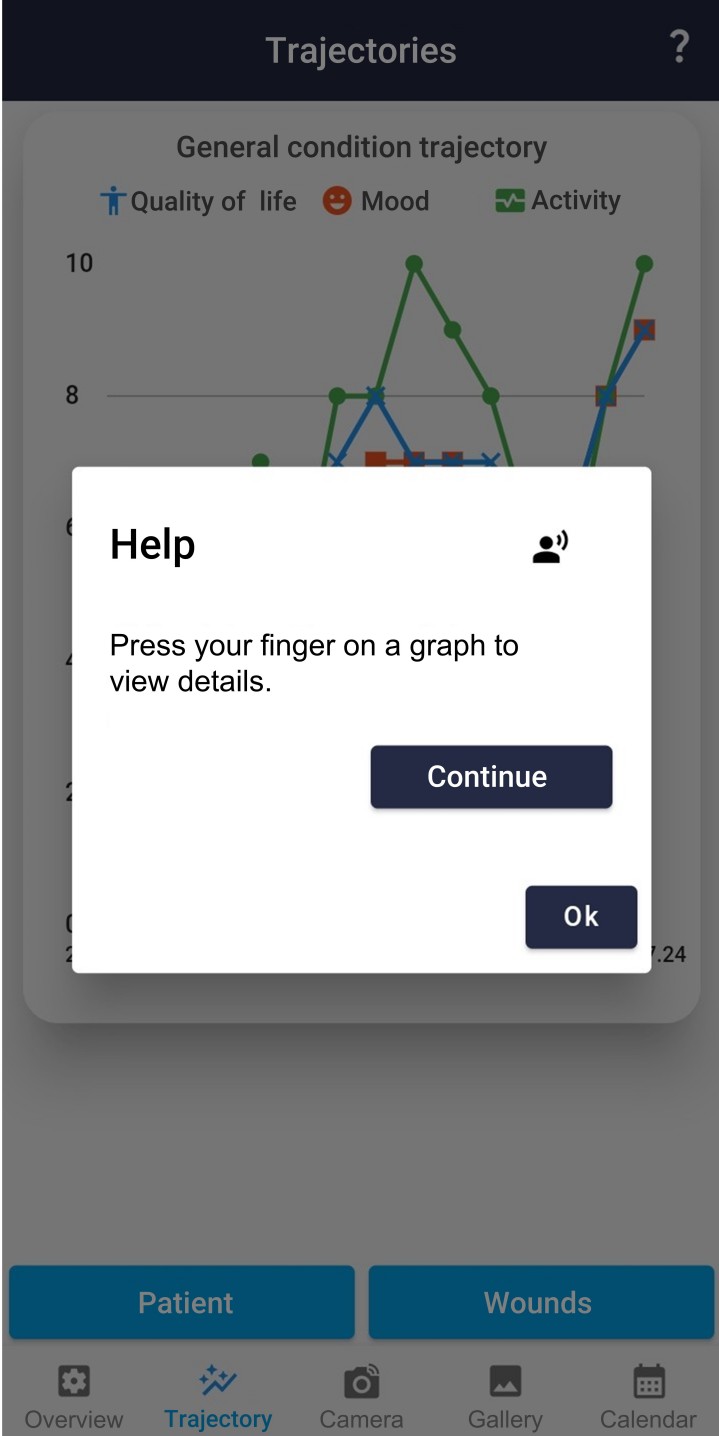}
        \caption{Built-in help function}
        \label{fig:help_function}
    \end{subfigure} \hfill
    
    \caption{Screenshots of the \AppName app: Informative screens as well as calendar and video chat functionalities.}
    \label{fig:screenshots_app:informative_screens_and_calendar}
    \Description{Eight app screenshots, including: two displaying graphs with different colors representing wound-specific parameters and general condition trajectories; one showcasing the image gallery with previously captured wound images; one presenting the patient overview, including details on the patient’s underlying conditions, allergies, medications, and wound dressings; two screenshots illustrating the calendar feature---one showing a detailed daily view with occupied, available, and requested appointment slots, and another showing an overview of the patient’s scheduled appointments; one screenshot displaying the video chat interface; and one example of the help function popup.}
\end{figure*}

\AppName also offers several features beyond wound documentation: It allows patients to track their wound healing and overall health using integrated progress curves (see Figure~\ref{fig:screenshots_app:informative_screens_and_calendar}a/b). Additional informative screens provide a history of captured images for long-term monitoring (see Figure~\ref{fig:screenshots_app:informative_screens_and_calendar}c) and an overview of patient details, including underlying conditions, allergies, medication, and current wound dressing (see Figure~\ref{fig:screenshots_app:informative_screens_and_calendar}d). 
The app also includes a calendar function for scheduling appointments and conducting video consultations with HCPs (see Figure~\ref{fig:screenshots_app:informative_screens_and_calendar}e-g). A help function is accessible via a question mark icon on all main screens (see Figure~\ref{fig:screenshots_app:informative_screens_and_calendar}h). 

To enhance usability and lower barriers for older adults, we followed design guidelines from Liu et al.~\cite{liu2021}, targeting age-related obstacles to mHealth adoption (see Section~\ref{sec:rw-mapps}). 
To address visual barriers, we used sans-serif fonts, high text–background contrast, and a limited, consistently applied color palette. 
Motor coordination issues were accommodated via simple gestural input (i.e., touch and drag) and a button- and slider-only navigation interface, mitigating physical barriers. 
Potential cognitive and memory deterioration was addressed by maintaining a consistent layout, simplifying navigation with a bottom bar, and limiting menu options. 
As recommended by Isaković et al. \cite{isakovic2016}, a help function is accessible via an icon on every screen to support motivation.
By combining advanced functionalities with intuitive design, \AppName delivers a complete telemedicine solution for at-home wound management under continuous clinical oversight. 
Accounting for age-related barriers explicitly tailors the app to older adults and aims to increase their willingness to use it, thereby counteracting reduced adoption.

\section{Overall System Overview and Qualitative Validation of the Mobile AI Component}
\label{sec:technical_components}

\subsection{System Architecture and Role of the Mobile AI Component} %
\label{ssec:technical_realization:overallSystem}

\begin{figure*}[ht]
\includegraphics[width=0.74\textwidth]{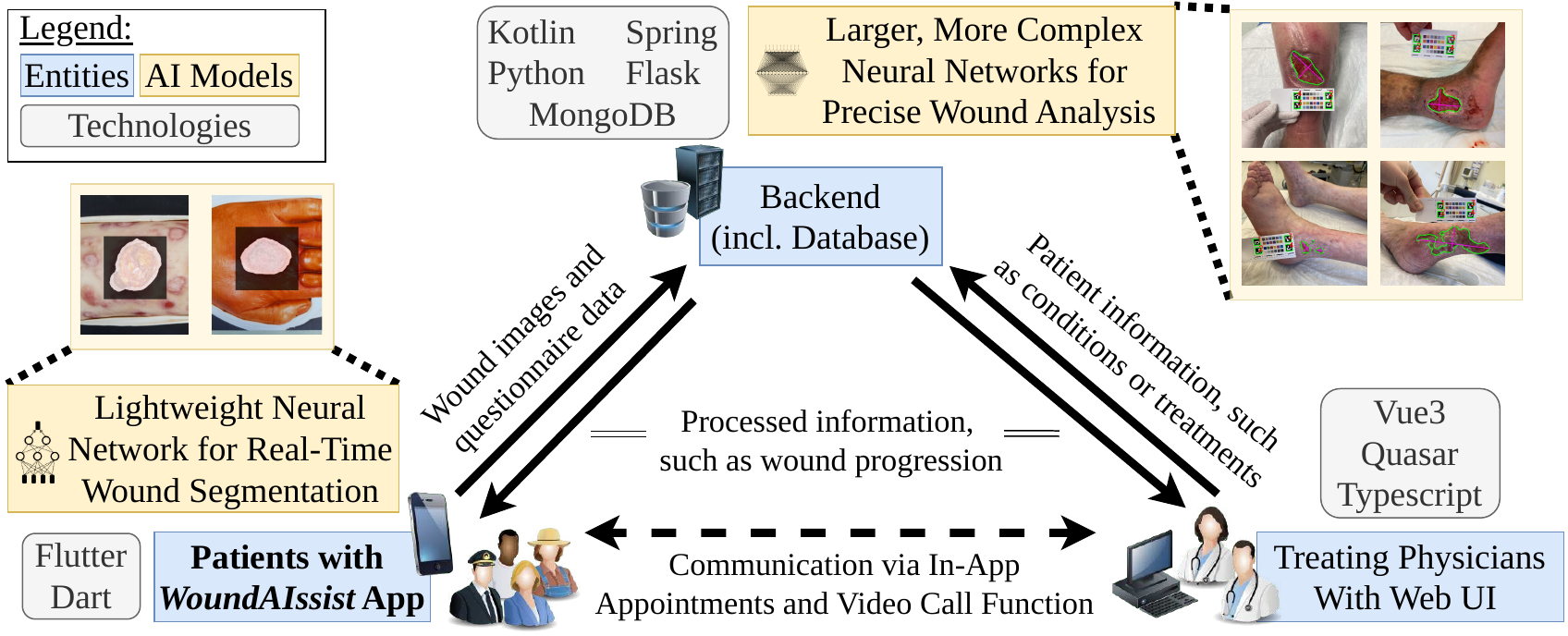}
\caption{Overall system design behind the \AppName smartphone app.}
\label{fig:overall_system_design}
\Description{A diagram illustrating the overall system architecture of \AppNameNoSpace, which consists of three main components: a smartphone application for patients, a backend system with AI models for analyzing wound images, and a web interface for \acp{hcp}. The mobile app enables patients to upload images of their wounds and complete questionnaires, where captured images are processed by the backend using advanced neural networks. Physicians can then access all patient information through the web interface to monitor the healing process. Additionally, the system facilitates smooth communication between patients and \acp{hcp} through in-app appointments and video calls.}
\end{figure*}

The system behind \AppName (see Figure~\ref{fig:overall_system_design}) comprises three components: a patient smartphone app, a physician web interface, and a backend for secure data storage, communication, and advanced AI-driven wound analysis. It supports both teledermatology consultation modalities identified by Trettel et al.~\cite{trettel2018telemedicine}: asynchronous ``store and forward'' exchange of images and questionnaires, and synchronous live video consultations.

Through the web dashboard, dermatologists can manage patient data, monitor wound progression using longitudinal images and questionnaire responses, and schedule appointments. The backend hosts advanced neural networks for detailed wound analysis, enabling automated, non-invasive wound size estimation based on AI-predicted areas and a reference object in captured images (see~\cite{borst2025woundambit}). Server-side processing supports the deployment of larger, more sophisticated models than feasible on mobile devices, thereby improving accuracy.

On-device AI complements the system by providing real-time segmentation feedback during wound image capture, reducing latency and enabling offline functionality, which is critical in low-connectivity settings. This immediate visual guidance allows patients to verify correct wound localization and adjust camera angle, lighting, or focus, facilitating the collection of consistent, clinically meaningful images. Combining mobile AI for user guidance with server-based AI for precise analysis balances interactivity and diagnostic accuracy. Although local processing could, in principle, enhance privacy by retaining images on-device and transmitting only derived measurements, the current version uploads images to ensure full physician access and advanced image analysis.

For guided image acquisition, we integrated an existing mobile-optimized AI model into \AppNameNoSpace. The chosen architecture, \textit{TopFormer-Tiny}~\cite{zhang2022topformer}, was selected based on our prior comparative evaluation of lightweight semantic segmentation (SS) models for wound analysis~\cite{borst2024early}. This hybrid model combines the robustness of Vision Transformers against distribution shifts~\cite{zhang2022delving} with the computational efficiency of Convolutional Neural Networks. Trained in our previous work~\cite{borst2024early}, the model is employed here without further fine-tuning or architectural modification.
Accordingly, this paper does not aim to develop a new segmentation model but instead focuses on integrating an existing model into a patient-centered wound care app to empirically evaluate its feasibility and user perception in a stakeholder-based usability study (\textit{Study~B}, Figure~\ref{fig:timeline_app_derivation}).
For transparency, Appendix~A details the model architecture, training and integration process, and its positioning within the landscape of mobile SS.

\subsection{Qualitative AI Validation on Contemporary and Patient Devices}
\label{ssec:technical_components:validation_contemporary_devices}

Because the deployed mobile AI model was not further optimized for this study, we performed a qualitative real-world validation prior to the user perception study to confirm that it relied on a reliably functioning AI component with adequate segmentation performance under practical conditions. This validation extends our earlier exploratory analysis~\cite{borst2024early}, which was limited to two online wound images displayed on a PC screen and controlled test scenes, primarily evaluated on an older Android device (OnePlus 7 Pro, 2019).
In contrast, the present assessment examines images captured under realistic conditions---i.e., real wounds from different patients rather than screen-based imitations---covering diverse wound shapes, anatomical locations, lighting conditions, and hardware devices.  All patients provided informed consent for use and publication of their images.

\begin{figure}[ht]
    \centering
    \includegraphics[trim={0 0 15 10 }, clip, height=2.8cm]{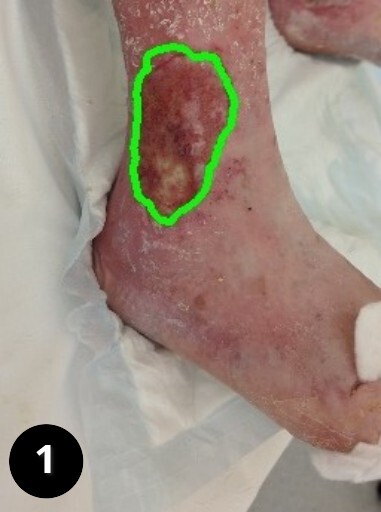}
    \hfill
    \includegraphics[trim={0 0 15 10 }, clip, height=2.8cm]{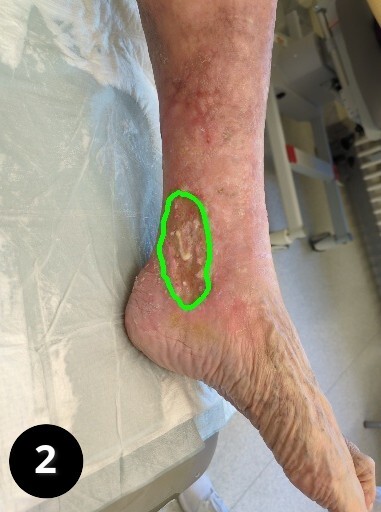}
    \hfill
    \includegraphics[trim={0 0 15 10 }, clip, height=2.8cm]{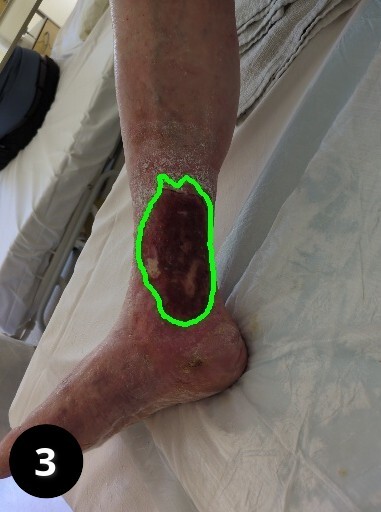}
    \hfill
    \includegraphics[trim={0 0 15 10 }, clip, height=2.8cm]{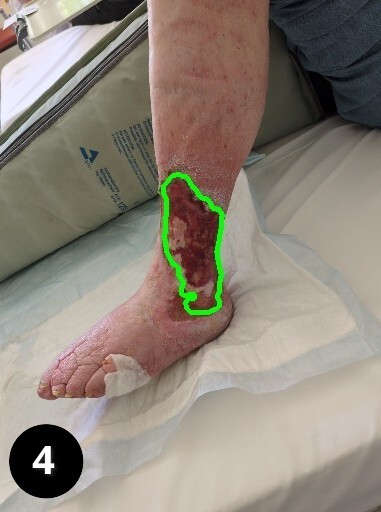}
    \hfill
    \includegraphics[trim={0 0 15 10 }, clip, height=2.8cm]{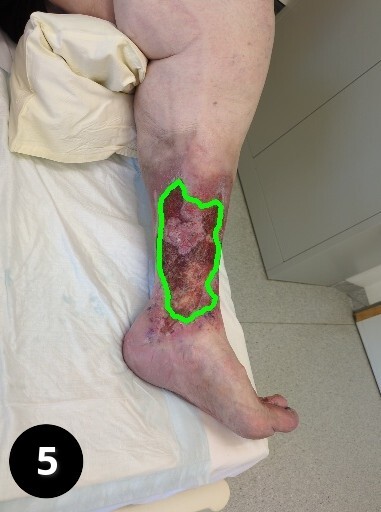}
    \hfill
    \includegraphics[trim={0 0 15 10 }, clip, height=2.8cm]{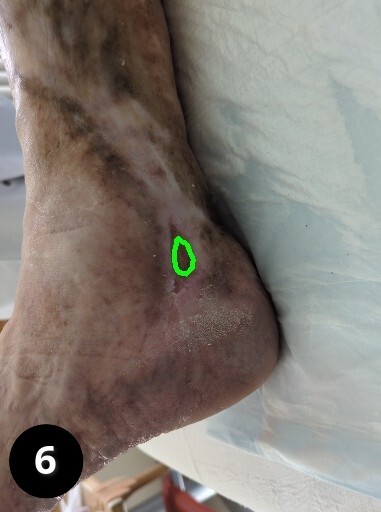}
    \hfill
    \includegraphics[trim={0 0 15 10 }, clip, height=2.8cm]{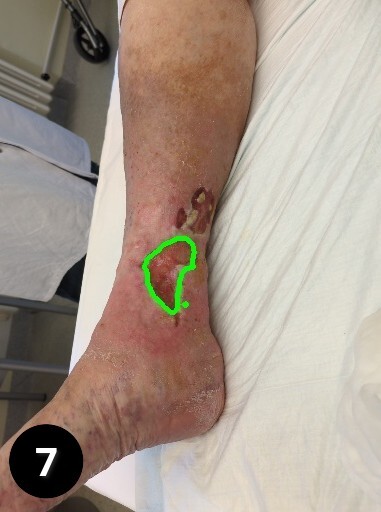}
\caption{Mobile AI-generated segmentation masks on images captured during clinical visits using a Google Pixel 8a.}
\label{fig:AI_validation:googlePixel}
\Description{The figure displays seven clinical photographs of lower leg wounds captured during in-clinic visits using a Google Pixel 8a smartphone. Each image includes a segmentation mask generated by a mobile AI model, showing wound boundaries highlighted directly on the wound area. The wounds differ in shape, size, and healing stage, demonstrating the AI model's segmentation performance across varied clinical presentations.}
\end{figure}

\begin{figure*}[ht]
    \centering
    \begin{subfigure}[t]{0.51\textwidth}
    \centering
    \footnotesize Four representative patients with chronic wounds\\
    \begin{minipage}[t]{0.05\textwidth}
        \rotatebox{90}{\footnotesize \hspace{0.8cm} Perspective 1}
    \end{minipage}%
    \begin{minipage}[t]{0.95\textwidth}
        \includegraphics[trim={0 0 0 15 }, clip, height=2.8cm]{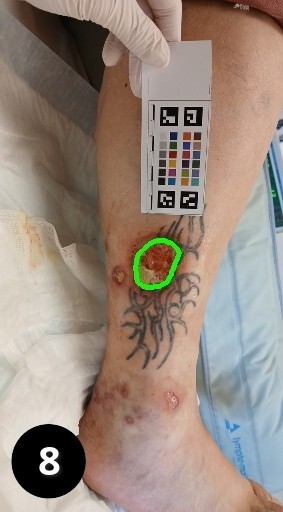}
        \hfill
        \includegraphics[trim={0 0 0 15 }, clip, height=2.8cm]{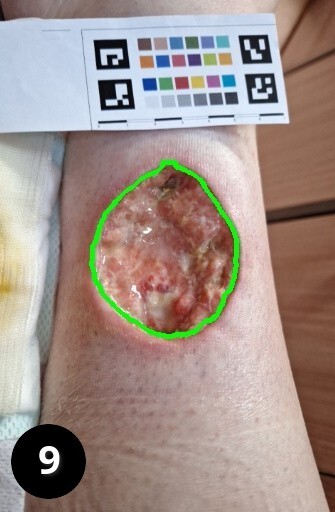}
        \hfill
        \includegraphics[trim={0 0 0 15 }, clip, height=2.8cm]{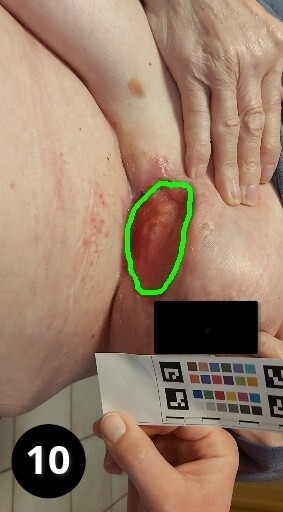}
        \hfill
        \includegraphics[trim={0 0 0 15 }, clip, height=2.8cm]{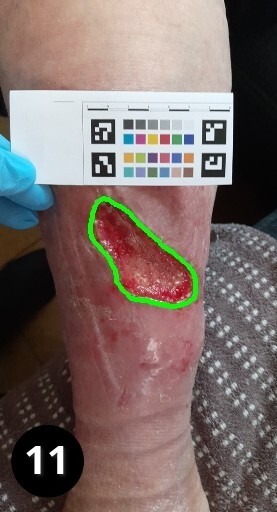}
    \end{minipage}\\[4pt]
    \begin{minipage}[t]{0.05\textwidth}
        \rotatebox{90}{\footnotesize \hspace{0.8cm} Perspective 2}
    \end{minipage}%
    \begin{minipage}[t]{0.95\textwidth}
        \includegraphics[trim={0 0 0 42 }, clip, height=2.6cm]{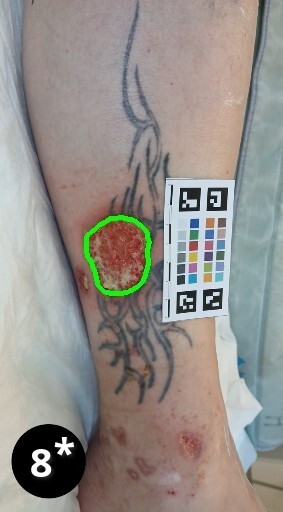}
        \hfill
        \includegraphics[trim={0 0 0 42 }, clip, height=2.6cm]{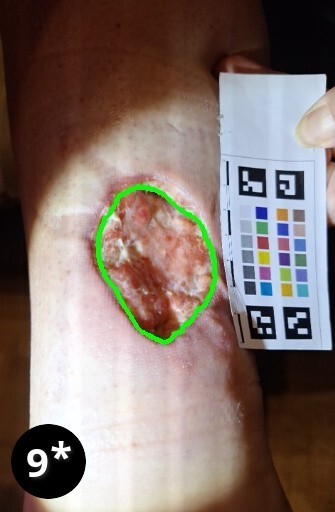}
        \hfill
        \includegraphics[trim={0 0 0 42 }, clip, height=2.6cm]{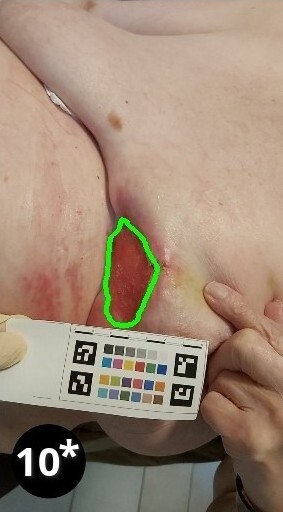}
        \hfill
        \includegraphics[trim={0 0 0 42 }, clip, height=2.6cm]{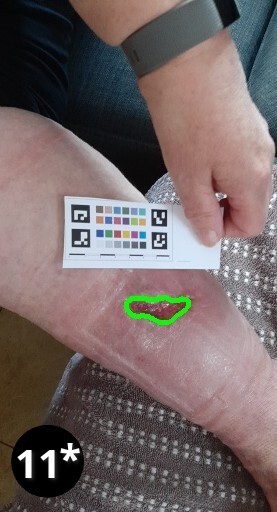}
    \end{minipage}
    \caption{Four wounds at different perspectives and healing states}
    \end{subfigure}
    \hfill
    \begin{subfigure}[t]{0.4\textwidth}
        \centering
        \footnotesize Six additional patients with differently shaped wounds\\
        \includegraphics[trim={0 0 0 15 }, clip, height=2.8cm]{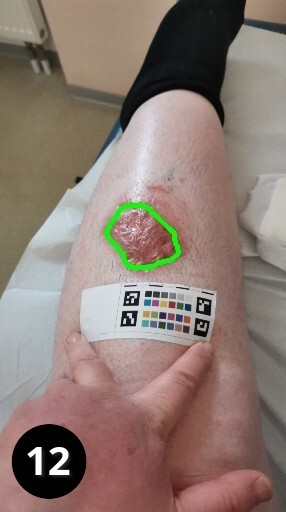}
        \hfill
        \includegraphics[trim={0 0 0 15 }, clip, height=2.8cm]{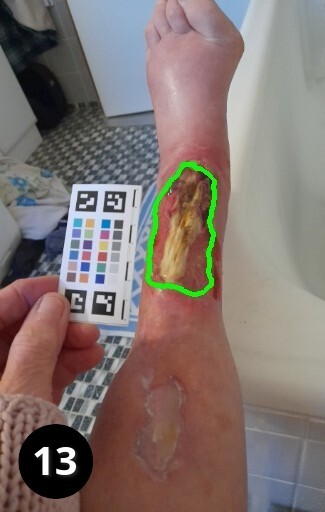}
        \hfill
        \includegraphics[trim={0 0 0 15 }, clip, height=2.8cm]{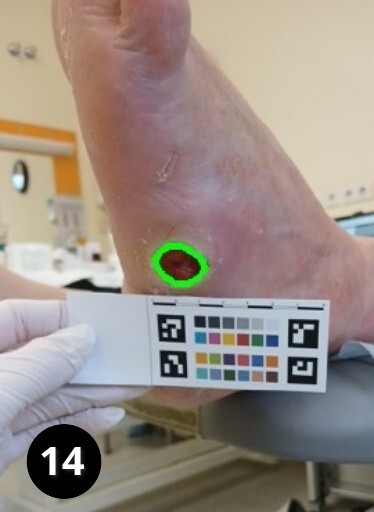}\\[4pt]
        \includegraphics[trim={0 0 0 42 }, clip, height=2.6cm]{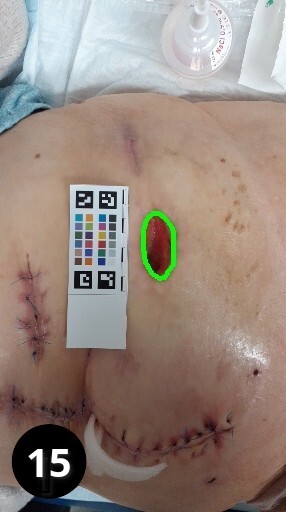}
        \hfill
        \includegraphics[trim={0 0 0 42 }, clip, height=2.6cm]{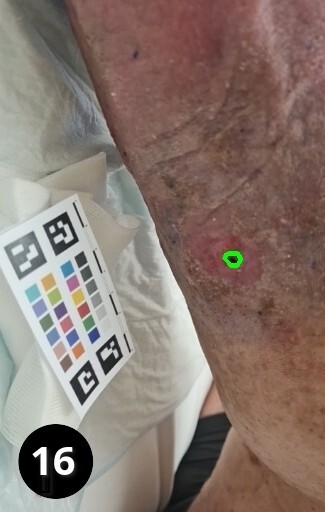}
        \hfill
        \includegraphics[trim={0 0 0 42 }, clip, height=2.6cm]{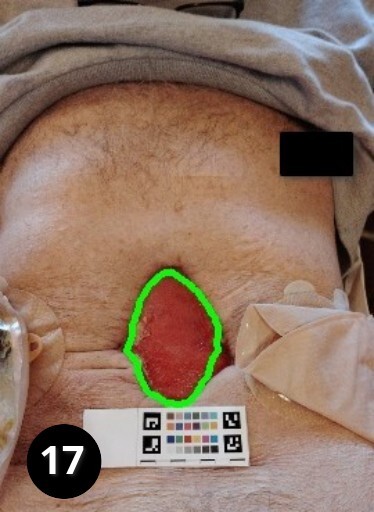}
        \caption{Further wounds at various anatomical locations}
    \end{subfigure}%
    \caption{Mobile AI-generated segmentation masks across diverse devices and wound shapes.}
    \label{fig:AI_validation:patients}
    \Description{The figure shows a set of wound images captured with various patient-owned smartphones. Subfigure (a) includes four different perspectives and healing stages from four representative patients. Subfigure (b) presents six more images from additional patients, depicting wounds at various anatomical locations. Each image is annotated with mobile AI-generated segmentation masks, demonstrating model performance across diverse conditions and user-acquisition contexts.}
\end{figure*}

To assess robustness and device independence beyond the previously tested OnePlus 7 Pro, we employed a newer smartphone (Google Pixel 8a, 2024). Images were acquired by the authors during routine clinical visits at the University Hospital Würzburg using \AppNameNoSpace, with segmentation results presented in Figure~\ref{fig:AI_validation:googlePixel}.
Figure~\ref{fig:AI_validation:patients} shows segmentation outputs for selected cases captured on various patient-owned smartphones. These images were acquired from patients with chronic wounds (or their treating \acp{hcp}) using \AppName on their personal devices.
\footnote{None of these patients participated in \textit{Study B}, ensuring a disjoint sample and minimizing potential bias.} 
Some images were acquired independently by patients at home; others were collected in the hospital.

\newcommand{\circledBlack}[1]{%
  \raisebox{-0.5ex}{\includegraphics[height=2.4ex]{figures/numbers/#1.pdf}}%
}

While not constituting formal generalization testing, qualitative observations suggest that \textit{TopFormer-Tiny} produces adequate segmentation across a range of real-world imaging conditions, devices, and patient anatomies (e.g.,~\circledBlack{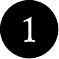},~\circledBlack{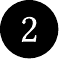},~\circledBlack{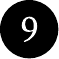}–\circledBlack{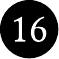}). 
Nevertheless, certain limitations remain, notably in precise wound boundary delineation. The model may slightly overestimate contours (e.g.,~\circledBlack{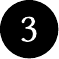}) or generate overly smooth boundaries that fail to capture irregular morphologies (e.g.,~\circledBlack{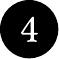}–\circledBlack{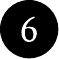},~\circledBlack{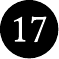}). Performance may also decrease when multiple wounds appear in a single image (e.g.,~\circledBlack{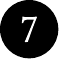},~\circledBlack{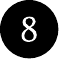}).
Despite these issues, results are encouraging, especially given that the model was not further optimized during this study. Even in complex cases, such as wounds on tattooed skin (e.g.,~\circledBlack{8}), the model performs reasonably well, showing some degree of robustness across anatomical locations and patient-operated devices. 
Importantly, perfect segmentation is not required for its intended role of guiding image capture and indicating image quality, as final wound size estimation is performed server-side (see Section~\ref{sec:technical_components}). 
Overall, these findings indicate that the AI model was sufficiently reliable for inclusion in the user perception study.

\section{Low-Fidelity Prototype Design and Usability Evaluation}
\label{sec:methodology:stage1}

\subsection{Design of a Digital Low-Fidelity Prototype}
\label{design_low_fidelity}

The development of the \textbf{initial low-fidelity prototype} of our mobile app followed an expert-driven, iterative design process within an interdisciplinary team. The core team consisted of two dermatologists, two computer scientists, and two HCI experts, and was supported by student assistants from computer science who contributed to software implementation.
Over more than 1.5 years, the interdisciplinary team engaged in regular meetings to iteratively refine the app’s concept, functionality, and user interface.
Initial sessions shaped the overall idea and high-level requirements, leading to a paper-based prototype, which was subsequently translated into a functional smartphone application. 
The digital prototype underwent multiple revisions informed by feedback from physicians and HCI specialists, progressively enhancing core functionality and interface design. For transparency, Table~\ref{tab:woundAIssistOverTime} summarizes the early development phase and outlines key design changes across four iterations, each separated by approximately four months.
Since not all intermediate versions or subscreens were captured, the table presents the design changes textually and is complemented by a selection of representative screenshots in Figure~\ref{fig:appendix:woundAIssist_over_time}.

\begin{table*}[ht]
\caption{Summary of design changes across iterations of the low-fidelity prototype: (1) Main aspects, (2) Minor changes.}
\label{tab:woundAIssistOverTime}
\fontsize{8pt}{8pt}\selectfont
{
\begin{tabular}{p{0.35cm} p{1.8cm} p{3.7cm} p{4.6cm} p{3.6cm}}
\toprule
\multirow{2}{*}{\textbf{Cat.}}
 & \textbf{Iteration 1}\newline(until July 2022) 
 & \textbf{Iteration 2}\newline(until Nov.\ 2022) 
 & \textbf{Iteration 3}\newline(until March/April 2023) 
 & \textbf{Final Low-Fidelity Prototype}\newline(until July 2023) \\
\midrule

\textbf{(1)} &
Initial technical realization of the preceding paper prototype &
General design updates,\newline including the menu bar&
\begin{itemize}[leftmargin=*, nosep, before=\vspace{-0.6\baselineskip}]
  \item[--] Clear separation of wounds and general condition across questionnaires and trajectories (with novel questions)
  \item[--] Guided, sequential documentation flow for images and questionnaires
  \item[--] Three additional screens during the documentation process 
  \item[--] Calendar redesign with updated colors and day-specific details via pop-up
\end{itemize} &
Gallery updates (i.e., removal of the detailed per-wound views, arrow-based navigation instead of swiping)\\

\midrule

\textbf{(2)} &
N/A &
\begin{itemize}[leftmargin=*, nosep, before=\vspace{-0.6\baselineskip}]
  \item[--] \textbf{Localization}: Revised design (buttons, more information); new body with front and back 
  \item[--] \textbf{Questionnaires}: Updated questions; renamed button text (from ``+1'' to ``\checkmark'')  
  \item[--] \textbf{Calendar}: Removed the misleading green coloring for non-confirmed slots
  \item[--] \textbf{Trajectories}: New diagram for wound size and redness
\end{itemize} &
\begin{itemize}[leftmargin=*, nosep, before=\vspace{-0.6\baselineskip}]
  \item[--] \textbf{Localization}: Revised localization icon; reduced on-screen information; modified buttons
  \item[--] \textbf{Questionnaires}: Hide menu during the question answering 
  \item[--] \textbf{Calendar}: Switched appointment slots from Thursday to Monday
  \item[--] \textbf{Trajectories}: Pain, itching, and exudate instead of wound size and redness 
  \item[--] \textbf{Gallery}: Minor design updates
\end{itemize} &
\begin{itemize}[leftmargin=*, nosep, before=\vspace{-0.6\baselineskip}]
  \item[--] \textbf{Localization}: Fixed a bug in wound coordinate translation; adapted button text
  \item[--] \textbf{Summary}: Replaced technical identifiers with meaningful names
  \item[--] \textbf{Calendar}: Removed the messaging option for appointment requests
  \item[--] \textbf{Trajectories}: Introduced tooltips for the curve coloring
\end{itemize}
\\\bottomrule
\end{tabular}
} 
\end{table*}

Following the first technical implementation in July 2022---approximately seven months after the initial paper-based sketches---two intermediate iterations were completed in November 2022 and March/April 2023, culminating in the final low-fidelity prototype released in July 2023.
While some screens (e.g., the patient information view) underwent only minor aesthetic refinements, other interface components were iteratively revised based on insights from regular interdisciplinary project meetings. 

The final set of requirements emerged from this iterative design process and was 
shaped by physicians’ insights into clinical workflows, their understanding of patient needs, and the anticipated benefits of the app for patients:

\begin{itemize}[leftmargin=*, align=left]
\item \textit{REQ1: Support for Wound Image Capture}: The app shall enable patients to capture and upload images of their wounds in a guided manner, using anatomical illustrations to indicate the correct localization.
\item \textit{REQ2: Verification of Image Quality}: The system shall prompt patients to confirm the quality of each captured image prior to uploading.
\item \textit{REQ3: Consistency in Image Acquisition}: The system shall assist patients in capturing consistent wound images over time by providing visual guidance mechanisms, such as overlays of previously captured images or live feedback based on the system’s detection of the wound region.
\item \textit{REQ4: Automated Assessment of Quantitative Wound Parameters}: The system shall automatically analyze wound characteristics, such as size and redness, using suitable image processing techniques (e.g., AI-based).
\item \textit{REQ5: Collection of Patient-Reported Outcome Measures (PROMs)}: The app shall allow patients to complete questionnaires related to wound-specific parameters (e.g., pain, exudate, itching) and overall health aspects (e.g., mood, activity impact, quality of life) using numeric rating scales (NRS).
\item \textit{REQ6: Temporal Tracking}: 
The system shall record and visualize longitudinal trends in PROs and wound progression, with distinct trajectories for individual wounds and overall health indicators.
\item \textit{REQ7: Scheduling of Telemedicine Appointments}: The system shall include a calendar interface that enables patients to book and manage remote consultations within predefined time slots.
\item \textit{REQ8: Video Consultation Integration}: The system shall enable patients to initiate video consultations with clinicians directly through the app.
\item \textit{REQ9: Access to Past Wound Images}: The system shall offer a gallery feature that allows patients to view previously captured wound images in chronological order.
\item \textit{REQ10: Display of Key Medical Information}: The system shall present an overview of critical patient data, including known allergies, current medications, and wound dressing details.
\end{itemize}

\subsection{Usability Evaluation of the Low-Fidelity Prototype}
\label{usability_study_low_fidelity}
Following the process outlined in Section~\ref{design_low_fidelity}, we iteratively implemented a digital low-fidelity app prototype using Flutter (see screenshots in Figure~\ref{fig:appendix:low_fidelity:screenshots_app}).
Notably, image capturing used a visual wound contour overlay for guidance but did not yet include AI functionality. To evaluate the app's usability regarding the intended target group, we conducted a usability study with affected wound patients using this prototype (\textit{Study~A}). In a mixed-methods approach, we collected both quantitative and qualitative data to address the following RQs:
\begin{itemize}
    \item \textit{RQ-A1:} Are the core features of the app prototype easy to use and intuitive?
    \item \textit{RQ-A2:} How do patients perceive the app’s impact on care, feature scope, and potential for future use?
\end{itemize}

\subsubsection{Participants}
Thirteen patients with chronic wounds participated in the clinic-based usability study, which was conducted at the University Hospital Würzburg following consultation with the institutional ethics committee. All provided informed consent and watched a short video introducing the app’s core functionalities. 
Two participants withdrew after viewing the video. One reported feeling overwhelmed by the amount of information, while the other likely faced comprehension difficulties due to language barriers and concurrent circulatory issues.
The remaining $n=11$ patients ranged in age from 42 to 88 years (mean age $M = 67.00$, $SD = 15.97$), including five females and six males.  

\subsubsection{Measures}
In addition to collecting demographics (age, gender) and prior experience with mobile devices and mHealth apps, the German version~\cite{gao2020} of the System Usability Scale (SUS-DE)~\cite{brooke2013} was employed to evaluate the prototype's usability.
The SUS-DE assesses perceived usability on a five-point Likert scale, with results converted to a score between 0 and 100 for ease of reading. Scores above 68 are considered ``above average''~\cite{bangor2009}. 
A semi-structured interview with 32 questions was conducted and audio recorded. It covered participants’ overall impressions of the app, assessments of individual features, including usability and areas for improvement, and their intentions to continue app usage. All interview questions are provided in the Supplementary Material. 

\subsubsection{Procedure}
\label{eval1method}
Throughout the study, an experimenter was present to assist participants with the app and answer any questions.
After providing informed consent and watching the introductory video, participants received a fictional patient profile and completed predefined tasks
using the low-fidelity app prototype installed on a Samsung Galaxy A21s to familiarize themselves with its features. The task sheet is summarized below; full details are available in the Supplementary Material:
\begin{enumerate*}[label=(\Roman*)] 
    \item Locating information in the patient overview, 
    \item Finding wound trajectory values for a specific day and wound, 
    \item Navigating the image gallery, 
    \item Managing appointments (scheduling, canceling, and starting a video meeting), and 
    \item Documenting two wounds and the general condition of the fictitious patient using the camera and questionnaires.
\end{enumerate*}
Participants used photographs of wound replicas displayed on a tablet, ensuring the app was not used on their own wounds. 
During app interaction, they employed the think-aloud method, verbalizing their thoughts as they navigated features, while the experimenter observed and noted any issues. 
After approximately 10 minutes of hands-on use, participants completed the questionnaires and then participated in the semi-structured interview. The entire session lasted between 60 and 75 minutes.

\subsection{Results}

\begin{figure*}[ht]
\includegraphics[width=\textwidth]{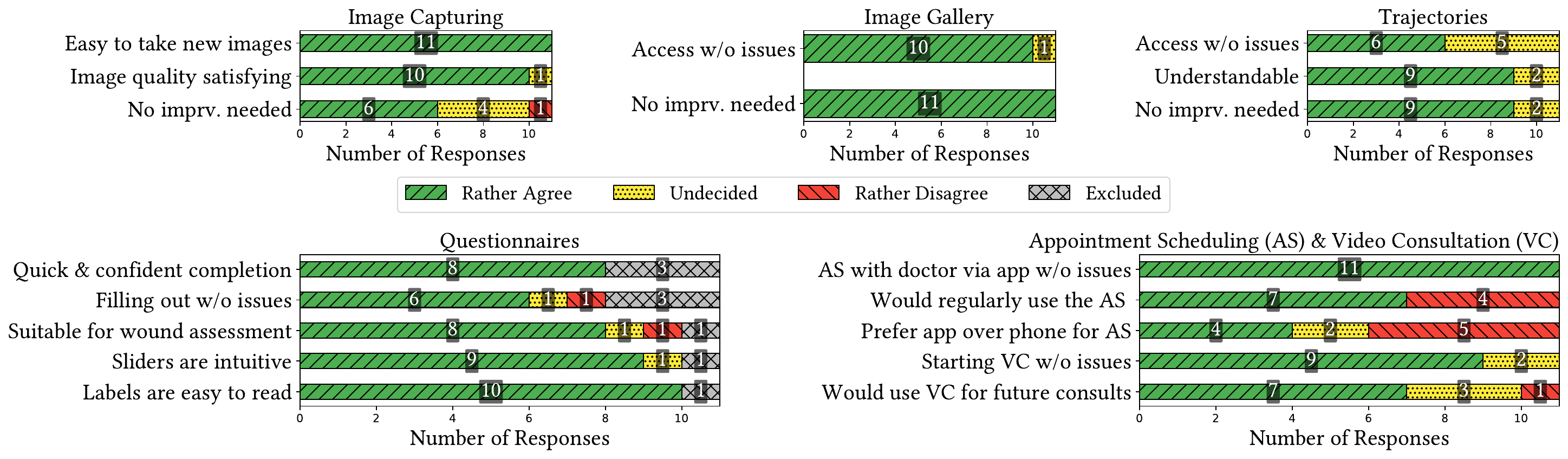}
\caption{Patient feedback on the the different functionalities of the low-fidelity app prototype.}
\label{fig:low_fidelity_qualitative_feedback:functionalities}
\Description{Stacked horizontal bar charts showing participant responses (n=11) across five app functionalities: Image Capturing, Image Gallery, Trajectories, Questionnaires, and Appointment Scheduling & Video Consultation. Bars represent levels of agreement using color-coded segments: Rather Agree (green), Undecided (yellow), Rather Disagree (red), and Excluded (gray). Most responses indicate positive attitudes towards the different functionalities.}
\end{figure*}

\begin{figure*}[ht]
\includegraphics[width=\textwidth]{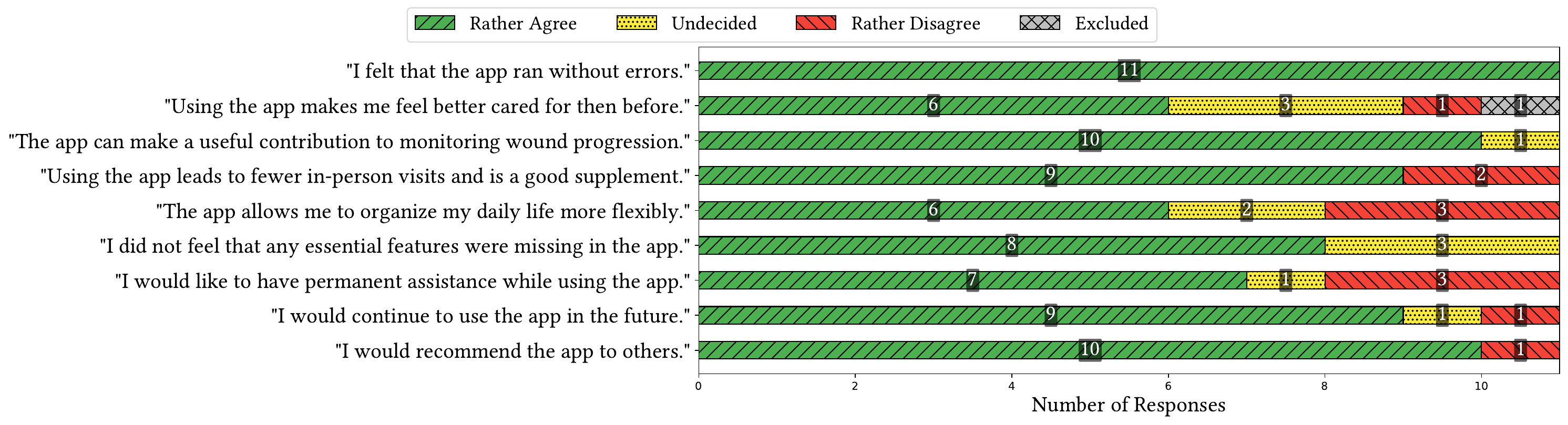}
\caption{Patient feedback on the overall impression of the of the low-fidelity app prototype.}
\label{fig:low_fidelity_qualitative_feedback:overall_perception}
\Description{
Stacked horizontal bar charts showing participant responses (n=11) to nine statements about the app, including  its contribution to care, functional scope, and potential for future use. Bars represent levels of agreement using color-coded segments: Rather Agree (green), Undecided (yellow), Rather Disagree (red), and Excluded (gray). Overall, the chart suggests that patients perceive the app positively and consider it to be useful.
}
\end{figure*}

To address the research questions more effectively, qualitative interview responses were categorized into four attitudinal groups: ``Rather agree'', ``Undecided'', ``Rather disagree'', and ``Excluded''. Responses concerning specific app functionalities are illustrated in Figure~\ref{fig:low_fidelity_qualitative_feedback:functionalities}, while Figure~\ref{fig:low_fidelity_qualitative_feedback:overall_perception} presents overall app perception. 
For consistent visualization and interpretation, interview questions were reformulated post hoc so that ``Rather agree'' indicates positive sentiment and ``Rather disagree'' negative sentiment.
``Excluded'' denotes responses that could not be clearly assigned to the remaining three categories. These edge cases are discussed below to ensure transparency.

\textbf{Prior experience with mobile devices.}
Most participants (9/11) owned a smartphone, with a subset ($n = 4$) also using a tablet, while two reported not owning any mobile device. Regular app use in daily life was limited: the majority reported no app use ($n = 6$), four participants used 1–5 apps, and only one reported using 6–10 apps. None of the participants had prior experience with medical apps, indicating an overall low level of digital literacy.

\textbf{Individual functionalities.}
Below, participants’ perceptions of individual app functionalities are summarized.

\textit{Image capturing.}
Participants successfully completed the task, reporting ease of use and satisfaction with image quality. One patient noted lower quality but attributed this to personal handling rather than technical limitations.
No concrete suggestions for improvement were offered. However, four participants gave vague responses---one referred to future technological advancements, another noted that effective usage may depend on wound location and the ability to capture images independently, and one suggested that potential refinements might become apparent with continued use. One participant assumed improvements were possible but did not specify them.

\textit{Image gallery.}
All patients accessed the gallery without difficulty, and none suggested improvements. However, two found the arrow-based navigation somewhat unintuitive, leading to difficulties in systematically browsing all images (e.g., moving forward and directly backward), as reflected in one participant’s hesitant response.

\textit{Trajectories.}
Participants encountered terminology inconsistencies, as the app labeled the feature as ``Trajectory'' while the task referred to it as ``Statistics''. 
Several patients had difficulty accessing detailed NRS progress scores; five stated that they managed it ``with assistance'', three attributing this to initial unfamiliarity. Two did not fully understand the feature. 
Notably, most participants were unfamiliar with dragging along the curve and using a press-and-hold gesture to view detailed NRS values.
Two patients indicated that improvements could be made, but did not provide specific recommendations. One patient noted potential challenges for older users in interpreting graphical data, while the other cited insufficient interaction time as a limitation to offering concrete suggestions.

\textit{Questionnaires.}
One patient was excluded from the slider task due to severe visual impairments that hindered completion. Two others were excluded post hoc from the two items on questionnaire fill-out, as their responses referenced the SUS-DE---evident from remarks on item polarity and unfamiliar terminology---tough they provided valid responses to all other items.
Questionnaire completion was efficient for the remaining eight participants: six reported no issues, one experienced initial difficulty, and one required examiner assistance. The latter initially misused the slider by treating it as a set of discrete touch-points, resulting in selection errors, and found it intuitive only after receiving guidance.
Among the full sample ($n=10$), nine participants found the sliders intuitive, and all reported the font as easy to read. Eight considered the questionnaire items useful for assessing wound condition, one was uncertain due to perceived subjectivity, and one disagreed without offering alternatives.

\textit{Appointment scheduling and video consultations.}
In addition to terminological inconsistencies (e.g., ``Contact'' in the app vs. ``Appointment scheduling'' in the task), the calendar and video chat functions presented several challenges during app use. Some participants experienced difficulties opening or operating the video chat ($n=5$), while others were unable to schedule a new appointment without assistance ($n=6$).
Despite this, all participants consistently stated in the interviews that they would easily be able to make an appointment via the app, and most reported no problems initiating the video chat.
Seven participants indicated they would use the appointment function regularly. However, five preferred booking appointments by phone, and two had no preference. Those favoring the app cited benefits such as reduced waiting times ($n=2$), immediate visibility of availability ($n=1$), and alignment with the digitization of healthcare ($n=1$).
Most patients expressed willingness to continue using online consultations in the future. Three were uncertain---one noted the need for physical interaction (e.g., touch or smell), while two preferred in-person visits. One participant, without a smartphone, would not use the feature.

\textbf{Overall impression.}
The app functioned without errors for all participants ($n=11$). 
Six participants felt better cared for when using the app, with two attributing this to faster communication with physicians. Three were undecided, citing either no noticeable change or the need for longer-term use; one noted that communication requires engagement from both sides. One participant saw no added care value, and one was excluded due to contradictory responses.
Most participants ($n=10$) believed the app could support wound monitoring, and the same number felt it could reduce the need for in-person visits, serving as a valuable supplement to regular consultations.
Regarding increased daily life flexibility, six participants saw a benefit, two were uncertain, and three saw no advantage.
Eight participants reported that no essential features were missing. The remaining three were uncertain, citing limited long-term experience with the app or unfamiliarity with smartphones; however, none identified any specific missing functionalities.
Seven participants expressed interest in permanent in-app assistance---four for personal use, one considered it useful initially, and two felt it could benefit others.
Overall, nine participants indicated willingness to use the app in the future, and ten would recommend it to others.

\textbf{System usability.}
Two participants were excluded from the SUS-DE analysis due to selecting multiple answers or leaving an item unanswered, and another was excluded for highly inconsistent responses, likely due to misunderstanding the questions. For the remaining $n=8$ participants, the app prototype received an average SUS-DE score of $M = 75.30$ ($SD = 16.40$), indicating ``good'' usability according to Bangor et al.~\cite{bangor2009}.

\subsection{Discussion}
The instructional video appeared effective in conveying the app’s functionalities 
, although one patient suggested reducing information density to better emphasize key features. 
During task completion, most participants experienced at least occasional confusion, partly due to inconsistent feature labeling.
This stresses the importance of consistent terminology to support comprehension, particularly for users with limited digital literacy.
Regarding \textit{RQ-A1}, patient self-reporting features (i.e., image capture and questionnaires) were generally perceived as intuitive and easy to use. In contrast, the trajectory visualization and appointment management (i.e., scheduling and video calls) were more challenging and required greater assistance.
Nonetheless, participants expected to handle these functions more confidently with repeated use, reflected in comments such as ``I think I could do it better the second time,'' and in overall positive ratings for the respective items (see Fig.~\ref{fig:low_fidelity_qualitative_feedback:functionalities}).
With respect to \textit{RQ-A2}, overall app perception was positive (Fig.~\ref{fig:low_fidelity_qualitative_feedback:overall_perception}). Most participants recognized its potential to support wound monitoring, reduce in-person visits, and complement traditional care. While perceptions of improved care were more reserved, no missing features were identified, and a majority (9/11) expressed interest in continued use.
\section{Prototype Refinement and AI Extension}
\label{sec:methodology:stage2}

\subsection{Revision of the User Interface}
\label{ssec:method_iterative_development:revisionUI}
Based on findings from \textit{Study A}, targeted modifications were made to the app’s user interface to address participant feedback:
\begin{enumerate*}[label=(\Roman*)]
    \item \textbf{Menu:} The calendar icon and label were revised to more clearly indicate scheduling functionality. 
    \item \textbf{Trajectories}: Two improvements were made: 
    a legend was added to clarify trajectory colors, and the vertical reference line with NRS details was updated to remain visible after touchscreen release, preserving information for the last selected date.
    \item \textbf{Gallery}: An image counter was introduced to help users differentiate between similar images, reducing confusion when browsing wound history.
    \item \textbf{Questionnaires}: Sliders were updated to allow discrete selection by tapping directly on a scale value, in addition to standard sliding.
    \item \textbf{Built-in Help}: We introduced a question mark icon to the upper right corner of key screens, providing immediate assistance regarding the respective features. If desired, the help text can also be read aloud by activating the \includegraphics[height=2ex]{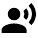} symbol. 
\end{enumerate*}
In addition, we modified the \textbf{Image Capturing} functionality. 
Specifically, the original camera screen overlay, designed to guide patients in taking consistent wound images, was replaced by variants of an AI-based guidance system (see Section~\ref{ssec:method_iterative_development:intergrationAI}) to evaluate user feedback on these alternatives. 
Due to technical constraints associated with the integration of the external library~\textit{Jitsi}, no changes could be made to the video chat feature.
Although almost half of the patients preferred phone-based scheduling, the in-app feature was retained as a non-compulsory alternative, since it aligns with broader digital health trends and does not interfere with the app’s primary function of enabling remote wound documentation for users who engage solely with image and questionnaire submission.

\subsection{Integration of a Mobile Wound Segmentation Model for User Guidance}
\label{ssec:method_iterative_development:intergrationAI}

\begin{figure*}[ht!]
    \centering
    \begin{subfigure}{0.32\textwidth}
        \centering
        \includegraphics[width=0.48\textwidth,clip]{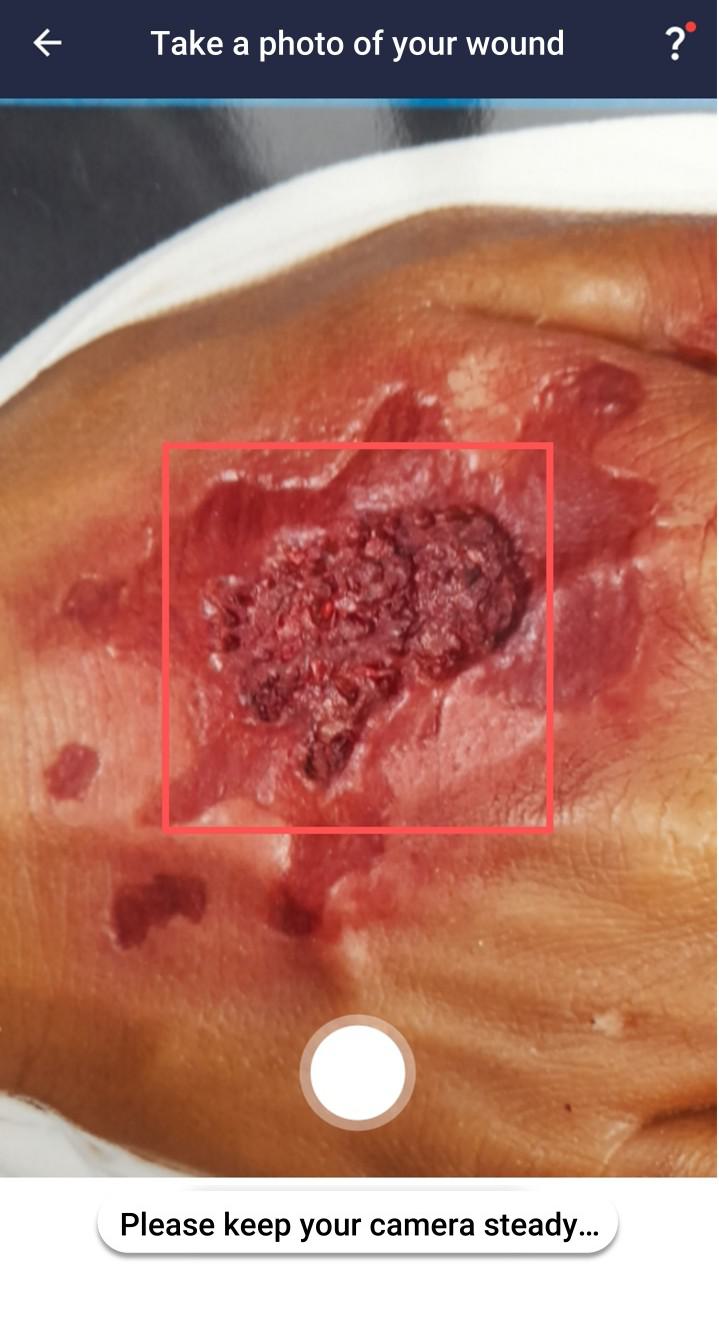}
        \includegraphics[width=0.48\textwidth,clip]{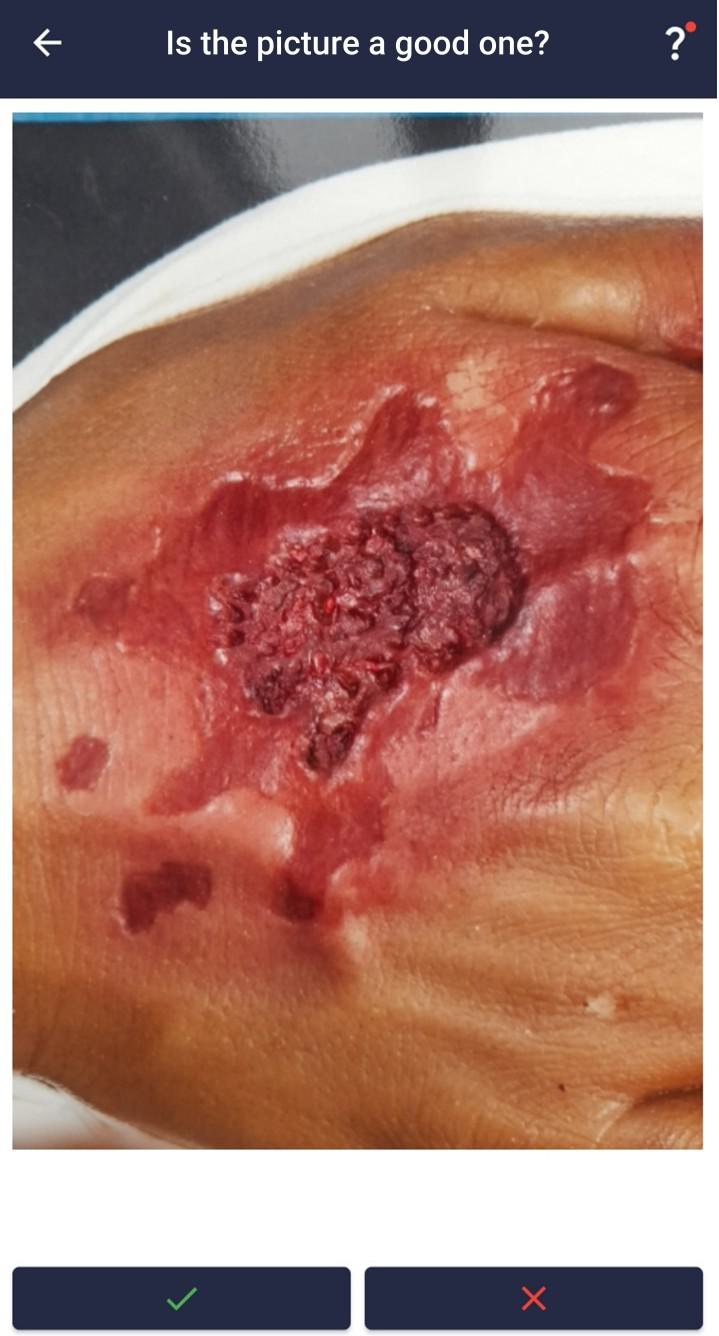}
        \caption{Basic variant without segmentation}
        \label{fig:variant1}
    \end{subfigure} 
    \hfill
    \begin{subfigure}{0.32\textwidth}
        \centering
        \includegraphics[width=0.48\textwidth,clip]{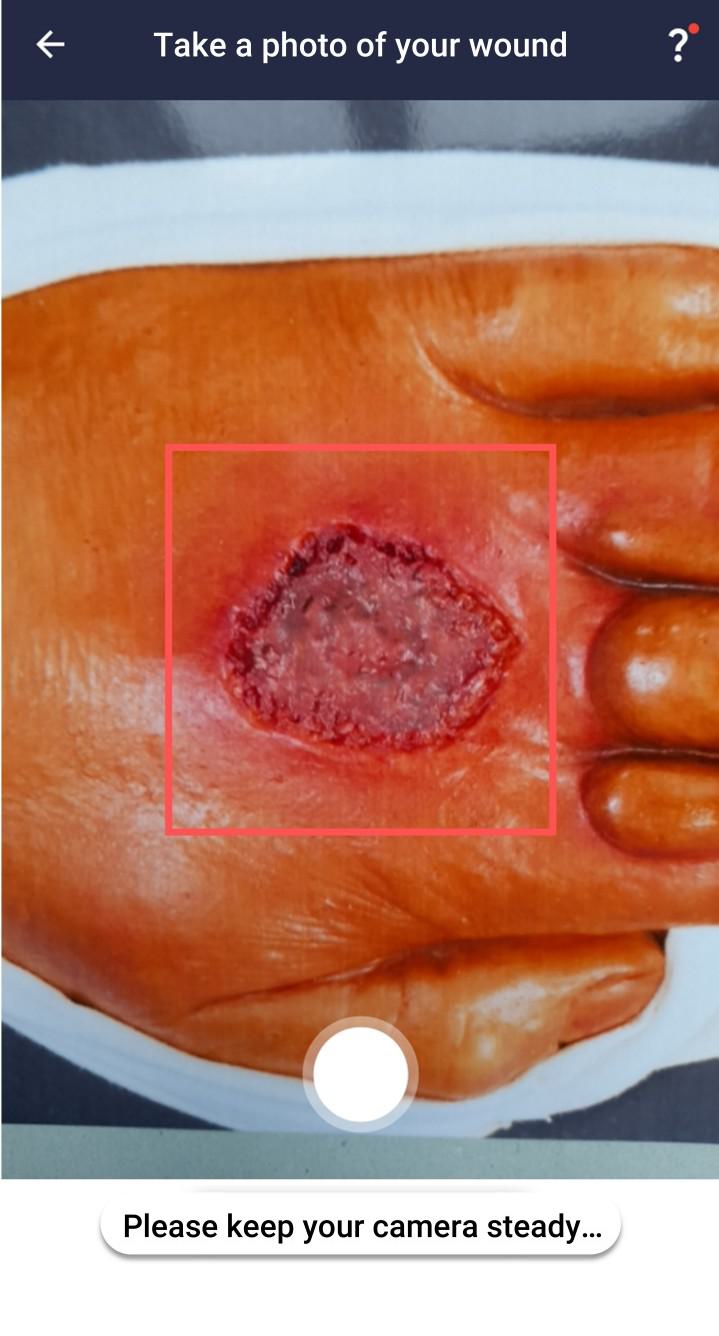}
        \includegraphics[width=0.48\textwidth,clip]{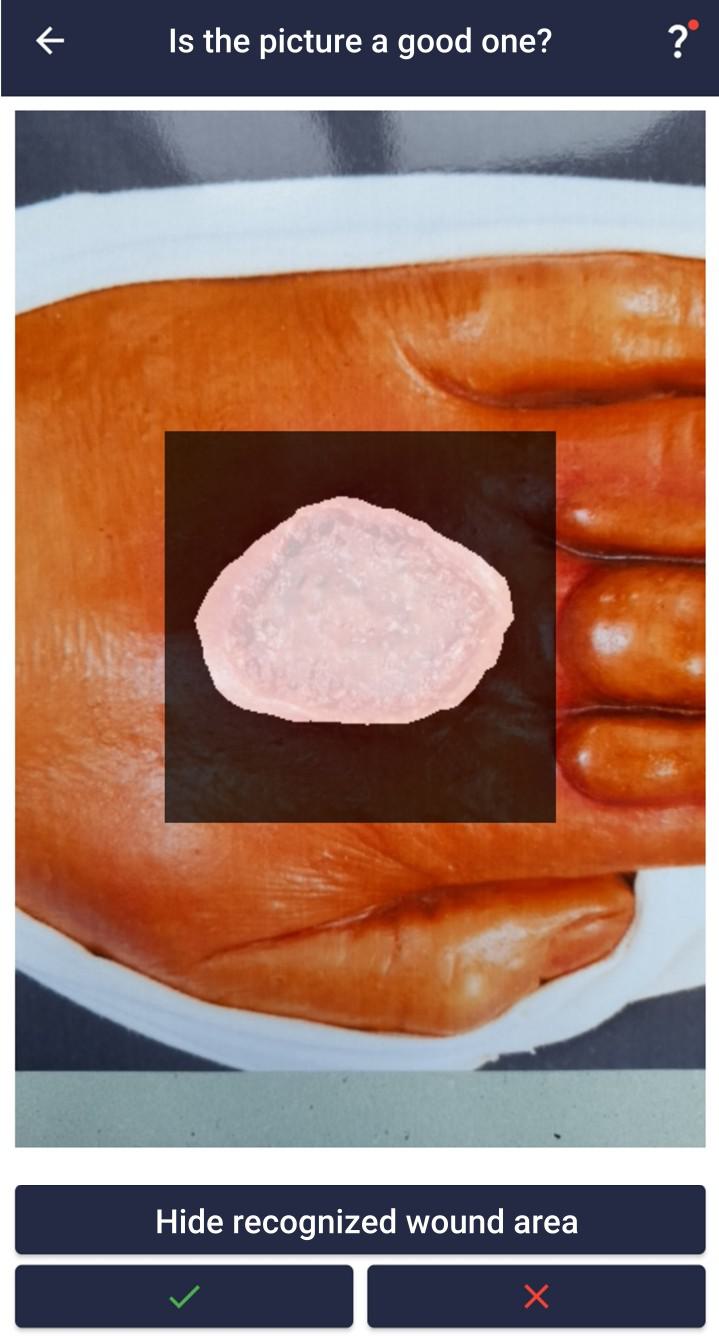}
        \caption{A posteriori segmentation}
        \label{fig:variant2}
    \end{subfigure} 
    \hfill
    \begin{subfigure}{0.32\textwidth}
        \centering
        \includegraphics[width=0.48\textwidth]{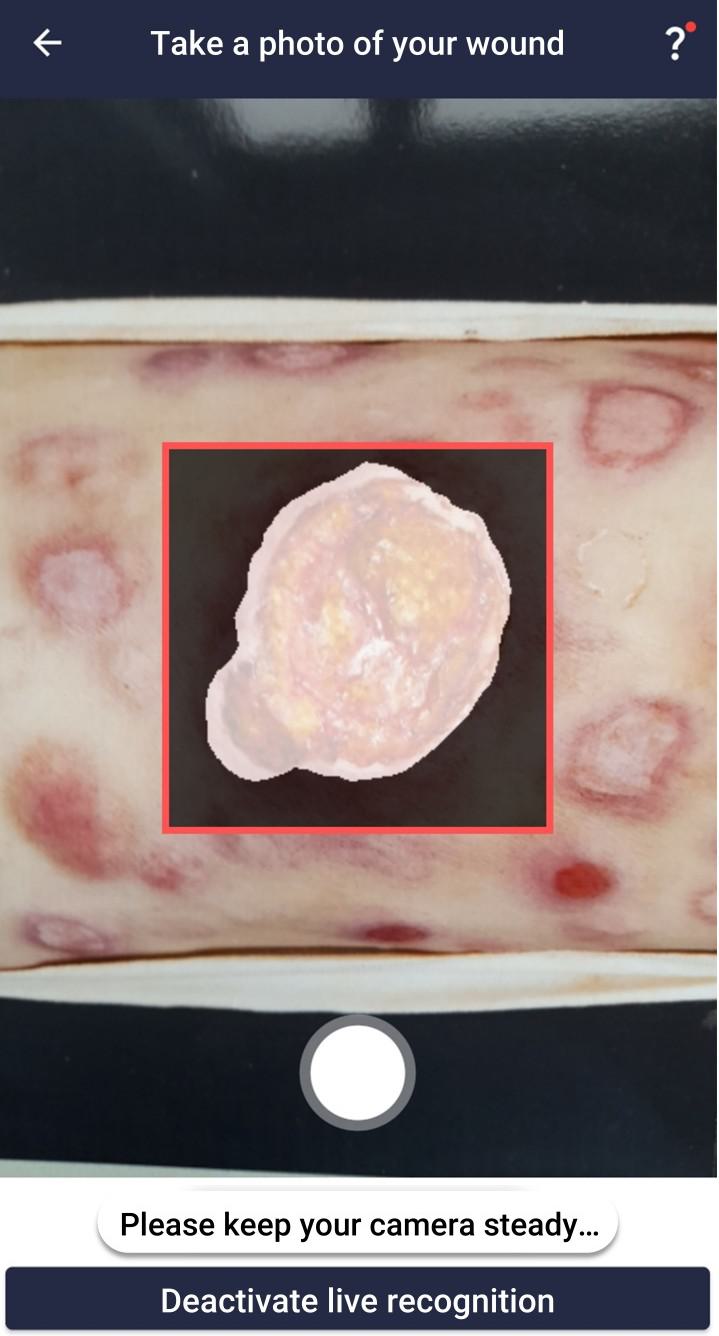}
        \includegraphics[width=0.48\textwidth]{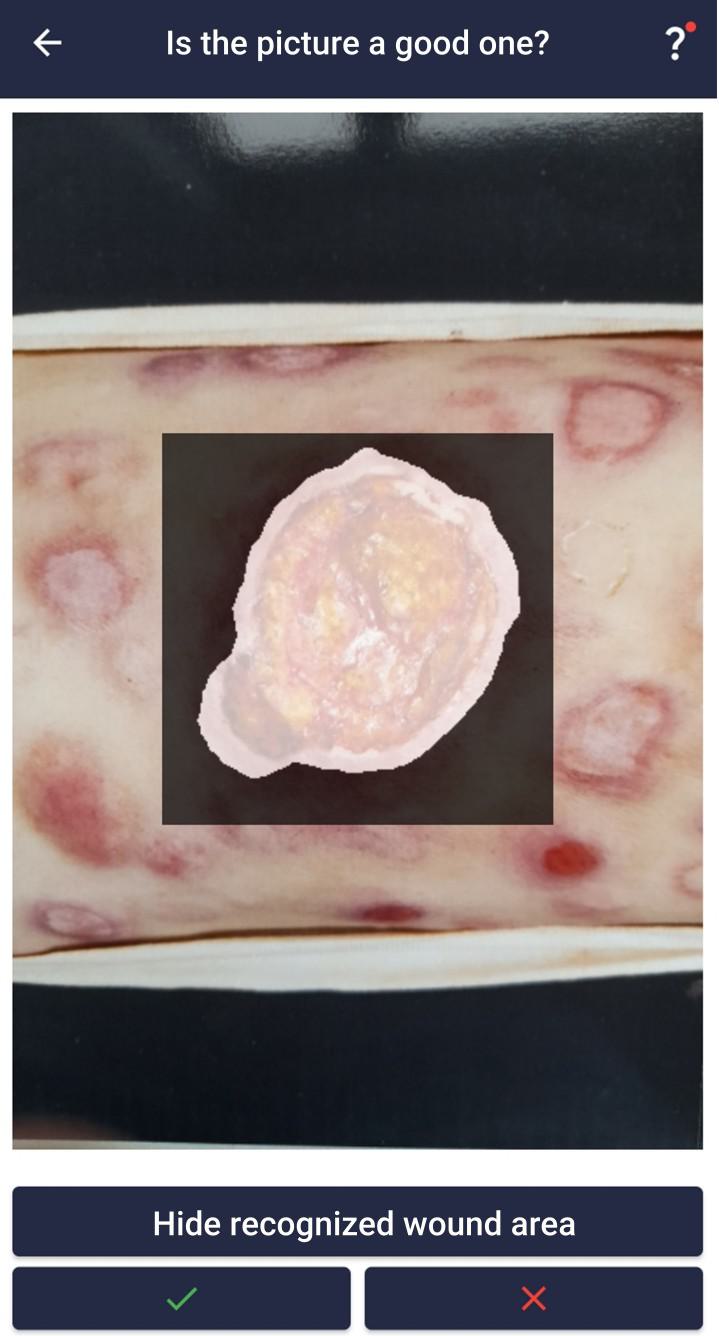}
        \caption{Live and a posteriori segmentation}
        \label{fig:variant3}
    \end{subfigure}

    \caption{Screenshots of the basic and the two wound segmentation variants.}
    \label{fig:screenshots_app:AI_variants}
    \Description{Six app screenshots illustrating the camera interface for capturing wounds at different locations, with two screenshots for each of the three variants: the basic variant without segmentation (right hand), the a posteriori segmentation variant showing the segmented wound after image capture (left hand), and the live segmentation variant displaying wound segmentation both during image capture and after completion (shin).}
\end{figure*}

To investigate how AI-based wound recognition influences user perception during image capture, we implemented three interface variants within the application, as summarized in Fig.~\ref{fig:screenshots_app:AI_variants}:
\begin{enumerate}[label=(\alph*), leftmargin=*, align=left, labelsep=0.2em]
    \item \textbf{Basic variant without segmentation}: Users take a photo of their wound without receiving any feedback. 
    \item \textbf{A posteriori segmentation}: Users take a photo without live feedback. However, in the image confirmation step, they have the option to view the wound area identified by the mobile AI algorithm.
    \item \textbf{Live and a posteriori segmentation}:  Users receive immediate visual feedback on the detected wound area in the live camera stream. After capturing, the recognized wound region can be reviewed during confirmation.
\end{enumerate}

These variants reflect increasing levels of AI visibility and user guidance. 
Variant (a), which lacks visible AI output, serves as a baseline and is included only in the overall preference ranking among the three. 
The core analysis focuses on variants (b) and (c), particularly on the individual effects of a posteriori and live segmentation-based feedback on perceived usefulness and ease of use.
Differing in the timing and extent of AI-based feedback, variant (b) aims to reduce potential distraction and cognitive load associated with real-time annotations, while variant (c) allows to assess whether live annotations are perceived as rather supportive or disruptive. 
Except for the overall preference selection, our evaluation considers the distinct impact of each feedback modality rather than treating the combined functionality of variant (c) as a holistic unit. This allows us to disentangle the specific contributions of real-time versus delayed feedback on user experience.
A \textbf{demo video} illustrating the variants (b) and (c) is included in the supplementary multimedia material. The revised application, incorporating the integrated segmentation model, served as our high-fidelity prototype, \AppNameNoSpace, for subsequent evaluation.

\section{High-Fidelity Prototype Assessment}
\label{sec:methodology:stage3}

\subsection{Stakeholder-Based Usability and Quality Assessment of \textit{WoundAIssist}}
To evaluate our high-fidelity prototype \AppName with the incorporated changes (see Section~\ref{sec:methodology:stage2}), we conducted a second usability study involving both patients and physicians---two key stakeholders in the wound care process. The study aimed to assess the app’s usability and quality from their perspectives, guided by the following RQs:

\begin{enumerate}[label=\textbf{(\Roman*)}, leftmargin=*, align=left, labelsep=-0.3em]

    \item \textbf{RQ-B1: How do stakeholders perceive the overall usability and app quality of \AppNameNoSpace?}

    \item \textbf{RQ-B2: How do stakeholders perceive AI-driven wound segmentation during image acquisition (live) and after image capture (a posteriori)?}
    
    \item \textbf{RQ-B3: Do patients and physicians differ in their attitudes toward the app and its AI component?}

\end{enumerate}

While RQ1 and RQ2 address different factors that likely influence potential adoption in remote wound care, RQ3 focuses on attitudinal differences between patients and physicians. 
The low-fidelity prototype was co-developed with dermatologists (see Section~\ref{design_low_fidelity}), 
and later refined based on patient feedback from \textit{Study A} (see Section~\ref{usability_study_low_fidelity}). 
Although physicians contributed clinical expertise and anticipated patient needs, RQ3 enables a direct comparison between these professional expectations and patients’ actual experiences with \AppNameNoSpace. 
This comparison is particularly relevant given demographic and experiential differences, such as age, clinical familiarity, and technological proficiency, which may influence app interaction and perception.
Besides addresssing the RQs, we compare \AppName with an existing patient-focused wound care app, \textit{WUND APP}~\cite{WUND_APP}, which received the highest average scores in our previous systematic review of patient-centered chronic wound applications~\cite{dege2024}.

\subsubsection{Participants}
Five patients with chronic wounds participated in the study (aged 21-82, $M = 58.00, SD = 22.48$), four of whom were older adults (aged $\ge 60$). Five physicians also took part (aged 24-45, $M = 31.20, SD = 8.23$), including two dermatologists involved in \AppNameNoSpace's iterative development process and three external physicians. 
There were three men and two women in each group. 
Informed consent was obtained from all participants, and the experiment was conducted after consultation with the local institutional ethics committee.

\subsubsection{Measures}
Alongside a brief questionnaire assessing demographic data (age, gender) and prior experience with mobile devices and mHealth apps, standardized instruments were administered to evaluate \AppName and its AI component. Exemplary items for all measures and details on reliability scores are provided in Section~\ref{appendix:ssec:additional_info_questionnaires}. A summary of references to the original English and German versions is included in the Supplementary Material.

\textbf{SUS-DE.}
To assess system usability, we again used the SUS-DE~\cite{brooke2013,gao2020} (see Section~\ref{usability_study_low_fidelity} for a detailed description). 
Scores above 68 are considered ``above average'', above 72 as ``good'', and above 85 as ``excellent'' usabilty~\cite{bangor2009}.

\textbf{MARS-G.}
We applied the German version of the Mobile App Rating Scale (MARS-G)~\cite{messner2020, stoyanov2015} to evaluate app quality. It comprises four primary subscales---\textit{Engagement}, \textit{Functionality}, \textit{Aesthetics}, and \textit{Information Quality}---each rated on a five-point Likert scale, whose mean represents overall app quality.
MARS-G also includes two subjective subscales: \textit{Subjective App Quality} and \textit{Perceived Impact}.
Physicians additionally completed the \textit{Therapeutic Gain} subscale of MARS-G~\cite{messner2020},  whereas patients used the simplified user version (uMARS)~\cite{stoyanov2016}.
Due to short interaction time, the \textit{Perceived Impact} subscale was omitted for both groups. 
One item from \textit{Subjective App Quality} (willingness to pay) was interpolated to a five-point scale ($1\mapsto1$, $2\mapsto3$, $3\mapsto5$) due to limited German response options.
Mean scores can be interpreted as star ratings from 1 (``Inadequate'') to 5 (``Excellent''), with 4 indicating ``Good''~\cite{stoyanov2015}.

\textbf{TAM.}
To assess perceptions of the AI-based segmentation feature, we employed two six-item scales measuring \textit{Perceived Usefulness (PU)} and \textit{Perceived Ease of Use (PEU)} based on Davis' Technology Acceptance Model (TAM)~\cite{davis1989}. 
Items were rated on a seven-point Likert scale, translated into German by two independent researchers and cross-checked with a translation by Jockisch~\cite{jockisch2009}, 
with minor contextual adaptations for targeting a posteriori and live segmentation (see Supplementary Material).
Following Lewis’s recommendations~\cite{lewis2019}, response options were ordered by increasing agreement and scores transformed to a 0–100 scale to align with other usability measures.
As all items were positively worded, scores below 33 indicate negative, 50 neutral, and 67–100 increasingly positive attitudes.
Before completing the TAM questionnaire, participants received a brief explanation and illustrative images of both feedback modes (a posteriori and live) to aid recall. Then, they indicated their preferred option: no segmentation, a posteriori only, or both live and a posteriori feedback.

\textbf{ATI.}
We measured participants' willingness to engage with new technology using the Affinity for Technology Interaction (ATI) scale by Franke et al.~\cite{franke2019}. The German ATI scale consists of nine items rated on a six-point Likert scale from ``Completely disagree'' to ``Completely agree'', with higher scores indicating greater affinity.

\subsubsection{Procedure}
Patients were recruited from the University Hospital Würzburg. An experimenter was present throughout each session to assist with app use and answer questions. After providing informed consent, participants watched a brief video introducing the app's features, received a fictitious patient profile, and completed predefined tasks using \AppName on a Samsung Galaxy A21s.
These tasks, adapted from the initial usability study (see Section~\ref{eval1method}), included revised instructions aligned with the app menu (see Supplementary Material). Participants documented three wound replicas using printed images. For the second wound, they reviewed the a posteriori segmentation; for the third, they enabled both live and a posteriori segmentation. After approximately 10 minutes of app interaction, participants completed the questionnaires. Each session lasted 60 to 75 minutes.

Physicians participated independently without supervision. After providing informed consent, each received a Samsung Galaxy A21s with \AppName pre-installed, a printed study guide with detailed instructions, questionnaires, relevant materials, and contact information for support. They followed the same procedure as patients, without experimenter oversight. Upon finishing, they returned all materials and completed questionnaires.

\subsection{Results}

We present the results of the quality assessment of \AppNameNoSpace, involving dermatologists and patients, in Table~\ref{tab:eval:comparison:WUNDAPP} and Table~\ref{resTAM}. All analyses were conducted with JASP~\cite{jaspteam2021} and $\alpha = .05$.
We conducted a one-way multivariate analysis of variance (MANOVA) to compare SUS-DE scores, MARS-G overall scores, and TAM ratings for a posteriori and live segmentation between physicians and patients.
We compared participants' overall PU and PEU ratings between the two segmentation feature versions using Mann–Whitney U tests with Bonferroni–Holm correction. 
Patients’ SUS-DE scores between the low-fidelity and the high-fidelity version were compared using Student's t-tests, as were the patients' and physicians' ATI scores.
Finally, we compared SUS-DE, overall MARS-G, and ATI scores with those reported for the patient-centered WUND APP in Dege et al.~\cite{dege2024} using Student's t-tests. Detailed information on the assessment of statistical test assumptions is provided in Section~\ref{appendix:ssec:statistical_assumptions}.

\subsubsection{Prior experience with mobile devices.}

Most patients (4/5) owned a smartphone; one additionally used a tablet, and one reported using only a PC. Regular app use was generally low: one patient used no apps, two used 1–5 apps, one used 6–10 apps, and data were missing for one participant. None of the patients reported prior experience with medical apps, indicating limited digital literacy.
All physicians (5/5) regularly used a smartphone, and three additionally used a tablet. App use was higher than among patients: three used 1–5 apps, one used 6–10 apps, and one used more than 10 apps in daily live. All physicians reported prior experience with medical apps.

\subsubsection{System Usability, App Quality, and User Technical Affinity} 

Across all participants, \AppName achieved a mean SUS-DE score of $M = 87.00$ ($SD = 6.54$). The increase in patients’ SUS-DE scores from the low- to the high-fidelity version was not statistically significant, $t(11) = -1.36$, $p = .202$, $d = -0.77$. Nevertheless, \AppNameNoSpace's score rose by 15.5\% compared to the earlier version ($M = 75.30$).
App quality (MARS-G) yielded a mean score of $M = 4.04$ ($SD = 0.30$), with similarly high ratings across the four primary and \textit{Subjective Quality} subscales.
Physicians rated \textit{Therapeutic Gain} at $M = 4.15$ ($SD = 0.38$). 
MANOVA revealed no significant differences between patients and physicians in SUS-DE or overall MARS-G scores ($Wilks' \Lambdaup = .57$, $F(2, 6) = 2.27$, $p = .184$).
Participants’ technical affinity averaged $M = 4.05$ ($SD = 1.00$), with no significant difference between groups, $t(7) = 1.03$, $p = .337$, $d = 0.69$.

\begin{table*}[ht!]
\caption[]{Mean scores of \textit{WUND APP (WA)}~\cite{WUND_APP} and \textit{WoundAIssist (WAI)} for SUS-DE, MARS-G, and ATI (SD in parentheses). 
}
\label{tab:eval:comparison:WUNDAPP}
\centering
\fontsize{8pt}{9pt}\selectfont
\begin{tabular}{l@{\hskip 0.15in}l@{\hskip 0.15in}c@{\hskip 0.15in}l
@{\hskip 0.15in}c@{\hskip 0.12in}c@{\hskip 0.06in}c@{\hskip 0.06in}c@{\hskip 0.06in}c@{\hskip 0.06in}c
@{\hskip 0.15in}c}
\toprule
\multirow{2}{*}{Group}& \multirow{2}{*}{App}& \multirow{2}{*}{$n$}& \multirow{2}{*}{SUS-DE} & \multicolumn{6}{c}{MARS-G} & \multirow{2}{*}{ATI} \\
        \cmidrule(r{10pt}){5-10}
        & & & & Overall & Engmt. & Functionality & Aesthetics & Info. Quality & Subj. Quality & \\ 
\midrule\midrule
\multirow{2}{*}{Overall}    &  \textit{WA} $^{*}$      & 21 & 75.12 (20.35)          & 3.90 (0.52)                   & 3.41 (0.71)               & 4.31 (0.63)           & \textbf{4.13} (0.58)      & 3.74 (0.45)                       &  -                &  3.74 (1.19) \\
                            &  \textit{WAI}               & 10 & \textbf{87.00} (6.54)  & \textbf{4.04} (0.30)$^{\text{a}}$ & \textbf{3.58} (0.55)      & \textbf{4.45} (0.42)  & 3.93 (0.44)               & \textbf{4.32} (0.25)$^{\text{a}}$     &  3.93 (0.68)      &  4.05 (1.00)$^{\text{a}}$  \\
\midrule
\multirow{2}{*}{Physicians} &  \textit{WA} $^{*}$      & 10 & 80.50 (17.71)          & 3.89(0.65)                    & 3.36 (0.89)               & 4.38 (0.66)           & \textbf{4.13 (0.76)}      & 3.67 (0.57)                       &  -               &  3.88 (1.04) \\
                            &  \textit{WAI}               & 5  & \textbf{88.00} (6.47)  & \textbf{4.20} (0.20)          & \textbf{3.80} (0.32)      & \textbf{4.50} (0.31)  & 4.07 (0.28)               & \textbf{4.42} (0.23)              &  3.80 (0.65)     &  4.36 (0.84) \\ 
\midrule
\multirow{2}{*}{Patients}   &  \textit{WA} $^{*}$      & 11 & 70.23 (22.15)          & \textbf{3.90} (0.40)         & \textbf{3.45} (0.54)      & 4.25 (0.62)           & \textbf{4.12} (0.40)      & 3.80 (0.33)                       &  -                &  3.62 (1.35) \\
                            &  \textit{WAI}               & 5  & \textbf{86.00} (7.20)  & 3.83 (0.30)$^{\text{b}}$         & 3.36 (0.68)               & \textbf{4.40} (0.55)  & 3.80 (0.56)               & \textbf{4.19} (0.24)$^{\text{b}}$     &  4.05 (0.76)      &  3.67 (1.17)$^{\text{b}}$ \\
\bottomrule
\\[-2pt]
\multicolumn{11}{l}{%
$^{*}$ Values calculated based on the patients' and physicians' raw scores from the appendix of Dege et al.~\cite{dege2024}.\quad\quad
$^{\text{a}}$ $n=9$\quad\quad
$^{\text{b}}$ $n=4$
}

\end{tabular}
\end{table*}

\subsubsection{Perception of the AI Component}

A posteriori segmentation yielded a mean \textit{PU} of $M = 76.39$ ($SD = 16.63$) and a \textit{PEU} of $M = 83.89$ ($SD = 5.21$), while live segmentation scored $M = 70.83$ ($SD = 15.06$) for \textit{PU} and $M = 82.78$ ($SD = 8.57$) for \textit{PEU}. 
MANOVA showed no significant differences between patients and clinicians in their \textit{PU} and \textit{PEU} ratings ($Wilks' \Lambdaup = .41$, $F(4, 5) = 1.78$, $p = .270$).
Comparing the two segmentation variants across all ten participants separately for each subscale revealed no significant differences, neither for \textit{PU}  ($U = 37.00$, $p_{holm} = .676$, $r = -0.26$) nor for \textit{PEU} ($U = 47.00$, $p_{holm} = .845$, $r = 0.00$). 
For preferred options (no automated recognition, a posteriori, live), 80\% of physicians ($n=4$) chose a posteriori segmentation; one preferred live. Patient preferences were more evenly split: two favored live, two no segmentation, and one a posteriori.

\begin{table}[ht]
\caption{Mean TAM scores for patients and physicians (SD in parentheses) for both a posteriori and live segmentation.}
\label{resTAM}
\centering
\fontsize{8pt}{9pt}\selectfont
\begin{tabular}{l@{\hskip 0.25in}l@{\hskip 0.25in}lcl@{\hskip 0.25in}clc}
\toprule
\multirow{2}{*}{Group} & \multirow{2}{*}{$n$}  & \multicolumn{3}{c}{A Posteriori Segmentation} & \multicolumn{3}{c}{Live Segmentation} \\
\cmidrule(r{10pt}){3-5}\cmidrule(l{1pt}){6-8}
& & \multicolumn{1}{c}{PU} & PEU & Overall & PU & \multicolumn{1}{c}{PEU} & Overall\\
\midrule
Patients   & 5 & 69.44 (22.22)           & 82.78 (4.97)    & 76.11 (16.73)     & 72.78 (13.94)          & 81.67 (5.42)        &   77.22 (11.02)                                            \\
Physicians & 5 & 83.33 (2.78)            & 85.00 (5.76)    & 84.17 (4.35)      & 68.89 (17.50)          & 83.89 (11.52)       &   76.39 (16.05)                                             \\
\midrule
Overall    & 10 & 76.39 (16.63)          & 83.89 (5.21)    & 80.14 (12.59)     & 70.83 (15.06)          &  82.78 (8.57)     &  76.81 (13.41) \\
\bottomrule
\\[-2pt]
\multicolumn{8}{l}{%
\textit{Note}. PU = \textit{Perceived Usefulness}; 
PEU = \textit{Perceived Ease of Use}.
}
\end{tabular}
\end{table}

\subsubsection{Comparison to \textit{WUND APP}}
Welch’s t-test indicated a significant difference in perceived usability (SUS-DE) between WUND APP~\cite{WUND_APP} and \AppNameNoSpace, $t(26.82) = -2.43$, $p = .022$, $d = -0.79$. 
\AppName received higher usability ratings from patients and physicians ($M = 87.00$, $SD = 6.54$) than WUND APP ($M = 75.12$, $SD = 20.35$).

Regarding overall app quality (MARS-G), no significant difference was found between the mean app ratings across physicians and patients for the two apps, $t(27.73) = -0.79$, $p = .438$, $d = -0.27$. 
App quality was perceived descriptively similar for \AppName ($M = 4.04$, $SD = 0.30$) and \textit{WUND APP} ($M = 3.90$, $SD = 0.52$).

Technical affinity (ATI) also did not differ significantly between participants rating either \AppName or WUND APP, $t(28) = -0.68$, $p = .502$, $d = -0.27$. On average, patients and physicians interacting with \AppName reported slightly higher affinity levels ($M = 4.05$, $SD = 1.00$) compared to WUND APP users ($M = 3.74$, $SD = 1.19$).

\subsection{Discussion}
We evaluated the high-fidelity \AppName app through an empirical usability study using standardized assssment instruments. Findings indicated good overall usability and positive app quality ratings from both dermatologists and patients. The AI-driven features were also well received. Below, we address the research questions in detail:

\noindent\textbf{RQ-B1: How do stakeholders perceive the overall usability and app quality of \AppNameNoSpace?}

Both patients and physicians rated \AppName’s usability as ``excellent'' (SUS-DE: 86.00 vs. 88.00), indicating an intuitive, user-friendly interface.
The close alignment between patient and \acp{hcp} ratings suggests that the app effectively addresses usability requirements for both user populations.
%
Overall app quality (MARS-G) was rated as ``good'' with an average score of 4.04. 
Subscale scores were strongest in \textit{Functionality} (4.45) and \textit{Information Quality} (4.32), highlighting technical reliability and content relevance. 
Lower \textit{Engagement} (3.58) and \textit{Aesthetics} (3.93) scores indicate potential for enhancing interaction and visual design.
%
Compared with the low-fidelity prototype, patient SUS-DE scores increased by 10.7 percentage points (75.30 $\rightarrow$ 86.00). While not statistically significant---likely due to small sample size---this suggests design refinements improved usability.

\noindent\textbf{RQ-B2: How do stakeholders perceive AI-driven wound segmentation during image acquisition (live) and after image capture (a posteriori)?}

Both patients and physicians responded positively, with 80\% preferring one of the AI-based variants over no AI.
Physicians favored a posteriori segmentation on both \textit{PU} and \textit{PEU}. In contrast, patients slightly preferred live segmentation in terms of \textit{PU} and overall impression, although
rating a posteriori marginally higher on \textit{PEU}.
Overall, a posteriori segmentation scored slightly higher (\textit{PU}: 76.39 vs. 70.83; \textit{PEU}: 83.80 vs. 82.78), possibly reflecting reduced cognitive load during image capture.
Both variants received consistently favorable ratings (all $>67$), suggesting strong potential of AI-based wound segmentation in clinical contexts. While scores near 83 indicate a clearly positive user perception, those in the range of 69 to 76---though still favorable---suggest opportunities for further refinement.
%
No statistically significant differences were found between a posteriori and live segmentation variants 
on \textit{PU} and \textit{PEU}, suggesting both were well-received. 
Individual preferences may be influenced more by personal or contextual factors than by measurable differences in TAM ratings.

\noindent\textbf{RQ-B3: Do patients and physicians differ in their attitudes toward the app and its AI component?}

No significant differences emerged between physicians and patients in usability (SUS-DE) or app quality ratings (\mbox{MARS-G}), indicating \mbox{\AppName} is similarly experienced and meets both groups’ usability and quality expectations.  
This largely convergent perception across stakeholders is notable, given their differing roles, wound care expertise, and digital literacy. 
One plausible explanation lies in the app’s patient-centered yet clinically grounded design. 
While physicians guided the app’s functionality for clinically meaningful documentation, key interactions were iteratively refined using patient feedback from \textit{Study A}. Qualitative interview responses further indicated that patients valued \AppNameNoSpace’s potential to support wound monitoring and complement in-person visits, potentially contributing to the observed agreement in usability and app quality.

Both groups evaluated the AI component positively, showing no substantial divergence in \textit{PU} or \textit{PEU} ratings. This indicates broadly similar perceptions of AI-based wound segmentation, despite subtle differences in preferences. Physicians generally favored a posteriori segmentation, whereas patients were evenly split between a posteriori and live modes. This finding is somewhat unexpected given patients’ lower digital literacy, as live segmentation might have been perceived as distracting or intrusive.
Nonetheless, the overall convergence in perceptions may be explained by how the segmentation model was integrated into \AppNameNoSpace. For patients, the AI primarily functioned as a visual aid for image capture and interpretation, which was well received. Physicians, who were on average younger, may have had a generally more favorable disposition toward AI technologies, potentially contributing to the similarly positive evaluations observed across both groups.

Overall, these findings suggest that aligning expectations through iterative, multi-stakeholder involvement can foster shared acceptance, even among users with fundamentally different roles and levels of technical expertise.
\section{WoundAIssist in Context: Lessons Learned, Impact, and the Road Ahead}
\label{sec:final_discussion}

\subsection{Lessons and Experience-Based Design Patterns for Patient-Centered Remote Monitoring Apps}

The following summarizes key insights gained from developing \AppName and engaging with patients:

\textit{Reluctance and Social Desirability.}
In the initial usability study, some patients hesitated to give detailed feedback, often due to limited interaction time with the app. In a few cases, responses reflected social desirability rather than candid impressions. While not broadly representative, these observations highlight the need to repeatedly encourage honest, constructive feedback---especially in early-stage usability testing---to improve data quality.

\textit{Assistive App and Interaction Design for Older Adults.}
Certain gestures, such as press-and-hold, proved challenging for older adults, particularly those unfamiliar with such concepts. In contrast, buttons, sliders, and camera-based inputs were perceived as more intuitive. A built-in help function was broadly viewed as helpful, and a clear presentation (e.g., intuitive labels and icons) proved essential. 
Simplifications such as legends may improve comprehension of complex visualizations (e.g., progression curves), and older users may benefit from additional navigational and operational support, including visual cues like counters to facilitate multi-view browsing.

\textit{Digital Literacy.}
Despite potentially lower digital experience, \textit{Study~A} suggested older adults’ confidence with the app may increase with repeated exposure, as reflected in statements like: ``I think I could do better the second time around. If you use it more often, it becomes easier~[...],'', ``I only had a little problem at the beginning'', or ``[I made it] with your help, but in the future I can do that''.
Guided explanations during the interviews also improved comprehension (e.g., ``Well, [I understood the trajectories] with your explanation''), suggesting that sufficient introduction time and active support during initial use are critical for adoption. 
At the same time, \textit{Study~A} revealed that some older patients do not own or rarely use a smartphone, limiting their ability to use apps in daily life and potentially shaping their attitudes toward mHealth. 
Illustrative statements include: ``No, I don’t want to deal with that'' (appointment function), ``No, I don’t think so. Fortunately, my doctor is nearby'' (reduced in-person visits), or ``No, not at my age'' (future app use). Comments like ``I don’t know about that stuff because I don’t have it'' indicate that unfamiliarity with digital technology can reduce willingness to adopt app-based care.

\textit{mHealth as Feasible Complement to In-Person Care.}
Most patients expressed willingness to use \AppName in the future, suggesting mHealth can support continuous care and patient engagement. 
Over half welcomed video consultations, though in-person visits remained highly valued, and some preferred phone calls over app-based scheduling. 
Despite initial usability barriers, many anticipated growing comfort with the app, supporting mHealth as a feasible complement to traditional care, especially when paired with flexible communication channels.

\textit{Distillation of Design Patterns.}
Table~\ref{tab:design-principles} synthesizes lessons learned into experience-based design patterns for patient-centered remote disease monitoring tools, including apps for older adults. While not broadly generalizable due to the limited sample, these patterns draw on more than three years of iterative \AppName development, qualitative patient and clinician feedback, and the long-term clinical expertise of participating physicians.  
Rooted in wound care, the insights are framed to inform remote disease monitoring more broadly. 
They aim to guide the creation of accessible, user-centered, and clinically relevant mHealth solutions by outlining strategies to improve usability, promote patient engagement, and support clinical decision-making.

\begin{table}[p]
\centering
\caption{Experience-based patterns for the design of patient-centered remote disease monitoring applications.}
\label{tab:design-principles}
\fontsize{8.2pt}{9.4pt}\selectfont
\setlength{\tabcolsep}{0.15cm}
\begin{tabular}{p{2cm}p{4.95cm}p{8.35cm}} 
\toprule
\textbf{Pattern} & \textbf{Description} & \textbf{Supporting Findings}\\
\midrule

\textbf{Simplified app and interaction design} &
Intuitive interaction mechanisms, such as clearly labeled buttons, meaningful icons, and simplified input elements (e.g., sliders), can be beneficial for users with limited digital experience. Design should prioritize clarity through consistent navigation cues, visible legends, and built-in help functions, while avoiding complex gestures (e.g., dragging or prolonged touch). &
In \textit{Study A}\textsuperscript{1}, participants handled clearly labeled and visually simple interactions (e.g., image capture and questionnaires) with little difficulty. In contrast, gesture-based or less visible interactions---such as dragging plus press-and-hold actions in trajectory views or arrow-based gallery navigation---frequently caused confusion. Terminological ambiguities (e.g., the ``Contact''  button for appointment scheduling) further impeded task completion, underscoring the need for self-explanatory interaction elements and consistent labeling. Additionally, seven participants expressed interest in permanent in-app assistance.
\\\midrule

\textbf{Guided data\newline acquisition} &
Systems should incorporate user guidance to improve the quality and reliability of patient-generated health data (e.g., anatomical overlays, confirmation prompts). & 
This pattern is grounded in clinicians’ requirements (REQs~1--3), which emphasize structured guidance and image confirmation to ensure the completeness and consistency of patient-generated data. It is further supported by \textit{Study A}\textsuperscript{1}, where patients generally handled guided, step-by-step tasks (e.g., image acquisition and questionnaires) with fewer difficulties than less guided functionalities such as appointment scheduling and video consultations. 
\\\midrule 

\textbf{Longitudinal \newline monitoring}&
Digital tools should enable tracking of health indicators over time---via image comparisons and/or PROs---to support informed clinical decision-making and enhance patients' self-awareness. &
This pattern reflects clinicians’ requirements (REQs~1, 5, and 6) to monitor wound progression via visual documentation and patient-reported outcome measures (PROMs), consistent with current wound care practice and prior research demonstrating that PROMs can engage patients and improve care~\cite{lavallee2016incorporating}. In \textit{Study A}\textsuperscript{1}, most participants expressed willingness to continue using the app and considered it valuable for tracking wound progression, highlighting its relevance for both clinicians and patients.
\\\midrule  

\textbf{Reciprocal \newline communication} &
Applications should support synchronous (e.g., video consultations) and asynchronous (e.g., images, questionnaires) communication modalities between patients and \acp{hcp}, providing multiple channels for interaction (e.g., app and telephone for appointment scheduling) to accommodate diverse user preferences.&
This pattern is grounded in prior work, as a systematic review of telemedicine in dermatology identifies both asynchronous ``store-and-forward'' and synchronous communication as core teledermatology modalities~\cite{trettel2018telemedicine}. In \textit{Study A}\textsuperscript{1}, most patients considered questionnaires useful for wound assessment, indicating acceptance of asynchronous data exchange, while preferences for appointment scheduling varied, with some favoring the app-based approach and others preferring telephone contact. However, the majority of participants were willing to use video consultations and perceived benefits
(e.g., ''video chat for me is the easiest way of communicating with people''). 
\\\midrule

\textbf{Sustained physician\newline engagement}&
By keeping physicians actively engaged in the (communication) loop, systems can foster a sense of continuous care and strengthen patients’ perception of being well-supported throughout treatment. &
In \textit{Study A}\textsuperscript{1}, several participants reported feeling better cared for when using the app, attributing this to more immediate physician involvement (e.g., ``the doctor can see the wound right away and immediately assess whether it has improved or worsened'') and to the expectation of faster support in case of wound-related problems. One undecided participant, regarding feeling better cared for, emphasized the importance of communication (``Yes and no… it has to be communicated, because it always takes two''), further suggesting that timely feedback and sustained physician engagement are central to patients’ perception of continuous care.
\\\midrule

\textbf{Transparent access to\newline personal health trends} &
Patients should have clear and easy access to their self-reported health data, such as images and longitudinal trends. This transparency is essential for enhancing patient engagement, promoting a sense of control over their health, and sustaining app usage.&
This pattern was initially anticipated by the medical team based on their long-term experience with patients (REQs 6, 9, and 10). \textit{Study A}\textsuperscript{1} confirmed its relevance, as most participants (9/11) indicated they would continue using the app. While continued usage may not be solely due to transparent access to personal data, these findings suggest that patients value access to their self-reported health information and perceive tools such as PROMS (e.g., questionnaires) as useful for wound assessment.
\\

\bottomrule
\\[-2pt]
\begin{tabular}[c]{l}
\textit{\textsuperscript{1} Note}. 
For details and results of the low-fidelity usability evaluation (\textit{Study A}), see Section~\ref{usability_study_low_fidelity}.\\
\end{tabular}  
\end{tabular}
\end{table}

\subsection{Practical and Societal Impact of \AppName and the Underlying mHealth solution}

An AI-assisted teledermatology app like \AppName holds promise for improving patient outcomes, increasing self-care engagement, and reducing costs and workload for \acp{hcp}.
Our studies indicate promising user-friendliness and accessibility, suggesting potential for broader adoption in remote wound care.
%
Used over extended periods, \AppName can enhance patient convenience by supporting remote wound documentation while preserving clinical oversight. This is especially beneficial for individuals with limited mobility or in underserved areas, reducing the frequency of in-person visits.
Moreover, continuous monitoring can aid early complication detection and enable timely interventions, potentially improving healing outcomes.
%
Through AI-driven wound size estimation, our mHealth system aims to reduce variability in manual measurements~\cite{langemo2008measuring} and provide a reliable alternative in home environments, where laypersons cannot easily assess wounds.
\AppNameNoSpace's integrated AI can guide users during image acquisition in real-time to improve both quality and clinical relevance of remotely collected images. 
%
Looking ahead, \AppName may support public health by enabling aggregated, anonymized data collection for research to inform effective treatment protocols and refine AI models, ultimately improving wound care quality.

Although promising, our mHealth system entails some downsides and risks. 
Foremost, ensuring privacy and security of sensitive health data is paramount; but, even with robust encryption, secure storage and regulatory compliance, a residual data breach risk persists. 
%
Also, the reliability of AI components warrants thorough validation, particularly for patient-captured images in uncontrolled settings, where suboptimal conditions may compromise wound segmentation and 
user guidance. Similarly, errors in wound size estimation may negatively influence clinical decision-making.
Over-reliance on digital tools may lead patients to forgo necessary in-person consultations, assuming self-monitoring is sufficient, while \acp{hcp} may become less vigilant in critically verifying AI outputs.
%
Finally, while \AppName can bridge gaps in healthcare, it may inadvertently exacerbate the digital divide, limiting its utility for those with insufficient technological access or digital literacy.

\subsection{Limitations and Future Work}
While \AppName shows promise for chronic wound care, this study has limitations that warrant discussion.

\textit{Methodology.}
Two of the five physicians in the second usability study were involved in the app’s development, which may partly explain their slightly higher average ratings compared to patients.
Age differences and varying levels of digital competence may have further influenced these evaluations. 
In addition, the small sample size for both patients and physicians limits the generalizability of the findings. Moreover, some reliability estimates, particularly for the SUS-DE scale, were low (see Table~\ref{quesi}), possibly indicating inconsistencies in user feedback. 

\textit{Ecological Validity.}
The evaluation was conducted as a controlled, lab-based usability study using wound replicas, and therefore does not capture several factors relevant to real-world use, such as patients’ motivation over time, variable lighting and backgrounds, self-positioning during image capture, or integration of app usage into daily routines. Consequently, the findings primarily reflect initial user interaction with the system rather than sustained, correct use in long-term domestic settings. 
Some factors are expected to affect system performance to varying degrees. Minor occlusions of elements of the calibration reference object are likely to have limited impact, as server-side analysis can operate with partial visual information (see~\cite{borst2025woundambit}), and live wound segmentation does not require these elements. In contrast, partially visible wounds, insufficient image quality, or reduced adherence during long-term use may substantially affect performance and cannot be reliably assessed in laboratory settings. Their prevalence and practical significance therefore require investigation through real-world deployment. Finally, redness quantification is still under development and relies on a color reference in the images. Its integration into combined size and redness trajectories remains future work and will require further real-world validation.

\textbf{Open Research Directions.}
Key areas for future investigation include: 
\begin{enumerate*}[label=(\Roman*)]

    \item \textit{Longitudinal study.}   
    A long-term clinical trial has been initiated to evaluate \AppName’s effectiveness under real-world conditions, including image capture in home settings and its impact on patient adherence and wound care outcomes. The study will also enable investigation of factors difficult to capture in laboratory settings, such as fluctuating motivation and environmental variability. Subsequently, a more representative usability and app quality assessment will be conducted.

    \item \textit{Mobile AI validation and improvement.}
    Future efforts will further validate the \textit{TopFormer-Tiny} model across diverse conditions and rare cases to enhance reliability. 
    Data collected during the longitudinal study can be used to retrain the model on real-world images 
    to enhance robustness and generalizability.

    \item \textit{Wound size and redness trajectories.}
    Redness analysis will be integrated into the server-side pipeline, after which size and redness trajectories can be added to \AppName and the clinician interface.

    \item \textit{MDR-compliant app development.} After the pilot study, we plan to re-implement and publish the app in compliance with the Medical Device Regulation (MDR).
    
    \item \textit{Integration of a socially interactive agent.} 
    To further enhance user engagement, personalized guidance, and emotional support~\cite{bickmore2022}, a socially interactive agent will be developed and evaluated separately. 
    
    \item \textit{Multilingual support.} Adding languages will enhance accessibility for a broader patient population.
    
\end{enumerate*}

\section{Conclusion}
\label{sec:conclusion}
In this study, we introduced \AppNameNoSpace, a patient-centered mobile app for AI-assisted wound care at home. The app enables patients to document wounds through photographs and structured questionnaires, supports remote consultations with physicians, and integrates a lightweight AI model to guide patients during image capture via on-device wound segmentation. Although detailed wound analysis presently occurs server-side, leveraging on-device AI enables faster, privacy-preserving local inference and immediate user guidance.
Developed iteratively with direct input from \acp{hcp} and patients, \AppName addresses a critical gap in patient-oriented wound care tools. The initial low-fidelity prototype yielded promising usability results in a formative study with 11 patients, highlighting strong user intention for continued use, while also revealing areas for improvement. These insights informed the refined app version---\AppNameNoSpace---which achieved excellent usability and good overall quality ratings in an evaluation involving five patients and five physicians. Moreover, both stakeholders reported positive perceptions regarding the usefulness and ease of use of the AI-driven wound segmentation.
Lastly, grounded in over three years of interdisciplinary research, this work also distilled lessons learned for the design of patient-centered remote disease monitoring apps.
A longitudinal clinical study is currently underway to evaluate the app’s long-term impact on adherence and outcomes, followed by extended usability assessments with larger, more diverse cohorts.
Overall, \AppName shows strong potential to improve accessibility, patient engagement, and outcomes in chronic wound management, thereby contributing to efficient and scalable healthcare delivery.

\begin{acks}
The authors thank Maike Kübert and Luisa Deutzmann for their support in conducting the initial usability study (\textit{Study A}) and the final study (\textit{Study B}), respectively. We also thank the student assistants, in particular Timo Dittus, Julian Schmidt, and Simon Schwägerl, for their support in the technical implementation of the overall software system. We further acknowledge Hermann Mareth for his assistance with photographing and printing the wound moulage images.

\end{acks}

\bibliographystyle{ACM-Reference-Format}
\bibliography{main} 

@String{Computing = "Computing" }

@String{Computer = "{IEEE} Computer" }

@String{Springer = "Springer-Verlag" }

@inbook{bickmore2022,
author = {Bickmore, Timothy},
title = {Health-Related Applications of Socially Interactive Agents},
year = {2022},
publisher = {Association for Computing Machinery},
address = {New York, NY, USA},
booktitle = {The Handbook on Socially Interactive Agents},
pages = {403–-436}
}

@article{bangor2009,
author = {Bangor, Aaron and Kortum, Philip and Miller, James},
title = {Determining what individual SUS scores mean: adding an adjective rating scale},
year = {2009},
volume = {4},
number = {3},
journal = {Journal of Usability Studies},
pages = {114–-123}
}

@article{mao2005,
author = {Mao, Ji-Ye and Vredenburg, Karel and Smith, Paul W. and Carey, Tom},
title = {The state of user-centered design practice},
year = {2005},
volume = {48},
number = {3},
journal = {Communications of the {ACM}},
pages = {105–-109}
}

@article{wang2022usability,
 author = {Wang, Qiuyi and Liu, Jing and Zhou, Lanshu and Tian, Jing and Chen, Xuemei and Zhang, Wei and Wang, He and Zhou, Wanqiong and Gao, Yitian},
 year = {2022},
 title = {Usability evaluation of mHealth apps for elderly individuals: a scoping review},
 pages = {317},
 volume = {22},
 number = {1},
 journal = {BMC Medical Informatics and Decision Making},
}

@inproceedings{kirillov2023segment,
  author={Kirillov, Alexander and Mintun, Eric and Ravi, Nikhila and Mao, Hanzi and Rolland, Chloe and Gustafson, Laura and Xiao, Tete and Whitehead, Spencer and Berg, Alexander C. and Lo, Wan-Yen and Dollár, Piotr and Girshick, Ross},
  booktitle={IEEE/CVF International Conference on Computer Vision (ICCV)}, 
  title={Segment Anything}, 
  year={2023},
  pages={3992--4003},
  publisher={IEEE}
}

@article{ma2024segment,
  title={Segment anything in medical images},
  author={Ma, Jun and He, Yuting and Li, Feifei and Han, Lin and You, Chenyu and Wang, Bo},
  journal={Nature Communications},
  volume={15},
  number={1},
  pages={654},
  year={2024},
}

@article{kucs2024medsegbench,
  title={MedSegBench: A comprehensive benchmark for medical image segmentation in diverse data modalities},
  author={Ku{\c{s}}, Zeki and Aydin, Musa},
  journal={Scientific Data},
  volume={11},
  number={1},
  pages={1283},
  year={2024},
}

@article{qureshi2023medical,
  title={Medical image segmentation using deep semantic-based methods: A review of techniques, applications and emerging trends},
  author={Imran Qureshi and Junhua Yan and Qaisar Abbas and Kashif Shaheed and Awais Bin Riaz and Abdul Wahid and Muhammad Waseem Jan Khan and Piotr Szczuko},
  journal={Information Fusion},
  volume={90},
  pages={316--352},
  year={2023}
}

@inproceedings{zhang2022delving,
  title={Delving deep into the generalization of vision transformers under distribution shifts},
  author={Zhang, Chongzhi and Zhang, Mingyuan and Zhang, Shanghang and Jin, Daisheng and Zhou, Qiang and Cai, Zhongang and Zhao, Haiyu and Liu, Xianglong and Liu, Ziwei},
  booktitle={IEEE/CVF Conference on Computer Vision and Pattern Recognition (CVPR)},
  pages={7267--7276},
  year={2022},
  publisher={IEEE}
}

@article{liu2021,
 author = {Liu, Na and Yin, Jiamin and Tan, Sharon Swee-Lin and Ngiam, Kee Yuan and Teo, Hock Hai},
 year = {2021},
 title = {Mobile health applications for older adults: a systematic review of interface and persuasive feature design},
 pages = {2483--2501},
 volume = {28},
 number = {11},
 journal = {Journal of the American Medical Informatics Association}
}

@article{broekhuis2019,
 author = {Marijke Broekhuis and Lex {van Velsen} and Hermie Hermens},
 year = {2019},
 title = {Assessing usability of eHealth technology: A comparison of usability benchmarking instruments},
 pages = {24--31},
 volume = {128},
 journal = {International Journal of Medical Informatics},
}

@article{isakovic2016,
 author = {Isakovi{\'c}, Ma{\v{s}}a and Sedlar, Urban and Volk, Mojca and Be{\v{s}}ter, Janez},
 year = {2016},
 title = {Usability Pitfalls of Diabetes mHealth Apps for the Elderly},
 pages = {1604609},
 volume = {2016},
 journal = {Journal of Diabetes Research}
}

@article{Weichbroth2020,
 author = {Weichbroth, Paweł},
 year = {2020},
 title = {Usability of Mobile Applications: A Systematic Literature Study},
 pages = {55563--55577},
 volume = {8},
 journal = {IEEE Access},
}

@techreport{ISO9241-11,
type = {Standard},
key = {ISO 9241-11:2018},
year = {2018},
title = {{Ergonomics of human-system interaction — Part 11: Usability: Definitions and concepts}},
volume = {2018},
institution = {ISO}
}

@article{rubin2021,
 author = {Rubin, Mark},
 year = {2021},
 title = {When to adjust alpha during multiple testing: a consideration of disjunction, conjunction, and individual testing},
 pages = {10969--11000},
 volume = {199},
 number={3},
 journal = {Synthese},
}

@article{ates2019,
 author = {Ate{\c{s}}, Can and Kaymaz, {\"O}zlem and Kale, H Emre and Tekindal, Mustafa Agah},
 year = {2019},
 title = {Comparison of Test Statistics of Nonnormal and Unbalanced Samples for Multivariate Analysis of Variance in terms of Type-I Error Rates},
 pages = {2173638},
 volume = {2019},
 number = {1},
 journal = {Computational and Mathematical Methods in Medicine}
}

@misc{jaspteam2021,
 author = {{JASP Team}},
 year = {2021},
 title = {JASP},
 url = {https://jasp-stats.org/}
}

@article{dege2024,
 author = {Dege, Tassilo and Glatzel, Bernadette and Borst, Vanessa and Gr{\"a}n, Franziska and Goller, Simon and Glatzel, Caroline and Goebeler, Matthias and Schmieder, Astrid},
 year = {2024},
 title = {Patient-Centered Chronic Wound Care Mobile Apps: Systematic Identification, Analysis, and Assessment},
 pages = {e51592},
 volume = {12},
 journal = {JMIR Mhealth Uhealth}
}

@article{lewis2019,
 author = {Lewis, James R.},
 year = {2019},
 title = {Comparison of four TAM item formats: effect of response option labels and order},
 pages = {224--236},
 volume = {14},
 number = {4},
 journal = {Journal of Usability Studies}
}

@article{hamine2015impact,
  title={Impact of mHealth chronic disease management on treatment adherence and patient outcomes: a systematic review},
  author={Hamine, Saee and Gerth-Guyette, Emily and Faulx, Dunia and Green, Beverly B and Ginsburg, Amy Sarah},
  journal={Journal of Medical Internet Research},
  volume={17},
  number={2},
  pages={e52},
  year={2015},
}

@article{anderson2016mobile,
  title={Mobile health apps to facilitate self-care: a qualitative study of user experiences},
  author={Anderson, Kevin and Burford, Oksana and Emmerton, Lynne},
  journal={PLOS ONE},
  volume={11},
  number={5},
  pages = {1--21},
  year={2016},
}

@article{franke2019,
 author = {Thomas Franke and Christiane Attig and Daniel Wessel},
 year = {2019},
 title = {A Personal Resource for Technology Interaction: Development and Validation of the Affinity for Technology Interaction {(ATI)} Scale},
 pages = {456--467},
 volume = {35},
 number = {6},
 journal = {International Journal of Human-Computer Interaction},
}

@article{messner2020,
 author = {Messner, Eva-Maria
        and Terhorst, Yannik
        and Barke, Antonia
        and Baumeister, Harald
        and Stoyanov, Stoyan
        and Hides, Leanne
        and Kavanagh, David
        and Pryss, R{\"u}diger
        and Sander, Lasse
        and Probst, Thomas},
 year = {2020},
 title = {The German Version of the Mobile App Rating Scale (MARS-G): Development and Validation Study},
 pages = {e14479},
 volume = {8},
 number = {3},
 journal = {JMIR Mhealth Uhealth}
}

@article{stoyanov2016,
 author = {Stoyanov, Stoyan R and Hides, Leanne and Kavanagh, David J and Wilson, Hollie},
 year = {2016},
 title = {Development and Validation of the User Version of the Mobile Application Rating Scale (uMARS)},
 pages = {e72},
 volume = {4},
 number = {2},
 journal = {JMIR Mhealth Uhealth}
}

@article{stoyanov2015,
 author = {Stoyanov, Stoyan R and Hides, Leanne and Kavanagh, David J and Zelenko, Oksana and Tjondronegoro, Dian and Mani, Madhavan},
 year = {2015},
 title = {Mobile app rating scale: a new tool for assessing the quality of health mobile apps},
 pages = {e27},
 volume = {3},
 number = {1},
 journal = {JMIR Mhealth Uhealth}
}

@article{brooke2013,
 author = {Brooke, John},
 year = {2013},
 title = {SUS: a retrospective},
 urldate = {25.09.2024},
 pages = {29--40},
 volume = {8},
 number = {2},
 journal = {Journal of Usability Studies}
}

@article{gao2020,
 author = {Meiyuzi Gao and Philip Kortum and Frederick L. Oswald},
 year = {2020},
 title = {Multi-Language Toolkit for the System Usability Scale},
 pages = {1883--1901},
 volume = {36},
 number = {20},
 journal = {International Journal of Human–Computer Interaction}
}

@incollection{jockisch2009,
 author = {Jockisch, Maike},
 title = {Das Technologieakzeptanzmodell},
 pages = {233--254},
 publisher = {Gabler},
 address = {Wiesbaden},
 editor = {Bandow, Gerhard and Holzm{\"u}ller, Hartmut H.},
 booktitle = {{\textquotedbl}Das ist gar kein Modell!{\textquotedbl}},
 year = {2009},
}

@article{davis1989,
 author = {Fred D. Davis},
 year = {1989},
 title = {Perceived Usefulness, Perceived Ease of Use, and User Acceptance of Information Technology},
 pages = {319--340},
 volume = {13},
 number = {3},
 issn = {02767783},
 journal = {MIS Quarterly}
}

@article{kabir2024mobile,
  title={Mobile apps for wound assessment and monitoring: limitations, advancements and opportunities},
  author={Kabir, Muhammad Ashad and Samad, Sabiha and Ahmed, Fahmida and Naher, Samsun and Featherston, Jill and Laird, Craig and Ahmed, Sayed},
  journal={Journal of Medical Systems},
  volume={48},
  number={1},
  pages={80},
  year={2024}
}

@article{griffa2024artificial,
  title={Artificial Intelligence in Wound Care: A Narrative Review of the Currently Available Mobile Apps for Automatic Ulcer Segmentation},
  author={Griffa, Davide and Natale, Alessio and Merli, Yuri and Starace, Michela and Curti, Nico and Mussi, Martina and Castellani, Gastone and Melandri, Davide and Piraccini, Bianca Maria and Zengarini, Corrado},
  journal={BioMedInformatics},
  volume={4},
  number={4},
  pages={2321--2337},
  year={2024}
}

@article{athilingam2018mobile,
  title={Mobile phone apps to support heart failure self-care management: integrative review},
  author={Athilingam, Ponrathi and Jenkins, Bradlee},
  journal={JMIR Cardio},
  volume={2},
  number={1},
  pages={e10057},
  year={2018}
}

@article{trettel2018telemedicine,
  title={Telemedicine in dermatology: findings and experiences worldwide--a systematic literature review},
  author={Trettel, Arick and Eissing, Leah and Augustin, Matthias},
  journal={Journal of the European Academy of Dermatology and Venereology},
  volume={32},
  number={2},
  pages={215--224},
  year={2018},
}

@article{haggag2022large,
  title={A large scale analysis of mHealth app user reviews},
  author={Haggag, Omar and Grundy, John and Abdelrazek, Mohamed and Haggag, Sherif},
  journal={Empirical Software Engineering},
  volume={27},
  number={7},
  pages={196},
  year={2022},
}

@inproceedings{shenoy2018deepwound,
  title={Deepwound: Automated postoperative wound assessment and surgical site surveillance through convolutional neural networks},
  author={Shenoy, Varun N. and Foster, Elizabeth and Aalami, Lauren and Majeed, Bakar and Aalami, Oliver},
  booktitle={IEEE International Conference on Bioinformatics and Biomedicine (BIBM)},
  pages={1017--1021},
  year={2018},
  publisher={IEEE}
}

@misc{WUND_APP,
	author= {{Foundation for Innovation and Education in Wound-Management (FIEW)}},
	year  = {2023},
	title = {Die {WUND APP} - Zum Empowerment von Menschen mit chronischen Wunden},
	note  = {\url{https://web.archive.org/web/20260309060733/https://wundapp.at/}, Accessed: 2026-05-06},
}

@misc{imitoWound,
	author= {{imito AG}},
	year  = {2024},
	title = {{imitoWound}},
	note  = {\url{https://imito.io/imitowound}, Accessed: 2024-09-20},
}

@misc{swiftWoundCareApp,
	author= {{Swift Medical Inc.}},
	year  = {2024},
	title = {Swift Skin and Wound - The New Standard for Wound Care},
	note  = {\url{https://swiftmedical.com/}, Accessed: 2024-10-07},
}

@misc{tissueAnalytics,
    author= {{Net Health}},
	year  = {2025},
	title = {The Essential Wound Imaging Solution for Consistently Accurate Wound Assessments},
	note  = {\url{https://www.nethealth.com/tissue-analytics/}, Accessed: 2025-04-25},
}

@misc{woundVision,
	author= {{WoundVision™}},
	year  = {2024},
	title = {{WoundVision}},
	note  = {\url{https://woundvision.com/}, Accessed: 2025-05-26},
}

@misc{cares4wounds,
	author= {{Tetsuyu Healthcare}},
	year  = {2024},
	title = {{CARES4WOUNDS}},
	note  = {\url{https://tetsuyuhealthcare.com/solutions/wound-management-system/}, Accessed: 2026-05-06},
}

@article{yeo2019woundEducation,
  title={Wound Education App - Providing digitized wound care information on-the-go.},
  author={Yeo, Keith Jun-Yi and Faroukh, Khalisah and Harding, James and Harding, Keith and Yong, Kenneth Yu-Wei},
  journal={APD Trove},
  volume={2},
  number={3},
  year={2019},
  pages={1--5}
}

@inproceedings{borst2025woundambit,
  author="Borst, Vanessa and Dittus, Timo and Dege, Tassilo and Schmieder, Astrid and Kounev, Samuel",
  title="WoundAmbit: Bridging State-of-the-Art Semantic Segmentation and Real-World Wound Care",
  booktitle="European Conference on Machine Learning and Knowledge Discovery in Databases. Applied Data Science Track",
  year="2025",
  publisher="Springer Nature Switzerland",
  address="Cham",
  pages="285--303",
}

@article{wu2019APDSkinMonitoring,
  title={The {APD} Skin Monitoring App for wound monitoring: {Image} processing, area plot, and colour histogram},
  author={Wu, Wei-Ling and Yong, Kenneth Yu-Wei and Federico, Montero Aeole Jasmine and Gan, Samuel Ken-En},
  journal={Scientific Phone Apps and Mobile Devices },
  volume={5},
  number={3},
  year={2019},
  pages={1--9}
}

@article{geirhos2022standardized,
  title={Standardized evaluation of the quality and persuasiveness of mobile health applications for diabetes management},
  author={Geirhos, Agnes and Stephan, Mirjam and Wehrle, Mara and Mack, Chiara and Messner, Eva-Maria and Schmitt, Andreas and Baumeister, Harald and Terhorst, Yannik and Sander, Lasse B.},
  journal={Scientific Reports},
  volume={12},
  number={1},
  pages={3639},
  year={2022}
}

@article{molloy2021engagement,
  title={Engagement with mobile health interventions for depression: A systematic review},
  author={Anthony Molloy and Page L. Anderson},
  journal={Internet Interventions},
  volume={26},
  pages={100454},
  year={2021}
}

@article{zaidan2018review,
  title={A review on smartphone skin cancer diagnosis apps in evaluation and benchmarking: coherent taxonomy, open issues and recommendation pathway solution},
  author={Zaidan, Aos Alaa and Zaidan, Bilal Bahaa and Albahri, Osamah Shihab and Alsalem, Mohammed and Albahri, Ahmed Shihab and Yas, Qahtan and Hashim, Mashitoh},
  journal={Health and Technology},
  volume={8},
  pages={223--238},
  year={2018}
}

@article{lull2022german,
  title={German mobile apps for patients with psoriasis: systematic search and evaluation},
  author={Lull, Christian and von Ahnen, Jan Alwin and Gross, Georg and Olsavszky, Victor and Knitza, Johannes and Leipe, Jan and Schmieder, Astrid},
  journal={JMIR Mhealth Uhealth},
  volume={10},
  number={5},
  pages={e34017},
  year={2022}
}

@article{pereira2016expert,
  title={Expert involvement predicts mHealth app downloads: multivariate regression analysis of urology apps},
  author={Pereira-Azevedo, Nuno and Os{\'o}rio, Lu{\'i}s and Cavadas, Vitor and Fraga, Avelino and Carrasquinho, Eduardo and Cardoso de Oliveira, Eduardo and Castelo-Branco, Miguel and Roobol, Monique J.},
  journal={JMIR Mhealth Uhealth},
  volume={4},
  number={3},
  pages={e86},
  year={2016}
}

@article{pereira2015mhealth,
  title={mHealth in urology: a review of experts’ involvement in app development},
  author={Pereira-Azevedo, Nuno and Carrasquinho, Eduardo and Cardoso de Oliveira, Eduardo and Cavadas, Vitor and Os{\'o}rio, Lu{\'i}s and Fraga, Avelino and Castelo-Branco, Miguel and Roobol, Monique J.},
  journal={PLOS ONE},
  volume={10},
  number={5},
  pages={1--14},
  year={2015}
}

@article{collado2016smartphone,
  title={Smartphone applications for cancer patients; what we know about them?},
  author={Collado-Borrell, Roberto and Escudero-Vilaplana, Vicente and Ribed-S{\'a}nchez, Almudena and Ib{\'a}{\~n}ez-Garc{\'\i}a, Sara and Herranz-Alonso, Ana and Sanjurjo-S{\'a}ez, Mar{\'\i}a},
  journal={Farmacia Hospitalaria},
  volume={40},
  number={1},
  pages={25--35},
  year={2016}
}

@article{luo2019mobile,
  title={Mobile apps for individuals with rheumatoid arthritis: a systematic review},
  author={Luo, Dee and Wang, Penny and Lu, Fengxin and Elias, Josephine and Sparks, Jeffrey A. and Lee, Yvonne C.},
  journal={Journal of Clinical Rheumatology},
  volume={25},
  number={3},
  pages={133--141},
  year={2019}
}

@incollection{or2022human,
  title={Human factors engineering and user-centered design for mobile health technology: Enhancing effectiveness, efficiency, and satisfaction},
  author={Or, Calvin Kalun and Holden, Richard J. and Valdez, Rupa S.},
  booktitle={Human-Automation Interaction:  Mobile Computing},
  pages={97--118},
  editor={Duffy, Vincent G. and Ziefle, Martina and Rau, Pei-Luen Patrick and Tseng, Mitchell M.},
  year={2023},
  publisher={Springer},
  address={Cham},
}

@inproceedings{borst2024early,
  title={Early Explorations of Lightweight Models for Wound Segmentation on Mobile Devices},
  author={Borst, Vanessa and Dittus, Timo and M{\"u}ller, Konstantin and Kounev, Samuel},
  booktitle={German Conference on Artificial Intelligence},
  pages={282--291},
  year={2024},
  address={Cham},
  publisher="Springer Nature Switzerland"
}

@article{nichols2015wound,
  title={Wound assessment part 1: how to measure a wound},
  author={Nichols, Elizabeth},
  journal={Wound Essentials},
  volume={10},
  number={2},
  pages={51--55},
  year={2015}
}

@article{langemo2008measuring,
  title={Measuring wound length, width, and area: which technique?},
  author={Langemo, Diane and Anderson, Julie and Hanson, Darlene and Hunter, Susan and Thompson, Patricia},
  journal={Advances in Skin {\&} Wound Care},
  volume={21},
  number={1},
  pages={42--45},
  year={2008}
}

@incollection{shaw2011wound,
  title = {Wound Measurement in Diabetic Foot Ulceration},
  author={Shaw, Julia and Bell, Patrick M.},
  booktitle = {Global Perspective on Diabetic Foot Ulcerations},
  year={2011},
  publisher = {IntechOpen},
  address = {London},
  editor = {Thanh Dinh},
  pages={71--82}
}

@article{foltynski2015wound,
  title={Wound area measurement with digital planimetry: Improved accuracy and precision with calibration based on 2 rulers},
  author={Foltynski, Piotr and Ladyzynski, Piotr and Ciechanowska, Anna and Migalska-Musial, Karolina and Judzewicz, Grzegorz and Sabalinska, Stanislawa},
  journal={PLOS ONE},
  volume={10},
  number={8},
  pages={1--13},
  year={2015}
}

@article{jorgensen2016methods,
  title={Methods to assess area and volume of wounds--a systematic review},
  author={J{\o}rgensen, Line Bisgaard and S{\o}rensen, Jens A. and Jemec, Gregor BE. and Yderstr{\ae}de, Knud B.},
  journal={International Wound Journal},
  volume={13},
  number={4},
  pages={540--553},
  year={2016}
}

@article{mayrovitz2009wound,
  title={Wound areas by computerized planimetry of digital images: accuracy and reliability},
  author={Mayrovitz, Harvey N. and Soontupe, Lisa B.},
  journal={Advances in Skin {\&} Wound Care},
  volume={22},
  number={5},
  pages={222--229},
  year={2009},
}

@inproceedings{zhang2022topformer,
  title={{TopFormer}: Token Pyramid Transformer for Mobile Semantic Segmentation},
  author={Zhang, Wenqiang and Huang, Zilong and Luo, Guozhong and Chen, Tao and Wang, Xinggang and Liu, Wenyu and Yu, Gang and Shen, Chunhua},
  booktitle={IEEE/CVF Conference on Computer Vision and Pattern Recognition (CVPR)},
  pages={12083--12093},
  year={2022},
  publisher={IEEE}
}

@article{dhar2024fusegnet,
  title={{FUSegNet}: A deep convolutional neural network for foot ulcer segmentation},
  author={Dhar, Mrinal Kanti and Zhang, Taiyu and Patel, Yash and Gopalakrishnan, Sandeep and Yu, Zeyun},
  journal={Biomedical Signal Processing and Control},
  volume={92},
  pages={106057},
  year={2024}
}

@inproceedings{sandler2018mobilenetv2,
  title={{MobileNetV2}: Inverted residuals and linear bottlenecks},
  author={Sandler, Mark and Howard, Andrew and Zhu, Menglong and Zhmoginov, Andrey and Chen, Liang-Chieh},
  booktitle={IEEE/CVF Conference on Computer Vision and Pattern Recognition (CVPR)},
  pages={4510--4520},
  year={2018},
  publisher={IEEE},
  address = {Los Alamitos, CA, USA}
}

@preprint{kendrick2022translating,
  title={Translating clinical delineation of diabetic foot ulcers into machine interpretable segmentation},
  author={Kendrick, Connah and Cassidy, Bill and Pappachan, Joseph M. and O'Shea, Claire and Fernandez, Cornelious J. and Chacko, Elias and Jacob, Koshy and Reeves, Neil D. and Yap, Moi Hoon},
  archivePrefix = {arXiv},
  eprint={2204.11618} ,
  year={2022},
}

@article{wang2024fuseg,
  title={{FUSeg}: The foot ulcer segmentation challenge},
  author={Wang, Chuanbo and Mahbod, Amirreza and Ellinger, Isabella and Galdran, Adrian and Gopalakrishnan, Sandeep and Niezgoda, Jeffrey and Yu, Zeyun},
  journal={Information},
  volume={15},
  number={3},
  pages={140},
  year={2024}
}

@inproceedings{oota2023wsnet,
  title={{WSNet}: Towards an effective method for wound image segmentation},
  author={Oota, Subba Reddy and Rowtula, Vijay and Mohammed, Shahid and Liu, Minghsun and Gupta, Manish},
  booktitle={IEEE/CVF Winter Conference on Applications of Computer Vision (WACV)},
  pages={3234--3243},
  year={2023},
  publisher={IEEE}
}

@inproceedings{song2012automated,
  title={Automated wound identification system based on image segmentation and artificial neural networks},
  author={Song, Bo and Sacan, Ahmet},
  booktitle={IEEE International Conference on Bioinformatics and Biomedicine (BIBM)},
  pages={1--4},
  year={2012},
  publisher={IEEE}
}

@inproceedings{liu2017framework,
  title={A framework of wound segmentation based on deep convolutional networks},
  author={Liu, Xiaohui and Wang, Changjian and Li, Fangzhao and Zhao, Xiang and Zhu, En and Peng, Yuxing},
  booktitle={International Congress on Image and Signal Processing, BioMedical Engineering and Informatics (CISP-BMEI)},
  pages={1--7},
  year={2017},
  publisher={IEEE}
}

@article{chino2020segmenting,
  title={Segmenting skin ulcers and measuring the wound area using deep convolutional networks},
  author={Daniel Y.T. Chino and Lucas C. Scabora and Mirela T. Cazzolato and Ana E.S. Jorge and Caetano Traina-Jr. and Agma J.M. Traina},
  journal={Computer Methods and Programs in Biomedicine},
  volume={191},
  pages={105376},
  year={2020}
}

@article{wang2016area,
  title={Area determination of diabetic foot ulcer images using a cascaded two-stage {SVM}-based classification},
  author={Wang, Lei and Pedersen, Peder C. and Agu, Emmanuel and Strong, Diane M. and Tulu, Bengisu},
  journal={IEEE Transactions on Biomedical Engineering},
  volume={64},
  number={9},
  pages={2098--2109},
  year={2016},
  publisher={IEEE}
}

@inproceedings{kolesnik2005multi,
  title={Multi-dimensional color histograms for segmentation of wounds in images},
  author={Kolesnik, Marina and Fexa, Ales},
  booktitle={International Conference on Image Analysis and Recognition (ICIAR)},
  pages={1014--1022},
  year={2005},
  publisher={Springer}
}

@inproceedings{goyal2017fully,
  title={Fully convolutional networks for diabetic foot ulcer segmentation},
  author={Goyal, Manu and Yap, Moi Hoon and Reeves, Neil D. and Rajbhandari, Satyan and Spragg, Jennifer},
  booktitle={IEEE International Conference on Systems, Man, and Cybernetics (SMC)},
  pages={618--623},
  year={2017},
  publisher={IEEE}
}

@article{li2018composite,
  title={A composite model of wound segmentation based on traditional methods and deep neural networks},
  author={Li, Fangzhao and Wang, Changjian and Liu, Xiaohui and Peng, Yuxing and Jin, Shiyao},
  journal={Computational Intelligence and Neuroscience},
  volume = {2018},
  number = {1},
  pages = {4149103},
  year={2018}
}

@article{wang2020fully,
  title={Fully automatic wound segmentation with deep convolutional neural networks},
  author={Wang, Chuanbo and Anisuzzaman, D. M. and Williamson, Victor and Dhar, Mrinal Kanti and Rostami, Behrouz and Niezgoda, Jeffrey and Gopalakrishnan, Sandeep and Yu, Zeyun},
  journal={Scientific Reports},
  volume={10},
  number={1},
  pages={21897},
  year={2020},
}

@conference{rocha2021woundarch,
author={Carlos Diego F. da Rocha and Bruno S. Carvalho and Vítor G. Marques and Bruno M. Silva},
title={{WoundArch}: A Hybrid Architecture System for Segmentation and Classification of Chronic Wounds},
booktitle={International Joint Conference on Biomedical Engineering Systems and Technologies (BIOSTEC) - Volume 5: HEALTHINF},
year={2021},
pages={651-658},
publisher={SciTePress},
}

@article{varma2016vision,
  title={Vision based pus segmentation and area estimation of wound using android application},
  author={Varma, Anudeep and Varma, Ashok and Jakkampudi, Hema Madhav and Raj, Alex Noel Joseph},
  journal={Indian Journal of Science and Technology},
  year={2016},
  volume={9},
  number={3},
  pages={1--7}
}

@article{ferreira2021experimental,
  title={Experimental study on wound area measurement with mobile devices},
  author={Ferreira, Filipe and Pires, Ivan Miguel and Ponciano, Vasco and Costa, M{\'o}nica and Villasana, Mar{\'\i}a Vanessa and Garcia, Nuno M. and Zdravevski, Eftim and Lameski, Petre and Chorbev, Ivan and Mihajlov, Martin and Trajkovik, Vladimir},
  journal={Sensors},
  volume={21},
  number={17},
  pages={5762},
  year={2021}
}

@article{cazzolato2021utrack,
  title={The UTrack framework for segmenting and measuring dermatological ulcers through telemedicine},
  author={Mirela T. Cazzolato and Jonathan S. Ramos and Lucas S. Rodrigues and Lucas C. Scabora and Daniel Y.T. Chino and Ana E.S. Jorge and Paulo Mazzoncini {de Azevedo-Marques} and Caetano Traina and Agma J.M. Traina},
  journal={Computers in Biology and Medicine},
  volume={134},
  pages={104489},
  year={2021},
  publisher={Elsevier}
}

@article{zhang2022surveyWoundAI,
  title={A survey of wound image analysis using deep learning: Classification, detection, and segmentation},
  author={Zhang, Ruyi and Tian, Dingcheng and Xu, Dechao and Qian, Wei and Yao, Yudong},
  journal={IEEE Access},
  volume={10},
  pages={79502--79515},
  year={2022},
  publisher={IEEE}
}

@article{mohammed2023implementing,
  title={Implementing Technology in Practice: Factors Associated with Clinicians’ Satisfaction with an AI Wound Assessment Solution},
  author={Mohammed, Heba Tallah and Cassata, Amy and Fraser, Robert D. J. and Mannion, David},
  journal={International Journal of Nursing and Health Care Research},
  volume={6},
  pages={1452},
  year={2023}
}

@article{ramachandram2022fully,
  title={Fully automated wound tissue segmentation using deep learning on mobile devices: Cohort study},
  author={Ramachandram, Dhanesh and Ramirez-GarciaLuna, Jose Luis and Fraser, Robert D J and Mart{\'i}nez-Jim{\'e}nez, Mario Aurelio and Arriaga-Caballero, Jesus E and Allport, Justin},
  journal={JMIR Mhealth Uhealth},
  volume={10},
  number={4},
  pages={e36977},
  year={2022}
}

@article{frey2023association,
  title={Association Between the Characteristics of mHealth Apps and User Input During Development and Testing: Secondary Analysis of App Assessment Data},
  author={Frey, Anna-Lena and Baines, Rebecca and Hunt, Sophie and Kent, Rachael and Andrews, Tim and Leigh, Simon},
  journal={JMIR Mhealth Uhealth},
  volume={11},
  pages={e46937},
  year={2023}
}

@article{thisanke2023semantic,
  title={Semantic segmentation using Vision Transformers: A survey},
  author={Hans Thisanke and Chamli Deshan and Kavindu Chamith and Sachith Seneviratne and Rajith Vidanaarachchi and Damayanthi Herath},
  journal={Engineering Applications of Artificial Intelligence},
  volume={126},
  pages={106669},
  year={2023}
}

@article{lateef2019survey,
  title={Survey on semantic segmentation using deep learning techniques},
  author={Lateef, Fahad and Ruichek, Yassine},
  journal={Neurocomputing},
  volume={338},
  pages={321--348},
  year={2019}
}

@inproceedings{long2015fully,
  title={Fully convolutional networks for semantic segmentation},
  author={Long, Jonathan and Shelhamer, Evan and Darrell, Trevor},
  booktitle={IEEE Conference on Computer Vision and Pattern Recognition (CVPR)},
  pages={3431--3440},
  year={2015},
  publisher={IEEE}
}

@inproceedings{girshick2014rich,
  title={Rich feature hierarchies for accurate object detection and semantic segmentation},
  author={Girshick, Ross and Donahue, Jeff and Darrell, Trevor and Malik, Jitendra},
  booktitle={IEEE Conference on Computer Vision and Pattern Recognition (CVPR)},
  pages={580--587},
  year={2014},
  publisher={IEEE}
}

@inproceedings{zheng2021rethinking,
  title={Rethinking semantic segmentation from a sequence-to-sequence perspective with transformers},
  author = {Zheng, Sixiao and Lu, Jiachen and Zhao, Hengshuang and Zhu, Xiatian and Luo, Zekun and Wang, Yabiao and Fu, Yanwei and Feng, Jianfeng and Xiang, Tao and Torr, Philip H.S. and Zhang, Li},
  booktitle={IEEE/CVF Conference on Computer Vision and Pattern Recognition (CVPR)},
  pages={6877-6886},
  year={2021},
  publisher={IEEE}
}

@inproceedings{liu2021swin,
  title={Swin transformer: Hierarchical vision transformer using shifted windows},
  author={Liu, Ze and Lin, Yutong and Cao, Yue and Hu, Han and Wei, Yixuan and Zhang, Zheng and Lin, Stephen and Guo, Baining},
  booktitle={IEEE/CVF International Conference on Computer Vision (ICCV)},
  pages={9992-10002},
  year={2021},
  publisher={IEEE}
}

@inproceedings{cheng2022masked,
  title={Masked-attention mask transformer for universal image segmentation},
  author={Cheng, Bowen and Misra, Ishan and Schwing, Alexander G. and Kirillov, Alexander and Girdhar, Rohit},
  booktitle={IEEE/CVF Conference on Computer Vision and Pattern Recognition (CVPR)},
  pages={1290--1299},
  year={2022},
  publisher={IEEE}
}

@inproceedings{wan2023seaformer,
  title={{SeaFormer}: Squeeze-enhanced Axial Transformer for Mobile Semantic Segmentation},
  author={Wan, Qiang and Huang, Zilong and Lu, Jiachen and Gang, YU and Zhang, Li},
  booktitle={International Conference on Learning Representations (ICLR)},
  year={2023},
  publisher={OpenReview.net}
}

@inproceedings{mehta2021mobilevit,
  title={{MobileViT}: Light-weight, General-purpose, and Mobile-friendly Vision Transformer},
  author={Sachin Mehta and Mohammad Rastegari},
  booktitle={International Conference on Learning Representations (ICLR)},
  year={2022},
  publisher={OpenReview.net}
}

@inproceedings{xu2023pidnet,
  title={{PIDNet}: A real-time semantic segmentation network inspired by {PID} controllers},
  author={Xu, Jiacong and Xiong, Zixiang and Bhattacharyya, Shankar P.},
  booktitle={IEEE/CVF Conference on Computer Vision and Pattern Recognition (CVPR)},
  pages={19529--19539},
  year={2023},
  publisher={IEEE}
}

@article{wang2022medical,
  title={Medical image segmentation using deep learning: A survey},
  author={Wang, Risheng and Lei, Tao and Cui, Ruixia and Zhang, Bingtao and Meng, Hongying and Nandi, Asoke K.},
  journal={IET Image Processing},
  volume={16},
  number={5},
  pages={1243--1267},
  year={2022}
}

@inproceedings{ronneberger2015unet,
  title={{U-net}: Convolutional networks for biomedical image segmentation},
  author={Ronneberger, Olaf and Fischer, Philipp and Brox, Thomas},
  booktitle={International Conference on Medical Image Computing and Computer-Assisted Intervention (MICCAI)},
  pages={234--241},
  year={2015},
  publisher={Springer}
}

@inproceedings{valanarasu2022unext,
  title={{UNeXt}: {MLP}-based rapid medical image segmentation network},
  author={Valanarasu, Jeya Maria Jose and Patel, Vishal M.},
  booktitle={International Conference on Medical Image Computing and Computer-Assisted Intervention (MICCAI)},
  pages={23--33},
  year={2022},
  publisher={Springer}
}

@preprint{chen2021transunet,
  author={Jieneng Chen and Yongyi Lu and Qihang Yu and Xiangde Luo and Ehsan Adeli and Yan Wang and Le Lu and Alan L. Yuille and Yuyin Zhou},
  title={{TransUNet}: Transformers make strong encoders for medical image segmentation},
  year={2021},
  archivePrefix = {arXiv},
  eprint={2102.04306}
}

@inproceedings{fan2021rethinking,
  title={Rethinking {BiSeNet} for real-time semantic segmentation},
  author={Fan, Mingyuan and Lai, Shenqi and Huang, Junshi and Wei, Xiaoming and Chai, Zhenhua and Luo, Junfeng and Wei, Xiaolin},
  booktitle={IEEE/CVF Conference on Computer Vision and Pattern Recognition (CVPR)},
  pages={9711-9720},
  year={2021},
  publisher={IEEE}
}

@article{najm2019mobile,
  title={Mobile health apps for self-management of rheumatic and musculoskeletal diseases: Systematic literature review},
  author={Najm, Aur{\'e}lie and Gossec, Laure and Weill, Catherine and Benoist, David and Berenbaum, Francis and Nikiphorou, Elena},
  journal={JMIR Mhealth Uhealth},
  volume={7},
  number={11},
  pages={e14730},
  year={2019},
}

@article{renner2017depression,
  title={Depression and quality of life in patients with chronic wounds: ways to measure their influence and their effect on daily life},
  author={Renner, Regina and Erfurt-Berge, Cornelia},
  journal={Chronic Wound Care Management and Research},
  pages={143--151},
  year={2017},
  volume={4}
}

@article{herberger2011quality,
  title={Quality of life and satisfaction of patients with leg ulcers--results of a community-based study},
  author={Herberger, Katharina and Rustenbach, Stephan J. and Haartje, O. and Blome, Christine and Franzke, Nadine and Sch{\"a}fer, Ines and Radtke, Marc and Augustin, Matthias},
  journal={Vasa},
  volume={40},
  number={2},
  pages={131--138},
  year={2011}
}

@article{purwins2010cost,
  title={Cost-of-illness of chronic leg ulcers in Germany},
  author={Purwins, Sandra and Herberger, Katharina and Debus, Eike Sebastian and Rustenbach, Stephan J and Pelzer, Peter and Rabe, Eberhard and Sch{\"a}fer, Elmar and Stadler, Rudolf and Augustin, Matthias},
  journal={International Wound Journal},
  volume={7},
  number={2},
  pages={97--102},
  year={2010},
}

@article{gould2015chronic,
  title={Chronic wound repair and healing in older adults: Current status and future research},
  author = {Gould, Lisa and Abadir, Peter and Brem, Harold and Carter, Marissa and Conner-Kerr, Teresa and Davidson, Jeff and DiPietro, Luisa and Falanga, Vincent and Fife, Caroline and Gardner, Sue and Grice, Elizabeth and Harmon, John and Hazzard, William R. and High, Kevin P. and Houghton, Pamela and Jacobson, Nasreen and Kirsner, Robert S. and Kovacs, Elizabeth J. and Margolis, David and McFarland Horne, Frances and Reed, May J. and Sullivan, Dennis H. and Thom, Stephen and Tomic-Canic, Marjana and Walston, Jeremy and Whitney, JoAnne and Williams, John and Zieman, Susan and Schmader, Kenneth},
  journal={Wound Repair and Regeneration},
  volume={23},
  number={1},
  pages={1--13},
  year={2015}
}

@article{chandan2023humanWounds,
author = {Chandan K. Sen},
title = {Human Wound and Its Burden: Updated 2022 Compendium of Estimates},
journal = {Advances in Wound Care},
volume = {12},
number = {12},
pages = {657-670},
year = {2023},
}

@inproceedings{he2016deep,
  title={Deep residual learning for image recognition},
  author={He, Kaiming and Zhang, Xiangyu and Ren, Shaoqing and Sun, Jian},
  booktitle={IEEE Conference on Computer Vision and Pattern Recognition (CVPR)},
  pages={770--778},
  year={2016},
  publisher={IEEE}
}

@inproceedings{vaswani2017attention,
  title={Attention is all you need},
  author = {Vaswani, Ashish and Shazeer, Noam and Parmar, Niki and Uszkoreit, Jakob and Jones, Llion and Gomez, Aidan N. and Kaiser, \L{}ukasz and Polosukhin, Illia},
  booktitle = {International Conference on Neural Information Processing Systems (NeurIPS)},
  pages = {6000–6010},
  year={2017},
  publisher = {Curran Associates, Inc.}
}

@inproceedings{dosovitskiy21ViT,
  author       = {Alexey Dosovitskiy and Lucas Beyer and Alexander Kolesnikov and Dirk Weissenborn and Xiaohua Zhai and Thomas Unterthiner and Mostafa Dehghani and Matthias Minderer and Georg Heigold and Sylvain Gelly and Jakob Uszkoreit and Neil Houlsby},
  title        = {An Image is Worth 16x16 Words: Transformers for Image Recognition at Scale},
  booktitle    = {International Conference on Learning Representations (ICLR)},
  year         = {2021},
  publisher    = {OpenReview.net}
}

@article{verhoeven2010asynchronous,
  title={Asynchronous and synchronous teleconsultation for diabetes care: A systematic literature review},
  author={Fenne Verhoeven and Karin Tanja-Dijkstra and Nicol Nijland and Gunther Eysenbach and Lisette van Gemert-Pijnen},
  journal={Journal of Diabetes Science and Technology},
  volume={4},
  number={3},
  pages={666--684},
  year={2010},
}

@inproceedings{howard2019searching,
  title={Searching for {MobileNetV3}},
  author = {Howard, Andrew and Sandler, Mark and Chen, Bo and Wang, Weijun and Chen, Liang-Chieh and Tan, Mingxing and Chu, Grace and Vasudevan, Vijay and Zhu, Yukun and Pang, Ruoming and Le, Quoc and Adam, Hartwig},
  booktitle={IEEE/CVF International Conference on Computer Vision (ICCV)},
  pages={1314--1324},
  year={2019},
  publisher={IEEE}
}

@incollection{elguerapaez2019,
 author = {Elguera Paez, Lesly and Zapata Del R\'{\i}o, Claudia},
 title = {Elderly Users and Their Main Challenges Usability with Mobile Applications: A Systematic Review},
 pages = {423--438},
 volume = {11583},
 publisher = {Springer},
 editor = {Marcus, Aaron and Wang, Wentao},
 booktitle = {Design, User Experience, and Usability. Design Philosophy and Theory},
 year = {2019},
 address = {Cham},
}

@article{ramdowar2024,
 author = {Ramdowar, Harshini and Khedo, Kavi Kumar and Chooramun, Nitish},
 year = {2024},
 title = {A comprehensive review of mobile user interfaces in mHealth applications for elderly and the related ageing barriers},
 pages = {1613--1629},
 volume = {23},
 number = {4},
 journal = {Universal Access in the Information Society},
}

@article{wildenbos2018,
 author = {Wildenbos, Gaby Anne and Peute, Lina and Jaspers, Monique},
 year = {2018},
 title = {Aging barriers influencing mobile health usability for older adults: A literature based framework (MOLD-US)},
 pages = {66--75},
 volume = {114},
 journal = {International Journal of Medical Informatics}
}

@article{lavallee2016incorporating,
  title={Incorporating patient-reported outcomes into health care to engage patients and enhance care},
  author={Lavallee, Danielle C. and Chenok, Kate E. and Love, Rebecca M. and Petersen, Carolyn and Holve, Erin and Segal, Courtney D. and Franklin, Patricia D.},
  journal={Health Affairs},
  volume={35},
  number={4},
  pages={575--582},
  year={2016}
}

@article{fischer2020importance,
  title={The importance of user involvement: a systematic review of involving older users in technology design},
  author={Fischer, Bj{\"o}rn and Peine, Alexander and {\"O}stlund, Britt},
  journal={The Gerontologist},
  volume={60},
  number={7},
  pages={e513--e523},
  year={2020},
}

@misc{anx_fast_facts_on_wound_care,
	author= {ANX Home Healthcare},
	year  = {2018},
	title = {Fast Facts on Wound Care (Infographic)},
	note  = {\url{https://www.anxlife.com/infographics/fastfactsonwoundcare/}, Accessed: 2026-05-06},
}

@misc{Healogics_wound_care_awareness,
	author= {Healogics, Inc.},
	year  = {2021},
	title = {2021 Wound Care Awareness Infographic},
	note  = {\url{https://www.healogics.com/wp-content/uploads/2021/05/2021_Wound-Care-Infographic_8.5x11_Healogics.pdf}, Accessed: 2025-12-22},
}

@misc{isla_NHS_statistics,
	author= {Isla Health},
	year  = {2024},
	title = {Transforming chronic wound care management in the NHS: scaling solutions for growing challenges},
	note  = {\url{https://isla.health/resource/transforming-chronic-wound-care-management-in-the-nhs/}, Accessed: 2025-12-22},
}

@article{sen2009human,
  title={Human skin wounds: A major and snowballing threat to public health and the economy},
  author={Sen, Chandan K. and Gordillo, Gayle M. and Roy, Sashwati and Kirsner, Robert and Lambert, Lynn and Hunt, Thomas K. and Gottrup, Finn and Gurtner, Geoffrey C. and Longaker, Michael T.},
  journal={Wound Repair and Regeneration},
  volume={17},
  number={6},
  pages={763--771},
  year={2009}
}

\clearpage
\appendix

\counterwithin{figure}{section} 
\counterwithin{table}{section} 
\renewcommand{\thefigure}{\thesection\arabic{figure}}
\renewcommand{\thetable}{\thesection\arabic{table}}

\vspace*{\baselineskip}
\noindent{\Huge{\textbf{- Appendix -}}}
\vspace*{3\baselineskip}
\section{Technical Background and Integration of the Mobile AI Model into \textit{WoundAIssist}}
\label{appendix:implementation_details}

This section provides a detailed, self-contained overview of the mobile AI model used for wound segmentation, including its derivation and integration into \AppNameNoSpace. While the main manuscript emphasizes real-world validation and usability evaluation, these technical details are included to document the foundational aspects of the AI model for readers seeking deeper technical context.

In Section~\ref{appendix:ssec:mobile_semantic_segmentation}, we review advances in general, mobile, and medical semantic segmentation, highlighting limitations specific to wound segmentation and lightweight model alternatives to contextualize the relevance of the AI component within \AppNameNoSpace. Section~\ref{appendix:ssec:topformer_architecture} provides an in-depth description of the employed model architecture, \textit{TopFormer-Tiny}. Section~\ref{appendix:ssec:topformer_training_and_eval} summarizes key elements of our prior study~\cite{borst2024early}, including the training and evaluation procedures used to develop the AI model. Finally, Section~\ref{appendix:ssec:topformer_integration_into_app} describes the technical integration of the derived mobile AI model into the Flutter-based \AppName smartphone application.

\subsection{(Mobile) Semantic Segmentation: Advances and Medical Applications} 
\label{appendix:ssec:mobile_semantic_segmentation}
Recent advances in semantic segmentation (SS) are largely driven by Convolutional Neural Networks (CNNs) alongside breakthroughs in Vision Transformers (ViTs)~\cite{dosovitskiy21ViT} such as \textit{SETR}~\cite{zheng2021rethinking}, \textit{Swin Transformer}~\cite{liu2021swin}, and \textit{Mask2Former}~\cite{cheng2022masked}. 
A comprehensive review of ViTs for SS is provided by Thisanke et al.~\cite{thisanke2023semantic}, while Lateef and Ruichek~\cite{lateef2019survey} focus on non-transformer-based Deep Learning techniques, including Fully Convolutional Networks~(FCN)~\cite{long2015fully} and Regional Convolutional Neural Networks~(R-CNN)~\cite{girshick2014rich}. 
%
Since many real-world SS tasks, such as scene understanding, require models to run with extremely low latency, including on resource-constrained mobile devices, 
various models have been developed for \textit{mobile SS}. Examples include CNN architectures using lightweight backbones, such as the \textit{MobileNet} series~\cite{sandler2018mobilenetv2,howard2019searching}, 
combined with different decoders, and approaches such as \textit{PIDNet}~\cite{xu2023pidnet} or the \textit{STDC} network~\cite{fan2021rethinking}. 
In addition, transformer-based methods such as \textit{MobileViT}~\cite{mehta2021mobilevit}, \textit{Seaformer}~\cite{wan2023seaformer}, and \textit{TopFormer}~\cite{zhang2022topformer} are becoming increasingly important for mobile segmentation tasks.

In medical segmentation, there has also been a notable shift towards Deep Learning, with Wang et al.~\cite{wang2022medical} providing an extensive survey on the subject. 
Prominent methods in medical applications include the seminal \textit{UNet}~\cite{ronneberger2015unet}, \textit{UNeXt}~\cite{valanarasu2022unext}, or the ViT-based \textit{TransUNet}~\cite{chen2021transunet}. Furthermore, foundation models like \textit{Segment Anything}~\cite{kirillov2023segment} have recently been adapted for medical images~\cite{ma2024segment}. Several application areas have gained significant attention, primarily due to the availability of extensive benchmarking datasets. For example, the MedSegBench benchmark combines 35 datasets with over 60,000 images from multiple medical imaging modalities, such as ultrasound, MRI, and X-ray, and covers a wide range of anatomical regions and pathologies, including lung, eye, and skin (dermoscopic images)~\cite{kucs2024medsegbench}.

However, despite significant advancements in both (mobile) SS and medical image segmentation, certain application areas, such as wound segmentation, remain insufficiently explored, with a limited number of relevant datasets and a relatively small body of research dedicated to this domain. A recent review indicated that most of the medical segmentation techniques focus on modalities such as X-ray, CT, MRI, and ultrasound~\cite{qureshi2023medical}, rather than smartphone-captured RGB images. In addition, robust wound segmentation methods remain underdeveloped in terms of their mobile applicability, as evidenced by the following key limitations: 
\begin{enumerate}[leftmargin=*]
    \item \textbf{Limited research on wound segmentation}: Early wound segmentation was focused on traditional feature-engineering-based Machine Learning and simple Artificial Neural Networks~\cite{song2012automated,wang2016area,kolesnik2005multi}. Over time, end-to-end Deep Learning approaches, such as \textit{WSNet}\cite{oota2023wsnet}, \textit{FUSegNet}~\cite{dhar2024fusegnet}, and other CNN-based techniques~\cite{wang2020fully,liu2017framework,chino2020segmenting,goyal2017fully} have emerged, including composite models that combine traditional methods with Deep Neural Networks~\cite{li2018composite}. Nonetheless, research specifically focused on wound segmentation remains relatively limited, particularly regarding the adoption of ViTs. Notably, a 2022 survey on Deep Learning for wound image analysis~\cite{zhang2022surveyWoundAI} did not mention any Transformer-based methods for segmentation. Given that ViTs have been shown to have a weaker background bias and generalize better than CNNs under most types of distribution shifts~\cite{zhang2022delving}, this observation is striking and highlights a significant gap in their application in the field.
    \item \textbf{Lack of lightweight models for mobile applications}:  
    Most wound segmentation methods continue to rely on large, complex architectures primarily designed for server-side inference~\cite{zhang2022surveyWoundAI}, with limited emphasis on developing lightweight models specifically optimized for mobile deployment. 
    While some studies suggest that certain architectures could be used in mobile environments, their evaluations typically focus on theoretical efficiency metrics such as FLOPs or parameter counts~\cite{wang2020fully,oota2023wsnet}. However, neither such metrics nor conventional accuracy measures (e.g., IoU) may accurately reflect real mobile performance. 
    In our prior work~\cite{borst2024early}, we demonstrated this discrepancy by comparing the applicability of several non-wound-specific architectures for mobile wound segmentation, including visual assessments within a minimally functional smartphone app. 
\end{enumerate}

\emph{
While our previous study~\cite{borst2024early} addressed some of these limitations, they also underscore the novelty of the mobile AI feature in \AppNameNoSpace. The combination of a lightweight, mobile-optimized segmentation model with structured PROs and direct connectivity to treating dermatologists represents, to our knowledge, a unique integration. Although the AI model was not trained as part of this study, its real-time integration for user-guided image capture allows evaluation of patients’ interaction with and perception of on-device wound segmentation. Furthermore, this integration establishes a foundation for future analyses within the ongoing long-term pilot study, which collects longitudinal data to evaluate AI performance under everyday conditions.
}

\subsection{TopFormer-Tiny Architecture}
\label{appendix:ssec:topformer_architecture}
\begin{figure*}[ht]
\includegraphics[width=\textwidth]{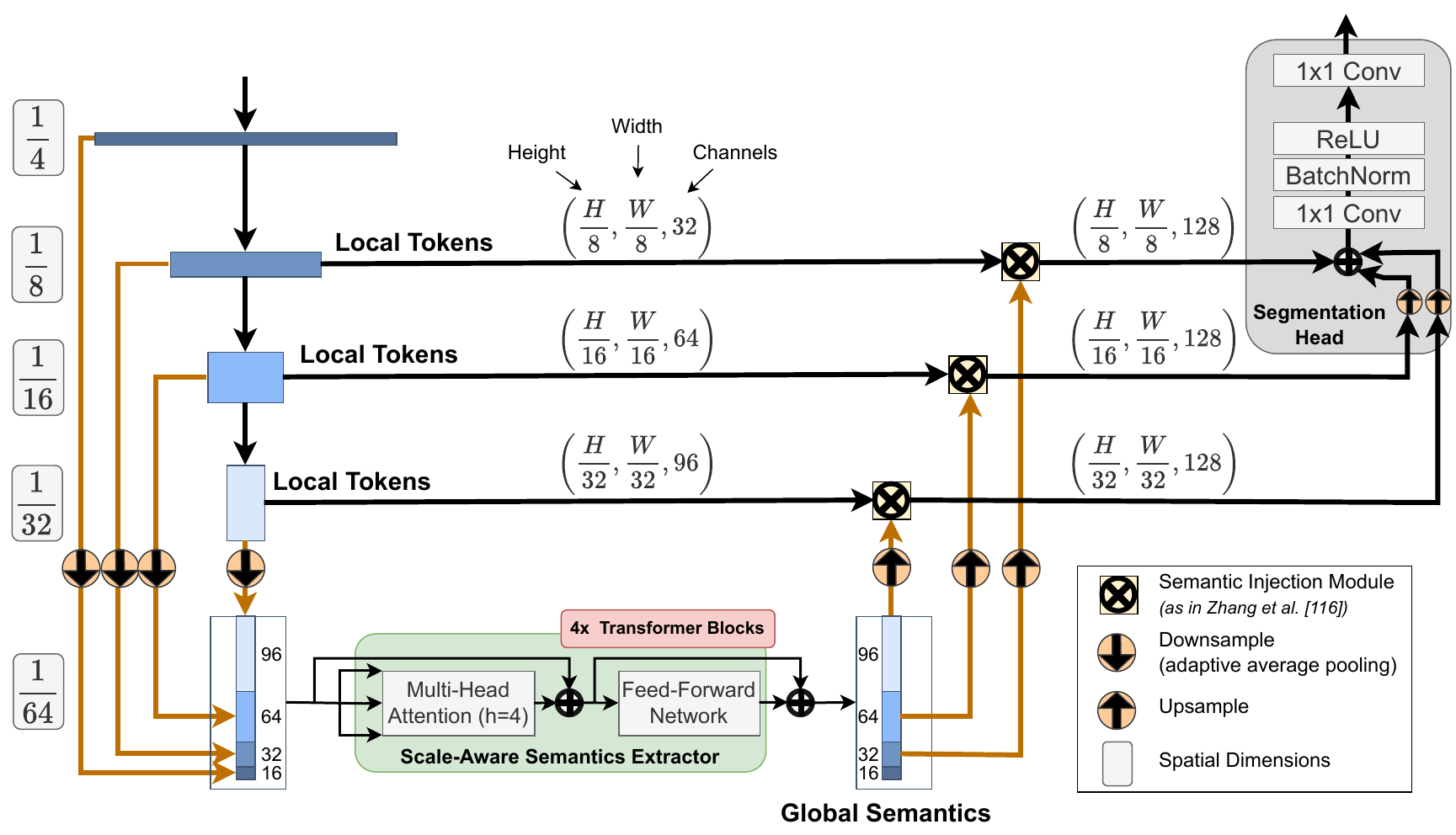}
\caption{Architecture of the TopFormer-Tiny model used in \AppName (custom adaptation from Zhang et al.~\cite{zhang2022topformer} that focuses on the properties of the \textit{tiny} variant and includes additional details regarding the token dimensions across scales).}
\label{fig:topformer}
\Description{A diagram illustrating the deep learning architecture of the TopFormer-Tiny model integrated into the \AppName app, as described in the main paper. The figure highlights the primary components and their roles in processing input data for segmentation tasks: the Token Pyramid Module (left), the Scale-Aware Semantics Extractor (center), the Semantic Injection Module (indicated only and referenced to the original paper), and the Segmentation Head (top-right). Arrows indicate data flow, including upsampling and downsampling operations, as well as the integration of local tokens into the Semantic Injection Modules. Spatial and channel dimensions are annotated for each scale to provide additional clarity. }
\end{figure*}

Building on our previous findings~\cite{borst2024early}, we selected the \textit{TopFormer-Tiny}---a lightweight variant of the \textbf{To}ken \textbf{P}yramid Vision Trans\textbf{former}~\cite{zhang2022topformer}---due to its compact architecture (1.39M trainable parameters) and strong segmentation capabilities on mobile platforms, including for wound analysis tasks.
\textit{TopFormer}, introduced by Zhang et al. at CVPR 2022~\cite{zhang2022topformer}, is a hybrid architecture tailored for mobile SS. Combining the strengths of CNNs and ViTs, it demonstrated superior performance compared to existing CNN- and ViT-based models across three segmentation benchmarks.
It employs a CNN-based \textit{Token Pyramid Module}, which is mainly composed of a few stacked \textit{MobileNetV2} blocks~\cite{sandler2018mobilenetv2} to capture local features from the input image. The tokens from different scales, downsampled to $\frac{1}{64 \times 64}$ of the original size via average pooling, are concatenated along the channels and then processed by a ViT-based semantics extractor to obtain scale-aware global semantics. 
Thereafter, the model's \textit{Semantic Injection Modules (SIMs)} process the retrieved global semantics, divided into channels of features from different scales. For each of the last three scales, the corresponding \textit{SIM} extrapolates its global semantic chunk to the required size and injects it into the local tokens of the corresponding scale to obtain powerful hierarchical features. 
Finally, the segmentation head upscales the low-resolution tokens to align with the high-resolution tokens, followed by element-wise summation. Two convolutional layers are applied to generate the final segmentation map, which is then upscaled to match the input size. The whole segmentation process is depicted in Figure~\ref{fig:topformer}, which we adapted from Zhang et al.~\cite{zhang2022topformer} for clarity. 

\subsection{Training and Evaluation of \textit{TopFormer-Tiny}}
\label{appendix:ssec:topformer_training_and_eval}

This work focuses on the integration of a mobile AI model into \AppName and examines the perceived benefits of AI-assisted wound segmentation from the perspectives of patients and \acp{hcp}. 
The complete training and evaluation workflow, including data pre-processing, model training, and quantitative results, is detailed in our previous study~\cite{borst2024early}. 
Here, rather than refining segmentation accuracy by further optimizing our existing AI model, we deploy and evaluate a fully operational end-to-end system, focusing on assessing user perceptions.
Nevertheless, we briefly summarize the training process for transparency: the \textit{TopFormer-Tiny} model was trained on a curated dataset of 2,887 annotated foot ulcer images, compiled from the FUSeg~\cite{wang2024fuseg} and DFUC2022~\cite{kendrick2022translating} datasets after duplicate removal. The data were partitioned into training (60\%), validation (20\%), and test (20\%) subsets. Then, the model was fine-tuned using standard augmentation and optimization techniques. The best-performing checkpoint, selected based on validation IoU, was evaluated on a hold-out test set and also qualitatively assessed during mobile deployment~\cite{borst2024early}.

\subsection{Integration of \textit{TopFormer-Tiny} into \AppNameNoSpace}
\label{appendix:ssec:topformer_integration_into_app}

As described in the main paper, the trained \textit{TopFormer-Tiny} architecture was selected for integration into \AppName based on our prior comparative study~\cite{borst2024early}. While the deployment procedure largely followed the approach briefly outlined in that work, we provide a more detailed description here, emphasizing additional implementation aspects not included in the previous minimal report.

\subsubsection{Deployment}
Following successful validation and testing on the corresponding splits of the public wound datasets (see~\cite{borst2024early}), we converted the trained \textit{TopFormer-Tiny} PyTorch model to TorchScript\footnote{\mbox{TorchScript: \url{https://pytorch.org/docs/stable/jit.html}}} using its \texttt{trace} method, which records operations during a forward pass with a sample input image.
TorchScript allows the model to be serialized and run outside the standard Python environment, such as on mobile platforms.
Next, we optimized the traced model using the \texttt{optimize\_for\_mobile} method from PyTorch Mobile\footnote{\mbox{PyTorch~Mobile: \url{https://pytorch.org/mobile/home/}}}, applying optimization passes tailored for mobile devices. 
Finally, the optimized model was saved in a format compatible with the PyTorch Lite interpreter, which is suitable for Flutter applications\footnote{\mbox{PyTorch Lite Flutter package: \url{https://pub.dev/packages/pytorch_lite/}}}.

\subsubsection{Inference}
During inference, the \textit{TopFormer-Tiny} model processes center-cropped images of size 224 × 224 pixels to enhance computational efficiency. 
Following the preliminary testing results from our prior work~\cite{borst2024early}, a prediction threshold of 0.75 is applied to generate binary masks and reduce jitter in the segmentation output. 
The lightweight architecture enables real-time wound segmentation directly on patients' smartphones, providing immediate feedback on the detected wound area in the live camera feed. 
To account for the variability in smartphone hardware, we do not impose a fixed frame rate; instead, a new image is processed as soon as the previous segmentation is complete, leveraging the device's maximum performance. The processing rate thus depends on the processing power of the smartphone.

\subsubsection{Validation of the Deployed AI Solution}
For initial validation of the deployed mobile wound segmentation approach, rigorous testing was conducted on two mid-range Android smartphones as part of our prior study~\cite{borst2024early}: a OnePlus 7 Pro (2019) and a Samsung Galaxy A21s (2020). 
The \textit{TopFormer-Tiny} model demonstrated fairly stable behavior on both devices, operating without interruption or major performance issues. 
Although perceptible delays in segmentation smoothness were noted when the device was in motion, the absence of stalling and the stabilizing segmentation masks at rest suggests the overall feasibility of real-time wound segmentation using Flutter, even on non-high-end mobile devices. 
In the context of the present study, we extended this validation to cover more recent devices and different wound shapes, anatomical locations, and lighting conditions. These results, presented in the main paper (Section~\ref{ssec:technical_components:validation_contemporary_devices}), provide qualitative real-world validation and further contextualize the suitability of the integrated AI component for the conducted user perception study.

\section{Screenshots and Evolution of the Low-Fidelity Prototype}
\label{appendix:screenshots_low_fidelity}

\begin{figure*}[ht!]
    \centering
    \begin{subfigure}{0.23\textwidth}
        \centering
        \includegraphics[width=\linewidth]{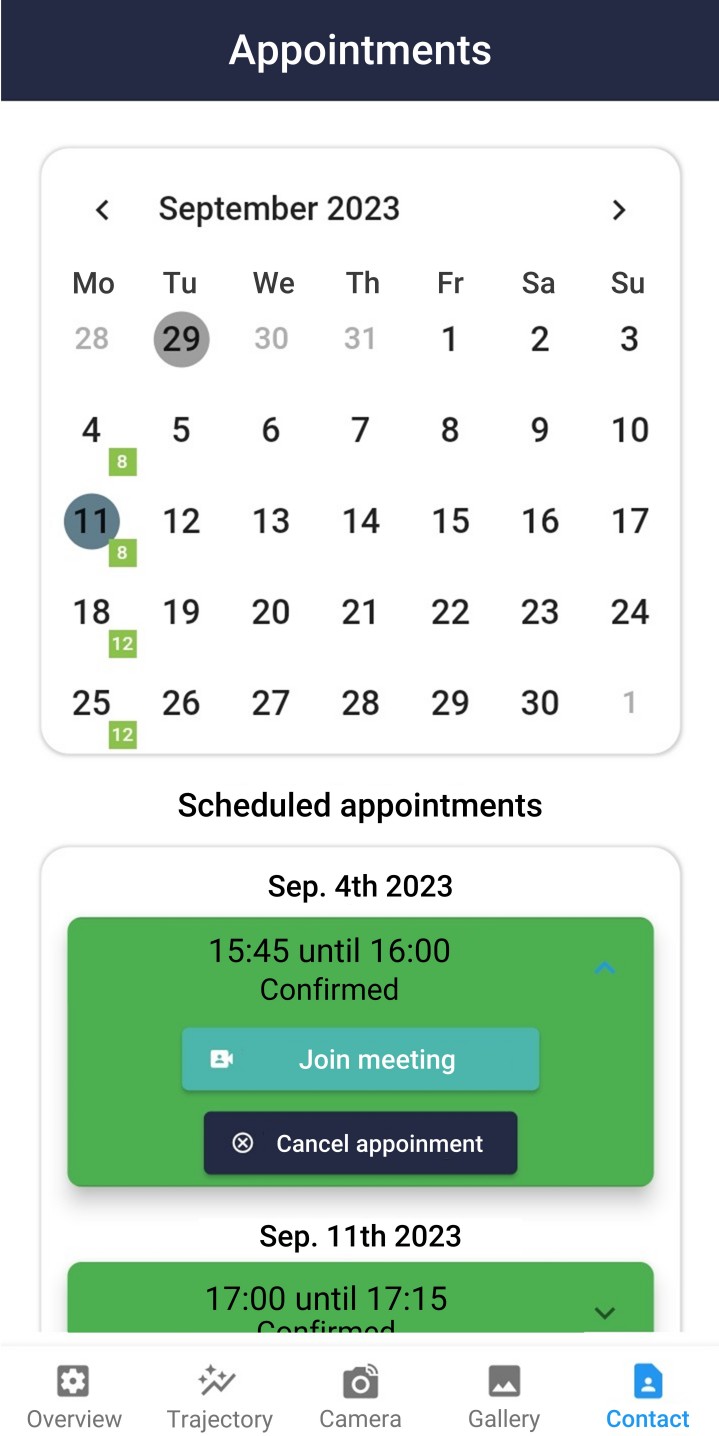}
        \caption{Calendar \textit{(misnamed)}}
        \label{fig:appendix:low_fidelity:calendar}
    \end{subfigure} \hfill
    \begin{subfigure}{0.23\textwidth}
        \centering
        \includegraphics[width=\linewidth]{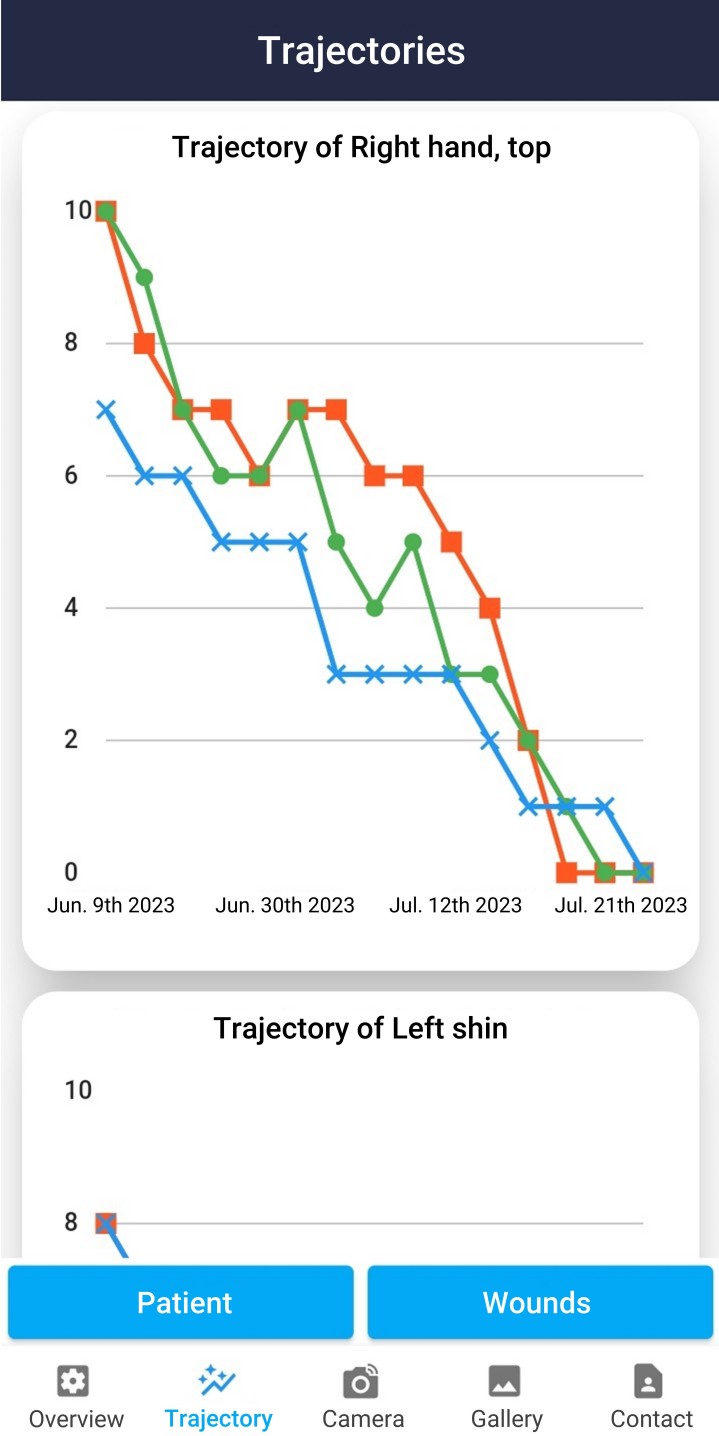}
        \caption{Wound trajectory}
        \label{fig:appendix:low_fidelity:wound_trajectory}
    \end{subfigure} \hfill
    \begin{subfigure}{0.23\textwidth}
        \centering
        \includegraphics[width=\linewidth]{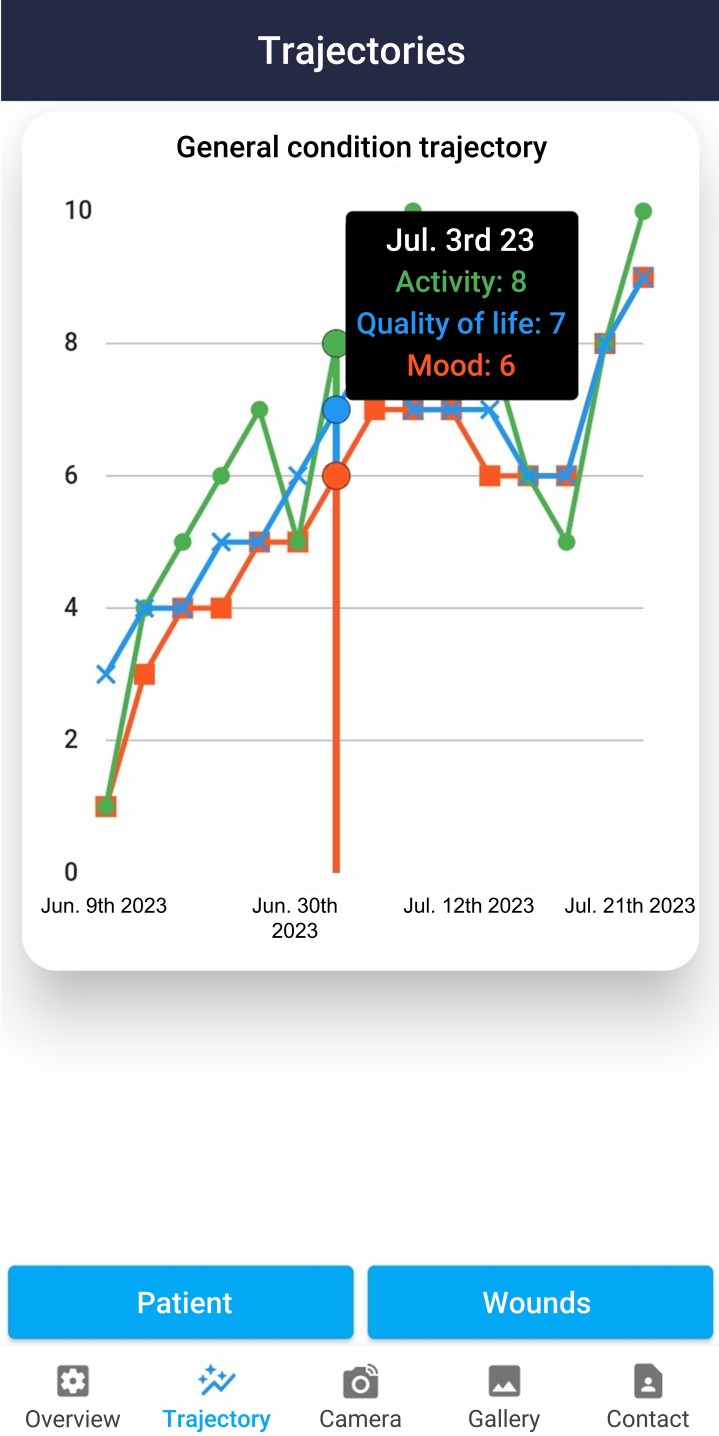}
        \caption{General health trajectory}
        \label{fig:appendix:low_fidelity:general_condition_trajectory}
    \end{subfigure} \hfill   
    \begin{subfigure}{0.23\textwidth}
        \centering
        \includegraphics[width=\linewidth]{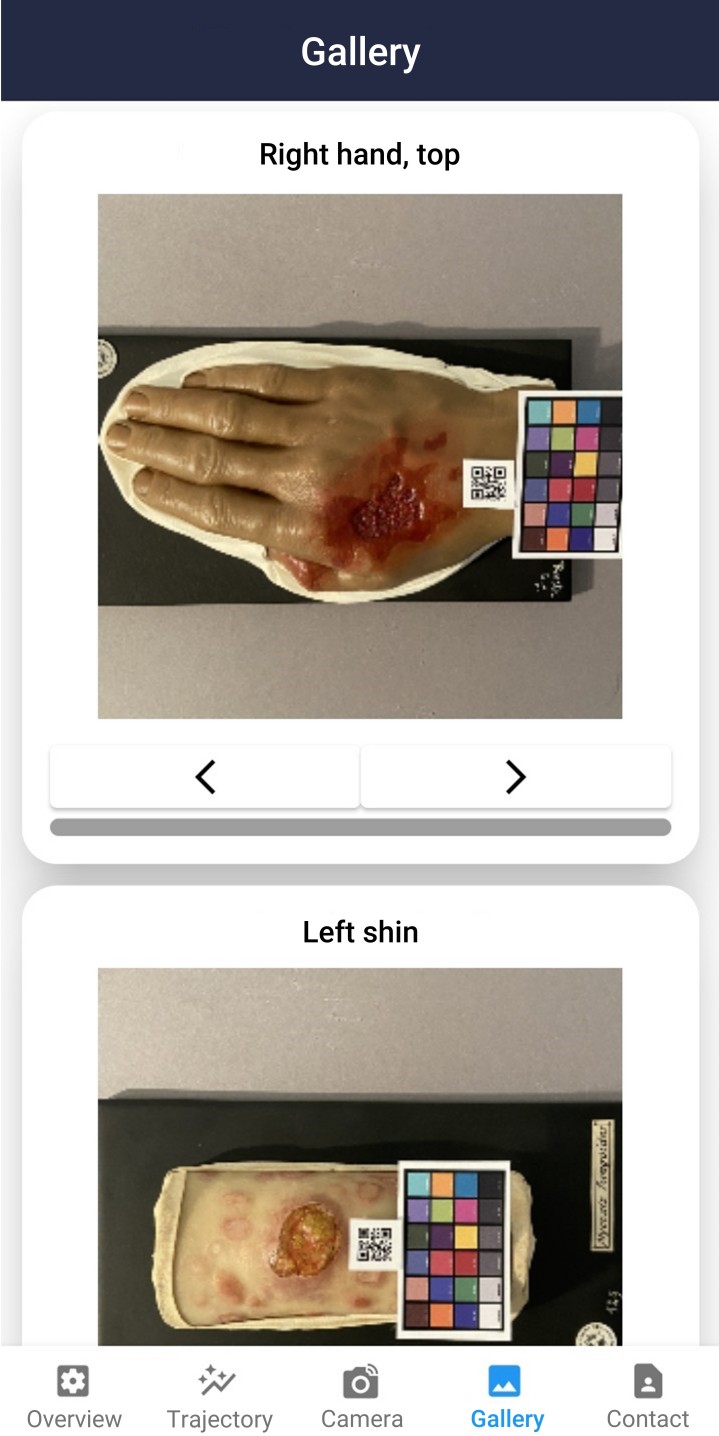}
        \caption{Gallery \textit{(w/o counter)}}
        \label{fig:appendix:low_fidelity:gallery}
    \end{subfigure} \hfill
    \\\vspace{0.5cm} 
    \begin{subfigure}{0.3\textwidth}
        \centering
        \includegraphics[width=0.76\linewidth]{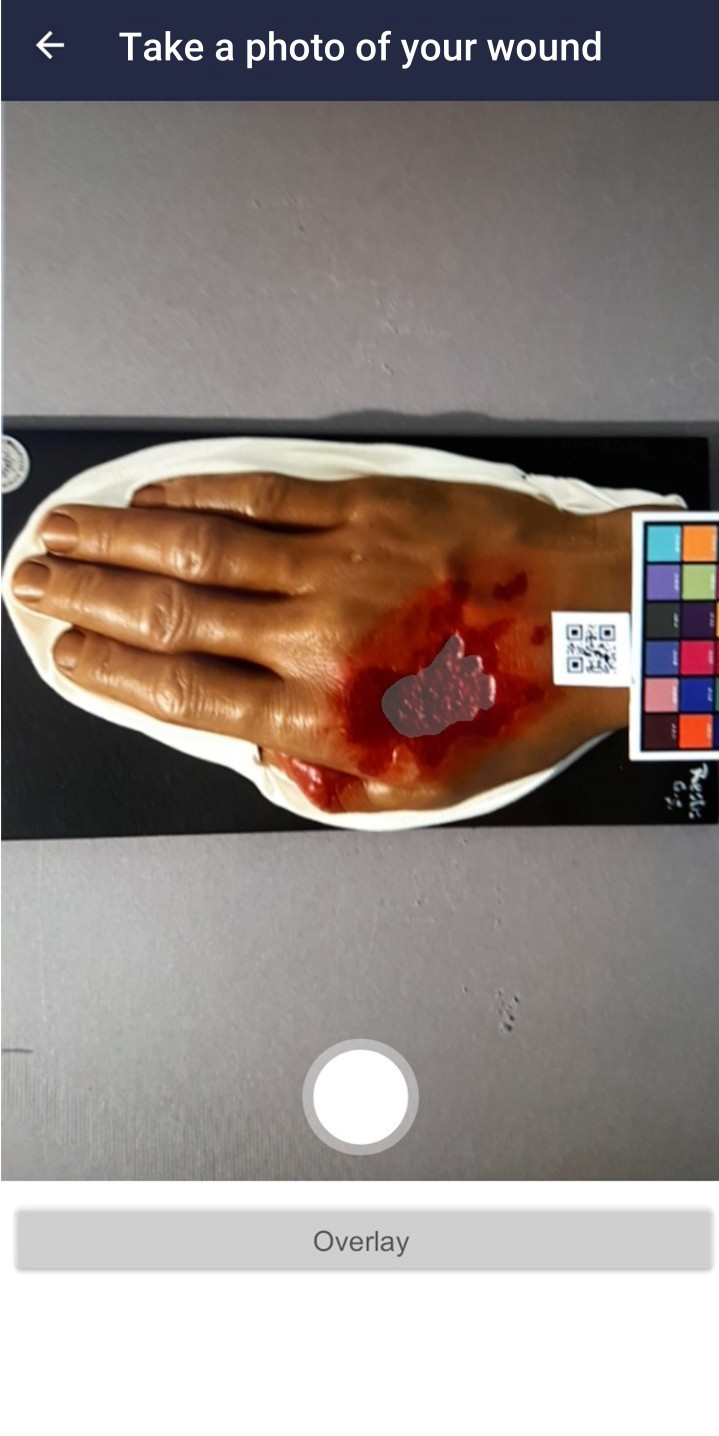}
        \caption{Camera screen with overlay}
        \label{fig:appendix:low_fidelity:camera_screen}
    \end{subfigure} \hfill
        \begin{subfigure}{0.3\textwidth}
        \centering
        \includegraphics[width=0.76\linewidth]{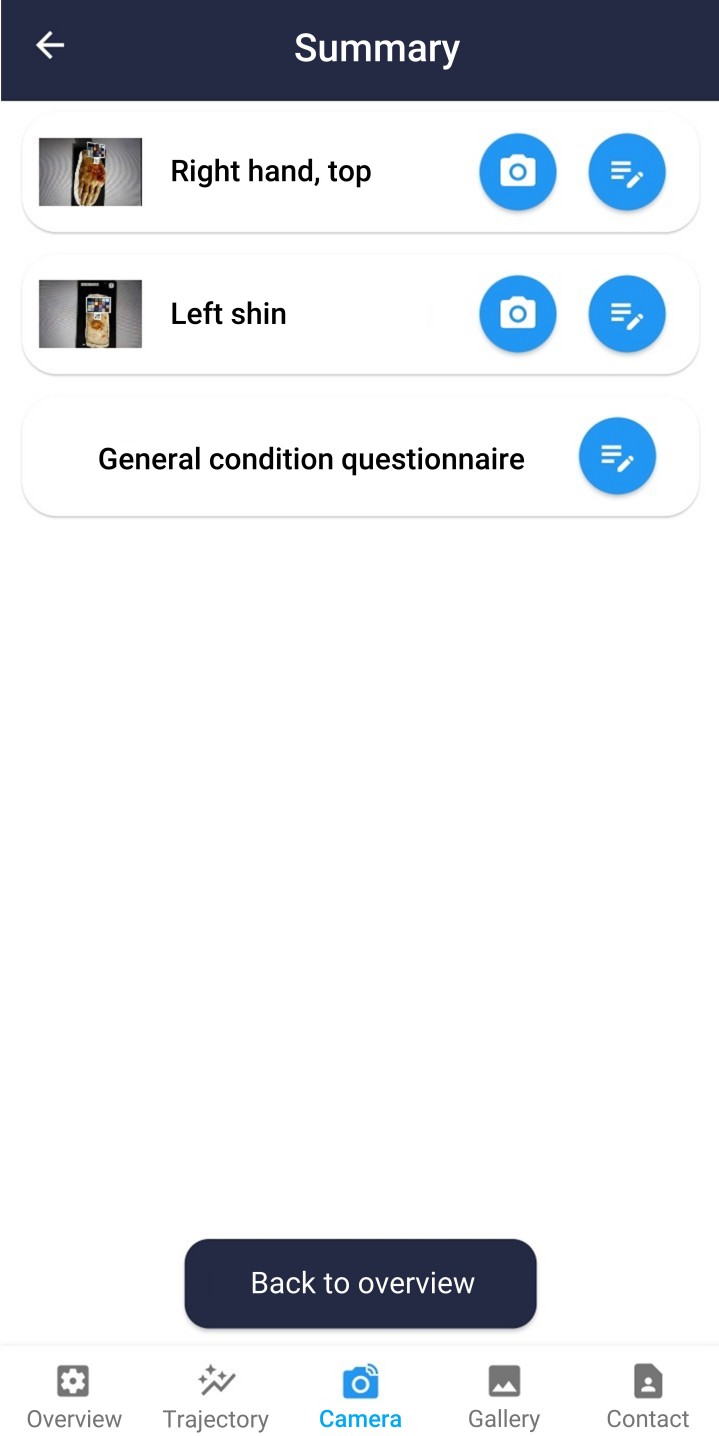}
        \caption{Summary \textit{(w/o help icon)}}
        \label{fig:appendix:low_fidelity:summary_view}
    \end{subfigure} \hfill
    \begin{subfigure}{0.3\textwidth}
        \centering
        \includegraphics[width=0.76\linewidth]{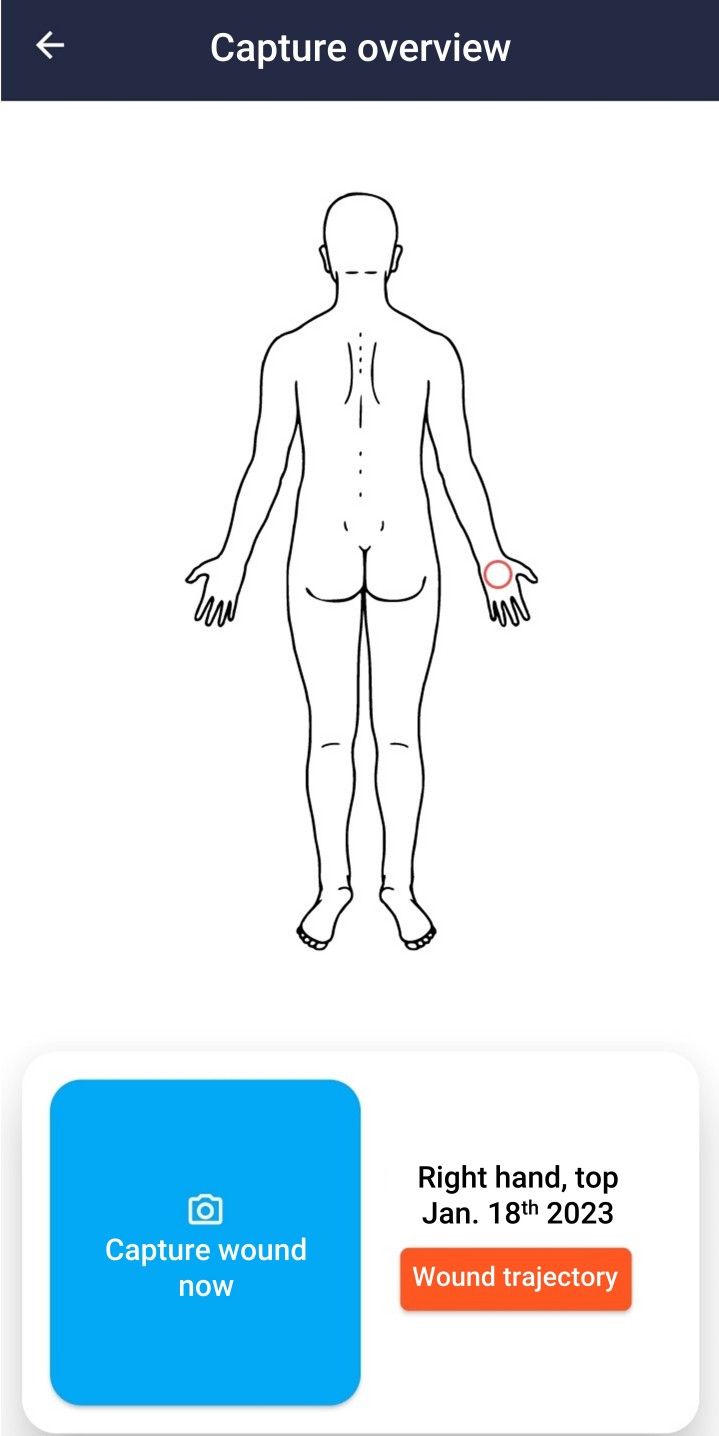}
        \caption{Wound localization (back)}
        \label{fig:appendix:low_fidelity:wound_localization}
    \end{subfigure} 
    \caption{Excerpt of screenshots of our low-fidelity prototype.}
    \label{fig:appendix:low_fidelity:screenshots_app}
    \Description{Seven app screenshots mainly showing functionalities of the low-fidelity prototype that underwent the most changes: the calendar function with scheduled appointments; separate trajectories for wound-specific parameters and overall condition without legends; the image gallery with previews of captured images but without counters; the previous camera interface with an overlay during image capture; the summary view shown after completing documentation (still without a help icon in this example); and the wound localization feature highlighting a wound on the back of the body.}
\end{figure*}

For completeness, we include screenshots of our digital low-fidelity prototype used in the initial usability study in Figure~\ref{fig:appendix:low_fidelity:screenshots_app}. We mainly focus on the functionalities that have changed the most, omitting some of the screens that remained largely unchanged in the transition from low-fidelity to high-fidelity prototype (e.g., overview, questionnaires, video chat). 

\begin{landscape}
    \begin{figure}[hp]
        \includegraphics[width=\linewidth,keepaspectratio,trim={0 10 20 0},clip]{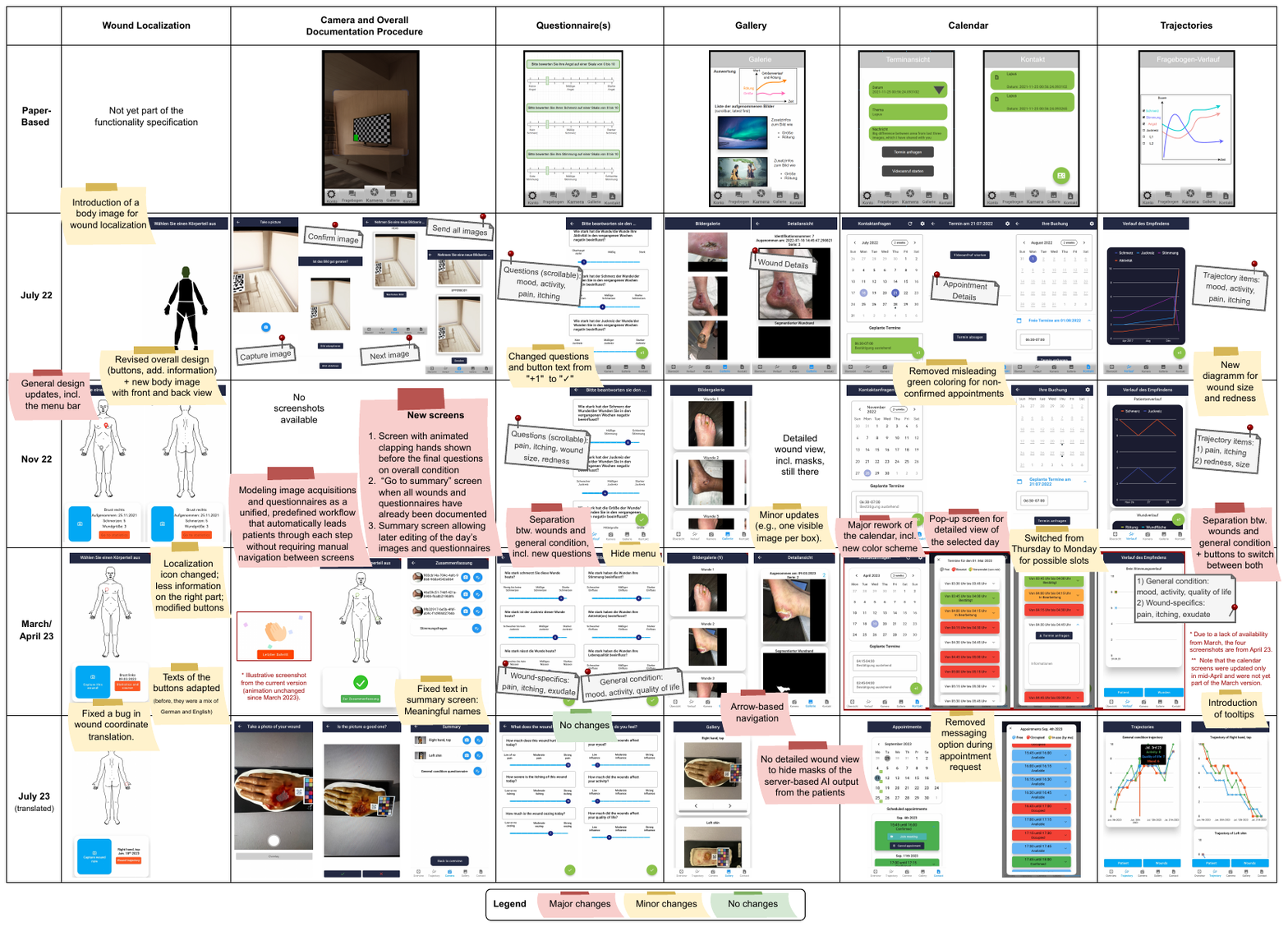}
        \caption{Evolution of the low-fidelity prototype. Due to the level of detail, digital viewing with zoom is recommended.}
        \label{fig:appendix:woundAIssist_over_time}
        \Description{
        Selected screenshots illustrating the iterative development of \AppName over an approximately 1.5-year early design phase, from initial paper-based concepts to the functional low-fidelity prototype completed in July~2023. The screenshots show the progressive refinement of functionality, user interface elements, and navigation structures. Screens are shown in German, except for the final prototype, which is presented in English; differently colored annotation boxes highlight key interface elements and changes.
        }
    \end{figure}
\end{landscape}

In addition to this, we present selected screenshots illustrating the iterative development of \AppName over the approximately 1.5-year early development phase in Figure~\ref{fig:appendix:woundAIssist_over_time}, complementing the textual description in Table~1 of the main article. The figures visualize the transition from initial paper-based concepts to the functional low-fidelity prototype finalized in July 2023. Across successive development stages, the screenshots illustrate how functionalities, user interface elements, and navigation structures were progressively refined to improve usability and better support patient workflows, informed by interdisciplinary feedback from dermatologists and HCI researchers.
All screenshots are presented in their original German version, with the exception of the final low-fidelity prototype from July 2023, which has been translated into English for clarity. To enhance readability, differently colored annotation boxes provide brief hints emphasizing the main elements and content of the corresponding German screen.

\let\cleardoublepage\clearpage

\let\cleardoublepage\clearpage
\newgeometry{top=2cm,bottom=2cm,left=2cm,right=2cm} 
\pagestyle{empty}    

\restoregeometry

\let\cleardoublepage\clearpage

\section{Additional information on administered questionnaires and statistical test requirements}
\label{appendix:additional_info_high_fidelity_eval}

\subsection{Additional information on the questionnaires used}
\label{appendix:ssec:additional_info_questionnaires}
Table \ref{quesi} provides detailed information on all measures, reliability scores, and exemplary items.

\begin{table*}[ht]
\caption{Rating scales, reliability scores, and exemplary items for all administered questionnaires.}\label{quesi}
\fontsize{8pt}{10pt}\selectfont
\begin{tabular}{llllll}
\toprule
\multicolumn{2}{l}{Questionnaire} & Scale & \multicolumn{2}{c}{Reliability}                                                                       & Exemplary Item                                                                                                               \\ \cmidrule(l{5pt}r{5pt}){4-5}
                                                                      &&                & Original & WAI
                                                                      &                                                                                                                                       \\ \midrule
\multicolumn{2}{c}{SUS-DE}                          & 1-5            & .74 - .88\textsuperscript{a}                                              & -.05\textsuperscript{a}          & {\tablesubsize ``I found the system easy to use.''}                                                                                                     \\\midrule
\multirow{7}{*}{\rotatebox{90}{MARS-G}}       & Engagement          & 1-5            & .85\textsuperscript{b}                                                    & .65\textsuperscript{a}           & {\tablesubsize ``Is the app fun/entertaining to use?''}                                                                                            \\
                              & Functionality       & 1-5            & .91\textsuperscript{b}                                                    & .57\textsuperscript{a}           & {\tablesubsize ``How easy is it to learn how to use the app?''}                                                                                   \\
                              & Aesthetics          & 1-5            & .93\textsuperscript{b}                                                    & .52\textsuperscript{a}           & {\tablesubsize ``How good does the app look?''}                                                                                        \\
                              & Information Quality & 1-5            & .72\textsuperscript{b}                                                    & .45\textsuperscript{a}           & {\tablesubsize ``Does the app contain what is described?''}                                                                                               \\
                              & Overall             & 1-5            & .82\textsuperscript{b}                                                    & .77\textsuperscript{a}           & -                                                                                                                                     \\
                              & Subjective Quality  & 1-5            & -                                                                         & .50\textsuperscript{a}           & {\tablesubsize ``Would you recommend this app to people who might benefit from it?''}                   \\
                              & Therapeutic Gain    & 1-5            & -                                                                         & .12\textsuperscript{a}           & {\tablesubsize ``Possible risks, side effects and harmful effects?''}                                                                                \\\midrule
\multirow{2}{*}{\rotatebox{90}{TAM}}          & PU  - A posterori   & 1-7            & .98\textsuperscript{a}                                                    & .87\textsuperscript{a}           & {\tablesubsize ``Using the display functionality would make it easier for me to assess the photo quality.''} \\
                              & PEU - A posterori   & 1-7            & .94\textsuperscript{a}                                                    & -.12\textsuperscript{a}          & {\tablesubsize ``It would be easy for me to learn how to use the display functionality.''}                   \\\midrule
\multirow{2}{*}{\rotatebox{90}{TAM}}          & PU - Live           & 1-7            & .98\textsuperscript{a}                                                    & .78\textsuperscript{a}           & {\tablesubsize ``Using the camera functionality would make it easier for me to assess the photo quality.''}   \\
                              & PEU - Live          & 1-7            & .94\textsuperscript{a}                                                    & .85\textsuperscript{a}           & {\tablesubsize ``It would be easy for me to learn how to use the camera functionality.''}                  \\\midrule
\multicolumn{2}{c}{ATI}                             & 1-6            & .83 - .94\textsuperscript{a}                                              & .85\textsuperscript{a}           & {\tablesubsize ``I like testing the functions of new technical systems.''}                                  \\
\bottomrule
\\[-2pt]
\multicolumn{6}{l}{
\begin{tabular}[c]{@{}l@{}l@{}l}
\textit{Note}. 
WAI = \AppName & PU = \textit{Perceived Usefulness} &  \hspace{1cm}  PEU = \textit{Perceived Ease of Use}. \\
\textsuperscript{a} Scale reliability assessed as Cronbach's $\alphaup$. \hspace{1cm} & \textsuperscript{b} Scale reliability assessed as $\omegaup$. & \\
\end{tabular}}                                                                                                                                                                       
\end{tabular}
\end{table*}

\subsection{Detailed assessment of statistical test assumptions}
\label{appendix:ssec:statistical_assumptions}
This section presents detailed information on how the assumptions underlying all statistical tests in the high-fidelity prototype evaluation were checked, along with the corresponding test procedures.

We conducted a one-way multivariate analysis of variance (MANOVA) to compare SUS-DE and MARS-G overall scores between dermatologists and patients, using the user version of the MARS-G for patient assessments.
Both multivariate normality (Shapiro-Wilk test: $p = .527$) and homogeneity of covariance matrices (Box's test: $p = .935$) were given. 
We also performed a MANOVA to compare TAM ratings of the a posteriori and live segmentation between physicians and patients. 
The Shapiro-Wilk test ($p < .001$) and Box's test ($p = .019$) indicated significance, however, MANOVA is considered relatively robust to violations of homogeneity of covariance matrices and normality assumptions given equal group sample size~\cite{ates2019}. 

In addition, we compared participants' overall \textit{Perceived Usefulness (PU)} and \textit{Perceived Ease of Use (PEU)} between the two segmentation feature versions using Mann-Whitney-U tests, applying Bonferroni-Holm correction. 

Furthermore, we compared patients' SUS-DE scores between the low-fidelity prototype 
and the high-fidelity version of \AppName using Student's t-test. Regarding the patients' and physicians' technical affinity, we also used a Student's t-test for comparison. For both Student's t-tests, requirements were fulfilled.

Using Student's t-tests, we compared our results on the SUS-DE, overall MARS-G, and ATI scales to the scores obtained by the patient-centered \textit{WUND APP} as reported in Dege et al.~\cite{dege2024}. 
Due to heterogeneous variances as indicated by Levene's test ($p = .015$), we applied Welch's t-test to compare SUS-DE scores. As noted by Rubin~\cite{rubin2021}, alpha adjustment is not necessary in the case of individual testing.

\end{document}